\documentclass[longauth]{aa}  

\usepackage{graphicx}
\usepackage{txfonts}
\usepackage[section]{placeins}

\usepackage{appendix}
\usepackage[font=small,labelfont=bf]{caption} 
\usepackage{float}
\usepackage{hyperref}  
\usepackage{xcolor}

\begin{document}

\title{ASTRODEEP-JWST: NIRCam-HST multi-band photometry and redshifts for half a million sources in six extragalactic deep fields}

\titlerunning{ASTRODEEP-JWST multi-band catalogue}
\authorrunning{Merlin et al.}

%\correspondingauthor{Emiliano Merlin}
%\email{emiliano.merlin@inaf.it}

\author{
E. Merlin \inst{1}
\and
P. Santini \inst{1}
\and
D. Paris \inst{1}
\and
M. Castellano \inst{1}
\and
A. Fontana \inst{1}
\and
T. Treu \inst{2}
\and
S. L. Finkelstein \inst{3}
\and
J. S. Dunlop \inst{4}
\and
P. Arrabal Haro \inst{5}
\and
M. Bagley \inst{3}
\and
K. Boyett \inst{6,7}
\and
A. Calabr\`o \inst{1}
\and
M. Correnti \inst{1,8}
\and
K. Davis \inst{9}
\and
M. Dickinson \inst{5}
\and
C. T. Donnan \inst{4}
\and
H. C. Ferguson \inst{10}
\and
F. Fortuni \inst{1}
\and
M. Giavalisco \inst{11} 
\and
K. Glazebrook \inst{12}
\and
A. Grazian \inst{13}
\and
N. A. Grogin \inst{10}
\and
N. Hathi \inst{10}
\and
M. Hirschmann \inst{14}
\and
J. S. Kartaltepe \inst{15}
\and
L. J. Kewley \inst{16}
\and
A. Kirkpatrick \inst{17}
\and
D. D. Kocevski \inst{18}
\and
A. M. Koekemoer \inst{10}
\and
G. Leung \inst{3}
\and
J. M. Lotz \inst{10}
\and
R. A. Lucas \inst{10}
\and
D. K. Magee \inst{19}
\and
D. Marchesini \inst{20}
\and
S. Mascia \inst{1}
\and
D. J. McLeod \inst{4}
\and
R. J. McLure \inst{4}
\and
T. Nanayakkara \inst{12}
\and
L. Napolitano \inst{1,21}
\and
M. Nonino \inst{22}
\and
C. Papovich \inst{23,24}
\and
L. Pentericci \inst{1}
\and
P. G. P\'erez-Gonz\'alez \inst{25}
\and
N. Pirzkal \inst{26}
\and
S. Ravindranath \inst{27,28}
\and
G. Roberts-Borsani \inst{29}
\and
R. S. Somerville \inst{30}
\and
M. Trenti \inst{6,7}
\and
J. R. Trump \inst{9}
\and
B. Vulcani \inst{13}
\and
X. Wang \inst{31,32,33}
\and
P. J. Watson \inst{13}
\and
S. M.~Wilkins \inst{34,35}
\and
G. Yang \inst{36,37}
\and
L. Y. A. Yung \inst{10}
}

\institute{
INAF -- Osservatorio Astronomico di Roma, Via Frascati 33, 00078 Monteporzio Catone, Rome, Italy\\ \email{emiliano.merlin@inaf.it} %1
\and
Department of Physics and Astronomy, University of California, Los Angeles, 430 Portola Plaza, Los Angeles, CA 90095, USA %2
\and
Department of Astronomy, The University of Texas at Austin, Austin, TX, USA %3 
\and
SUPA\thanks{Scottish Universities Physics Alliance}, Institute for Astronomy, University of Edinburgh, Royal Observatory, Edinburgh EH9 3HJ, UK %4
\and
NSF's National Optical-Infrared Astronomy Research Laboratory, 950 N. Cherry Ave., Tucson, AZ 85719, USA %5
\and
School of Physics, University of Melbourne, Parkville 3010, VIC, Australia %6
\and
ARC Centre of Excellence for All Sky Astrophysics in 3 Dimensions (ASTRO 3D), Australia %7
\and
Space Science Data Center, Italian Space Agency, via del Politecnico, 00133, Roma, Italy %8
\and
Department of Physics, 196A Auditorium Road, Unit 3046, University of Connecticut, Storrs, CT 06269, USA\thanks{NSF Graduate Research Fellow} %9
\and
Space Telescope Science Institute, Baltimore, MD, USA %10
\and
University of Massachusetts, 710 N. Pleasant St, LGRB-520, Amherst, MA 01003, USA %11
\and
Centre for Astrophysics and Supercomputing, Swinburne University of Technology, PO Box 218, Hawthorn, VIC 3122, Australia %12
\and
INAF -- Osservatorio Astronomico di Padova, Vicolo Osservatorio 5, 35122 Padova, Italy %13
\and
Institute of Physics, Lab for Galaxy Evolution, EPFL, Observatoire de Sauverny, Chemin Pegasi 51, 1290 Versoix, Switzerland %14
\and
Laboratory for Multiwavelength Astrophysics, School of Physics and Astronomy, Rochester Institute of Technology, 84 Lomb Memorial Drive, Rochester, NY 14623, USA %15
\and
Center for Astrophysics | Harvard \& Smithsonian, 60 Garden Street, Cambridge, MA 02138, USA %16
\and
Department of Physics and Astronomy, University of Kansas, Lawrence, KS 66045, USA %17
\and
Department of Physics and Astronomy, Colby College, Waterville, ME 04901, USA %18
\and
Department of Astronomy and Astrophysics, UCO/Lick Observatory, University of California, Santa Cruz CA 95064, USA %19
\and
Department of Physics \& Astronomy, Tufts University, MA 02155, USA %20
\and
Dipartimento di Fisica, Università di Roma Sapienza, Città Universitaria di Roma - Sapienza, Piazzale Aldo Moro, 2, 00185, Roma, Italy %21
\and
INAF -- Osservatorio Astronomico di Trieste, Via Tiepolo 11, I-34131 Trieste, Italy %22
\and
Department of Physics and Astronomy, Texas A\&M University, College Station, TX, 77843-4242 USA %23
\and
George P.\ and Cynthia Woods Mitchell Institute for Fundamental Physics and Astronomy, Texas A\&M University, College Station, TX, 77843-4242 USA %24
\and
Centro de Astrobiolog\'{\i}a (CAB), CSIC-INTA, Ctra. de Ajalvir km 4, Torrej\'on de Ardoz, E-28850, Madrid, Spain %25
\and
ESA/AURA Space Telescope Science Institute %26
\and
Astrophysics Science Division, NASA Goddard Space Flight Center, 8800 Greenbelt Road, Greenbelt, MD 20771, USA %27
\and
Center for Research and Exploration in Space Science and Technology II, Department of Physics, Catholic University of America, 620 Michigan Ave N.E., Washington DC 20064, USA %28
\and
Department of Astronomy, University of Geneva, Chemin Pegasi 51, 1290 Versoix, Switzerland %29
\and
Center for Computational Astrophysics, Flatiron Institute, 162 5th Avenue, New York, NY 10010, USA %30
\and
School of Astronomy and Space Science, University of Chinese Academy of Sciences (UCAS), Beijing 100049, China %31
\and
National Astronomical Observatories, Chinese Academy of Sciences, Beijing 100101, China %32
\and
Institute for Frontiers in Astronomy and Astrophysics, Beijing Normal University, Beijing 102206, China %33
\and
Astronomy Centre, University of Sussex, Falmer, Brighton BN1 9QH, UK %34
\and
Institute of Space Sciences and Astronomy, University of Malta, Msida MSD 2080, Malta %35
\and
Kapteyn Astronomical Institute, University of Groningen, P.O. Box 800, 9700 AV Groningen, The Netherlands %36
\and
SRON Netherlands Institute for Space Research, Postbus 800, 9700 AV Groningen, The Netherlands %37
}

\date{Received XXX; accepted XXX}

\newpage

\abstract
{}
{We present a set of photometric catalogues primarily aimed at providing the community with a comprehensive database for the study of galaxy populations in the high-redshift Universe. The set gathers data from eight \textrm{JWST} NIRCam observational programs, targeting the Abell 2744 (GLASS-JWST, UNCOVER, DDT2756, and GO3990), EGS (CEERS), COSMOS and UDS (PRIMER), and the GOODS North and South (JADES and NGDEEP) deep fields. This dataset covers a  total area  of $\simeq$0.2 sq. degrees.}
{ We obtained photometric estimates by means of well-established techniques, including tailored improvements designed to enhance the performance on the specific dataset. We also included new measurements from \textrm{HST} archival data, spanning 16 bands from 0.44 to 4.44 $\mu$m.} 
{A grand total of $\sim$530 thousand sources were detected on stacks of NIRCam 3.56 and 4.44 $\mu$m mosaics. 
We assessed the photometric accuracy by comparing fluxes and colours against archival catalogues. We also provide photometric redshift estimates, statistically validated against a large set of robust spectroscopic data.} 
{The catalogues are publicly available on the \textsc{Astrodeep} website.}

\keywords{galaxies: high redshift, galaxies: photometry}

\maketitle

\section{Introduction}
\label{sec:intro}

Two years since its first light, the \textit{James Webb} Space Telescope \citep[\textrm{JWST};][]{Gardner2006,Gardner2023} has provided a vast store of cutting-edge, quality data to many fields of astrophysical research. In particular, the study of the high-redshift Universe has benefitted from the joint effort of researchers designing tailored observational programmes and exploiting their outcomes, with an outburst of activity within the very first weeks after the arrival of the first data \citep[see e.g.][for a review]{Adamo2024}. 

The unmatched depth and resolution of \textrm{JWST} infrared imaging and spectroscopy has enabled a wealth of analysis at intermediate and high redshifts that were beyond the reach of previous instruments. Examples include the measurement of  optical rest-frame morphologies \citep[e.g.][]{Ferreira2022,Ferreira2023,Jacobs2023,Treu2023,Kartaltepe2023}, stellar masses \citep[e.g.][]{Santini2023,Weibel2024,Wang2024}, and star-formation histories \citep[e.g.][]{Dressler2023,Looser2023,Ciesla2024,Conselice2024} of galaxies up to $z\simeq10$, to the abundance and properties of red and optically dark sources \citep[e.g.][]{Glazebrook2023,Kirkpatrick2023,PerezGonzalez2023,Rodighiero2024} as well as quiescent galaxies \citep[e.g.][]{Carnall2023,Nanayakkara2024,Wright2024,Ward2024} up to $z\sim6$, the properties of compact red sources \citep{PerezGonzalez2024,Williams2024,Kokorev2024}, and the identification of extreme emission line galaxies in the epoch of reionization  \citep[e.g.][]{Davis2023}.
In particular, \textrm{JWST} has challenged our view of the early epochs of the cosmos. A number of exciting discoveries on the first phases of galaxy formation and evolution have resulted in more questions than answers, as we have yet to frame and understand our observations within a fully consistent theoretical framework.
The most consolidated result so far is the evidence of a striking overabundance of bright galaxies at $z\gtrsim 9-10$ compared to most predictions \citep[e.g.][]{Castellano2022,Castellano2023,Naidu2022,Finkelstein2023a,Finkelstein2024,McLeod2024,PerezGonzalez2023b,Carniani2024}. These distant sources have been mostly identified by means of colour selections or spectral energy distribution (SED) fitting, and the result is still a matter of debate, with numerous possible explanations proposed \citep[e.g.][]{Ferrara2023,Dekel2023,Mason2023,Trinca2024,Padmanabhan2023,Yung2024,Harvey2024}. While spectroscopic follow-up is crucial to understand the physical processes at play -- and, indeed, they currently seem to confirm the early results \citep[e.g.][]{ArrabalHaro2023,Harikane2024,Castellano2024}, photometric data still constitute the primary way to collect statistically significant samples, beyond providing targets for spectroscopy.

In this paper, we present our analysis of eight deep-sky NIRCam observational programmes: CEERS, DDT2756, GLASS-JWST, GO3990, JADES, NGDEEP, PRIMER, and UNCOVER. We used these data to  create a new set of photometric catalogues, mainly finalised to provide a consistent database for the study of the early phases of galaxy evolution in the high-redshift Universe ($z\geq3$). The programmes target six of the most well-known and studied areas of the sky, which have already been observed with \textrm{HST} and ground-based facilities \citep[in particular, the CANDELS and Frontier Fields campaigns; see][]{Grogin2011,Koekemoer2011,Lotz2017}, and have been subject of many ground-breaking studies in the past decades: (i) the GLASS-JWST, UNCOVER, DDT2756 and GO3990 programmes cover the Abell 2744 cluster of galaxies, also known as Pandora's cluster, and its surrounding area. For simplicity, we will refer to this extended region as ABELL2744; (ii) the CEERS survey overlaps with the Extended Groth Strip (EGS) field \citep[][]{Davis2007}; (iii-iv) the PRIMER programme covers two areas, one overlapping with the UKIDSS Ultra-deep Survey field \citep[UDS, ][]{Lawrence2007} and the other with the COSMOS field \citep[][]{Nayyeri2017}; (vi) JADES and NGDEEP overlap with the GOODS-North and ECDFS/GOODS-South regions
\citep[][]{Giacconi2002,Giavalisco2004}. 
The total area observed by these programmes is $\simeq$0.2 sq. degrees. While some catalogues for these observations are already available \citep[e.g.][]{Paris2023,Rieke2023,Weaver2024}, the aim of this new dataset is to provide a large, self-consistent database mainly aimed at the study of the high-redshift Universe. Therefore, we mostly focused our attention on the detection of faint sources when choosing detection parameters. Unsurprisingly, this choice can lead to some degree of tension when comparing our results to other catalogues (see Sect. \ref{sec:valid}).

The new catalogues are obtained with well-tested algorithms and techniques, largely building upon previous releases of our group \citep[][respectively M22 and P23 hereafter]{Merlin2022,Paris2023}, but with substantial improvements, which we describe in Sect.~\ref{sec:meth}. Up-to-date data and calibration files have been used whenever available (detailed information is provided). We also estimated photometric redshifts, which we obtained using the SED-fitting software packages \textsc{zphot}, first described in \citet{Fontana2000}, and \textsc{EAzY} \citep{Brammer2008}, for which we used three different sets of templates. Predictably, comparing the results from these four runs we find that they typically are in good agreement for high signal-to-noise (S/N) sources, while faint objects are less well constrained, often resulting in divergent fits.
Finally, we validated our results comparing them to the literature. We find substantial agreement with other photometric catalogues and good statistics when checking the photo-$z$  results against the spectroscopic data.

The paper is organised as follows. In Sect. \ref{sec:data} we describe the dataset. In Sect. \ref{sec:meth} we summarise the adopted techniques, referencing previous publications when useful and highlighting the differences with previous works. In Sect. \ref{sec:valid} we discuss the validation of our new catalogues against published data, considering direct comparisons of colours. In Sect. \ref{photoz}, we discuss the photometric redshifts. In Sect. \ref{sec:concl}, we make some final remarks and conclusions. The catalogue format is described in Appendix A (available on Zenodo; see Sect. \ref{dataav}). We use AB magnitudes \citep{Oke83} and we assume an $\Lambda$CDM cosmology ($H_0=70$ km/s/Mpc, $\Omega_m=0.27,$ and $\Omega_{\Lambda}=0.73$). %The catalogues are publicly available on the \textsc{Astrodeep} website\footnote{\texttt{https://astrodeep.eu/astrodeep-jwst-catalogs/}}.

\section{Dataset}
\label{sec:data}

In this section. we summarise the properties of the composite dataset we used to obtain the photometric catalogues presented in this work. For all the fields but ABELL2744, we collected the NIRCam 30mas mosaics created by the teams of each programme.
In most cases, ancillary HST images were also made available; when needed, we re-projected them on the same grid of the NIRCam mosaics, and checked for astrometric consistency (see Sect. \ref{align}). All of the images were also scaled to $\mu$Jy units, so that AB magnitudes can be obtained from the flux measurements applying a constant zero-point (ZP) of 23.9. In the following, we detail the process case by case; because of the composite nature of the dataset, which comprises images reduced by different teams at different times, the specific parameters of the reduction processes were unavoidably non-uniform. The different calibration files and pipeline versions used in this study are summarised in Table \ref{caltab}\footnote{See the official \textrm{JWST} pipeline web-page for detailed information on the keywords, \texttt{https://jwst-pipeline.readthedocs.io/en/latest/}}.

In order to create a formally homogeneous set of catalogues, we selected a fixed set of pass-band filters, thereby also facilitating the photo-$z$ estimates for which tailored libraries of models are required. The chosen bands are: \textrm{HST} ACS F435W, F606W, F775W and F814W; \textrm{HST} WFC3 F105W, F125W, F140W and F160W; and NIRCam F090W, F115W, F150W, F200W, F277W, F356W, F410M, and F444W. 
The main features of these pass-bands are summarised in Table \ref{tab1}. We point out that not all of the bands are available in all fields; we provide the relevant details in the following subsections. The areas given for each field are indicative, as many pass-bands have different sky coverage; the values refer to the F444W band area.

\begin{table}
\caption{Details of the pipeline versions and calibration files used for each dataset.} \label{caltab}
\centering
\begin{tabular}{lllll} 
 \hline
 \hline
 Dataset & DR & \texttt{CRDS\_VER} & \texttt{CRDS\_CTX} & \texttt{CAL\_VER} \\
 \hline
 ABELL2744\tablefootmark{a} & 2.0 & 11.17.2 & 1183 & 1.11.3\\ 
 CEERS DR0.5\tablefootmark{b} & 0.5 & 11.16.14 & 0989 & 1.7.2 \\ 
 CEERS DR0.6\tablefootmark{b} & 0.6 & 11.16.16 & 1023 & 1.8.5 \\ 
 JADES-GN & 1.0 & 11.17.6 & 1130 & 1.11.4 \\ 
 JADES-GS & 2.0 & 11.17.6 & 1132 & 1.11.4 \\ 
 NGDEEP SW & 0.2b & 11.16.15 & 1045 & 1.9.2 \\ 
 NGDEEP LW & 0.3 & 11.17.0 & 1084 & 1.10.2 \\ 
 PRIMER- & & & & \\
 COSMOS & 0.8 & 11.17.0 & 1123 & 1.10.2\\ 
 PRIMER-UDS & 0.6 & 11.17.0 & 1118\tablefootmark{c} & 1.10.2\\ 
 \hline
 \end{tabular}
 % \multicolumn{5}{l}{*GLASS-JWST, UNCOVER, DDT2756 and GO3390}\\
 % \multicolumn{5}{l}{**CEERS full field mosaics are constructed from two}\\
 % \multicolumn{5}{l}{ separate public data releases: DR0.5 for NIRCam}\\
 % \multicolumn{5}{l}{ pointings 1,2,3,6 and DR0.6 for NIRCam pointings }\\
 % \multicolumn{5}{l}{4,5,7,8,9,10 (DOI 10.17909/z7p0-8481)}\\
 % \multicolumn{5}{l}{***\texttt{pmap} 1117 for F356W, F410M, F444W}
\tablefoot{
\tablefoottext{a}{GLASS-JWST, UNCOVER, DDT2756, and GO3390.}
\tablefoottext{b}{CEERS full field mosaics are constructed from two separate public data releases: DR0.5 for NIRCam pointings 1,2,3,6 and DR0.6 for NIRCam pointings 4,5,7,8,9,10 (DOI 10.17909/z7p0-8481).}
\tablefoottext{c}{\texttt{pmap} 1117 for F356W, F410M, and F444W.}
} 
\end{table}

\begin{table}
\caption{Main features of the pass-band filters included in the catalogues.} \label{tab1}
\centering
\begin{tabular}{llllll} 
\hline\hline
Filter & $\lambda_{\mbox{ref}}$ & $\lambda_{\mbox{mean}}$ & $\lambda_{\mbox{eff}}$ & FWHM & $fr_{0.2}$ \\
\hline
\multicolumn{6}{c}{HST ACS} \\
%\hline
F435W & 432.9 & 436.0 & 434.2 & 0.112 & 0.657\\
F606W & 592.2 & 603.6 & 580.9 & 0.122 & 0.629\\
F775W & 769.3 & 773.1 & 765.2 & 0.111 & 0.638\\
F814W & 804.6 & 812.9 & 797.3 & 0.100 & 0.541\\
%\hline
\multicolumn{6}{c}{HST WFC3} \\
%\hline
F105W & 1055.0 & 1065.1 & 1043.1 & 0.162 & 0.371\\
F125W & 1248.6 & 1257.6 & 1236.4 & 0.181 & 0.359\\
F140W & 1392.3 & 1406.2 & 1373.5 & 0.178 & 0.320\\
F160W & 1537.0 & 1543.6 & 1527.8 & 0.182 & 0.310 \\
%\hline
\multicolumn{6}{c}{JWST NIRCam} \\
%\hline
F090W & 902.2 & 908.3 & 898.5 & 0.056 & 0.701\\
F115W & 1154.3 & 1162.4 & 1143.4 & 0.059 & 0.716\\
F150W & 1659.2 & 1786.6 & 1479.4 & 0.059 & 0.714\\
F200W & 1988.6 & 2002.8 & 1968.0 & 0.073 & 0.686\\
F277W & 2761.7 & 2784.5 & 2727.9 & 0.124 & 0.602\\
F356W & 3568.4 & 3593.4 & 3528.7 & 0.146 & 0.553\\
F410M & 4082.2 & 4088.7 & 4072.3 & 0.155 & 0.516\\
F444W & 4404.3 & 4439.4 & 4350.4 & 0.166 & 0.496\\
\hline
 \end{tabular}
\tablefoot{$fr_{0.2}$ is the fraction of the flux within a circular aperture of diameter 0.2" for point sources. Wavelengths are given in nanometers; FWHMs are given in arcseconds.}
\end{table}

\subsection{ABELL2744}
Imaging data from four programmes were combined in a single set of mosaics: GLASS-JWST \citep[ERS 1324, P.I. Treu; no F410M,][]{Treu2022}, UNCOVER \citep[GO 2561, P.I. Labb\'e; no F090W,][]{Bezanson2022}, DDT 2756 (P.I. Chen; no F090W and F410M), and GO 3990 (P.I. Morishita; no F410M). 
The resulting combined field of view (FoV) is centred on the galaxy cluster, and covers an area of $\simeq$45.7 sq. arcmin. With respect to P23, new data have been received and added: the GO 3990 images in the UNCOVER region and a new set of observations of the GLASS-JWST region acquired in July 2023 to correct the original 2022 images that were affected by a wing-tilt event in the short-wavelength bands (SW hereafter, i.e. F090W, F115W, F150W, and F200W; the redder bands are denoted long-wavelength, LW hereafter). 
The reduction has been re-done from scratch with new calibration files, following the procedure described in P23. 
The raw \texttt{uncal} images have been retrieved from the MAST archive\footnote{\texttt{https://mast.stsci.edu/}}, and combined to \texttt{cal} images by applying the first two stages of the official JWST calibration pipeline (\texttt{calwebb\_detector1} and \texttt{calwebb\_image2}) with the latest calibration and reference files available to date (see Table~\ref{caltab}). We then applied our modified version of the official pipeline using a number of custom procedures developed by our team to correct for defects: `snowballs', non-linear pixels, 1/$f$-noise, `wisps' and `claws'. We aligned the calibrated images to Gaia-DR3 astrometry, and finally we combined them into mosaics and ancillary weight and error (root mean square, RMS) maps, (see P23 Sect. 2.2.1 for details). We then complemented the NIRCam dataset with archival \textrm{HST} mosaics, reduced, and publicly released by G. Brammer\footnote{\texttt{https://s3.amazonaws.com/grizli-v2/JwstMosaics/\\v4/index.html}}. A global background subtraction on all the mosaics was performed  with \textsc{SExtractor} \citep{Bertin1996}.

We point out that this field is different from all the others because it is centred on a galaxy cluster. Working on this same region for the Frontier Fields campaign, \citet{Merlin2016a} developed a sophisticated technique to accurately subtract bright foreground galaxies and intra-cluster light from all the analysed bands. However, such a technique is complex and time-consuming, so we postpone its application to future work. As mentioned in P23, the global background and 1/$f$-noise subtraction techniques effectively remove most of the intra-cluster light from the images. Furthermore, we provide photometric estimates including local background subtraction (see Sect. \ref{photometry}), which effectively removes residual intra-cluster light while also mitigating spurious effects created by the global processing. Nevertheless, we warn  users that the sources close to the centre of the cluster should be treated with caution, since their photometry might be affected by the residual light of the cluster members or by artefacts caused by the global background subtraction (see also Sect. \ref{bkgsub}). Also, potential very-high-redshift objects magnified by (but close to) the cluster centre are currently impossible to detect.

\subsection{CEERS}
Data from the CEERS programme \citep[ERS 1345, P.I. Finkelstein; no F090W,][]{Finkelstein2022} cover an area of $\simeq$94.6 sq. arcmin. Images have been reduced by M. Bagley combining two epochs of observations \citep[see][]{Bagley2023}. The individual pointings are available from CEERS public releases 0.5 and 0.6\footnote{\texttt{https://ceers.github.io/dr05.html}, \texttt{https://ceers.github.io/dr06.html}} and at MAST as High Level Science Products via DOI 10.17909/z7p0-8481. We have drizzled all ten individual pointings from these two releases into single mosaics for this paper.
For \textrm{HST} we used the EGS dataset from CANDELS \citep[][no F435W and F775W data]{Stefanon2017}, with the addition of the F105W band from the CEERS HDR1 reduction\footnote{\texttt{https://ceers.github.io/releases\#hdr1}}. A new catalogue based on a more recent, improved reduction of the NIRCam data, and obtained applying slightly different techniques, is going to be published soon by the CEERS team (Cox et al., in preparation).

\subsection{JADES and NGDEEP}
Mosaics from the JADES programme \citep[GTO 1180, P.I. Eisenstein, and GTO 1210, P.I. Luetzgendorf,][]{Eisenstein2023}, including additional data from the FRESCO programme (PID 1895, P.I. Oesch), cover an area of $\simeq$83.0 sq. arcmin in the GOODS-North region (JADES-GN hereafter), and of $\simeq$84.5 sq. arcmin in the GOODS-South region (JADES-GS hereafter). The images are available to the public. We used the v2.0 version for JADES-GS and the v1.0 version for JADES-GN. 
We complemented the NIRCam data using the Hubble Legacy Fields images \citep{Illingworth2016}.

The NGDEEP programme \citep[P.I. Finkelstein; no F090W and F410M,][]{Bagley2024} adds an area of $\simeq$9.5 sq. arcmin from the outer ECDFS area, with a marginal overlap with the GOODS-South field. Since the observed FoV does not overlap with JADES-GS, we kept the two fields separated, creating two catalogues and analysing them individually. We used the same imaging data of \citet{Leung2023}, which only includes the first epoch of observation. This first epoch  suffered from a lack of depth due to the DEEP8 readout pattern with a small number of groups. An updated catalogue based on both significant improvements to the first epoch and including the second epoch will be published in the near future (Leung et al., in preparation).

\subsection{PRIMER-COSMOS and PRIMER-UDS}
Data is from the PRIMER programme (GO 1837, P.I. Dunlop; no F775W for PRIMER-COSMOS and no F775W and F105W for PRIMER-UDS). The COSMOS FoV has an area of $\simeq$141.8 sq. arcmin, and the UDS FoV has an area of $\simeq$251.2 sq. arcmin. The images have been reduced by D. Magee, with ancillary \textrm{HST} data re-reduced from the CANDELS imaging, UVCANDELS \citep{WangXin2024}, and supplemental programme 16872 (P.I. Grogin).

\section{Methods}
\label{sec:meth}

In this section, we summarise the techniques and algorithms we used to prepare the images and to extract photometric information.

\subsection{Alignment and astrometry} \label{align}

\begin{figure}
\center
\includegraphics[width=0.225\textwidth,height=0.3\textwidth]{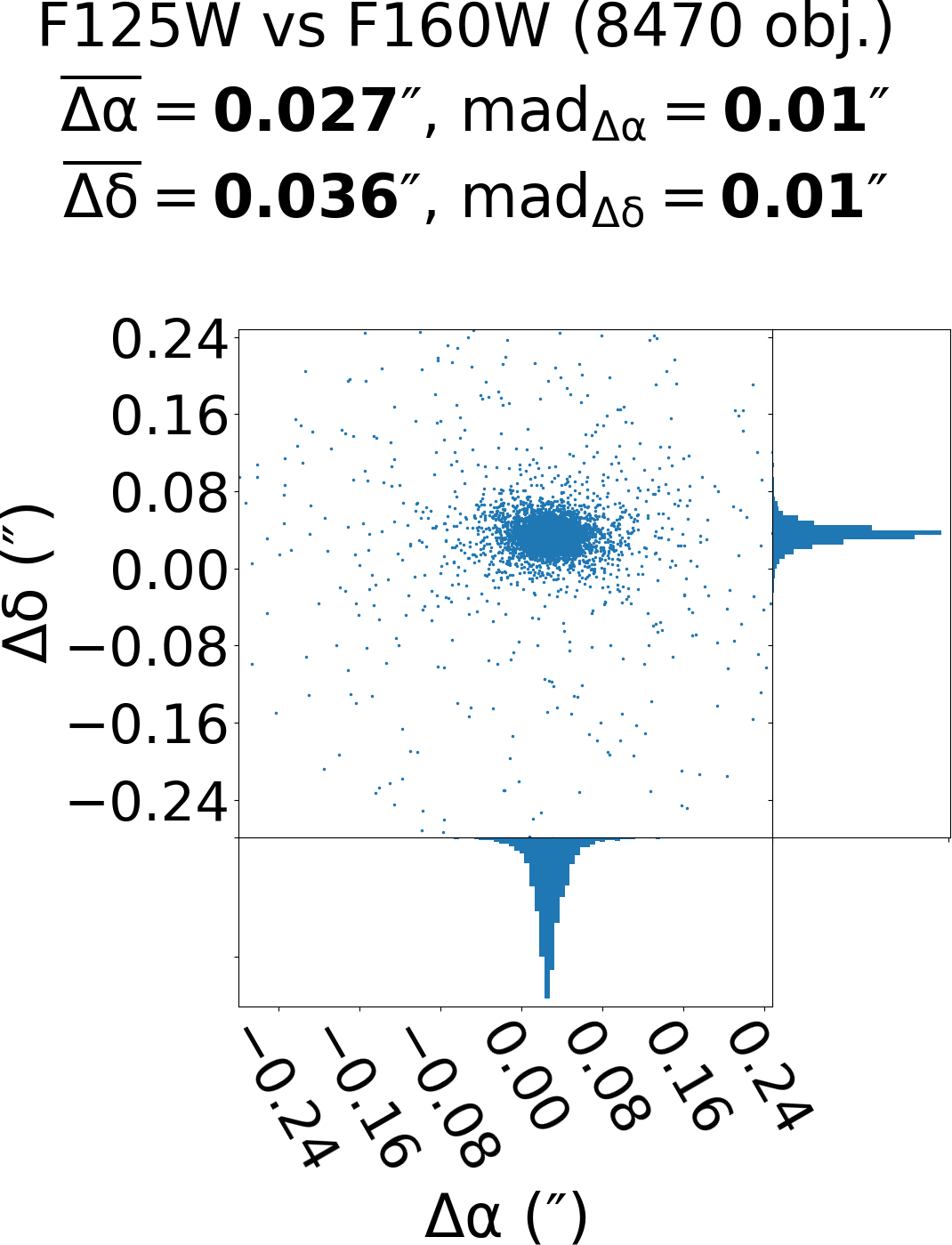}
\includegraphics[width=0.225\textwidth,height=0.3\textwidth]{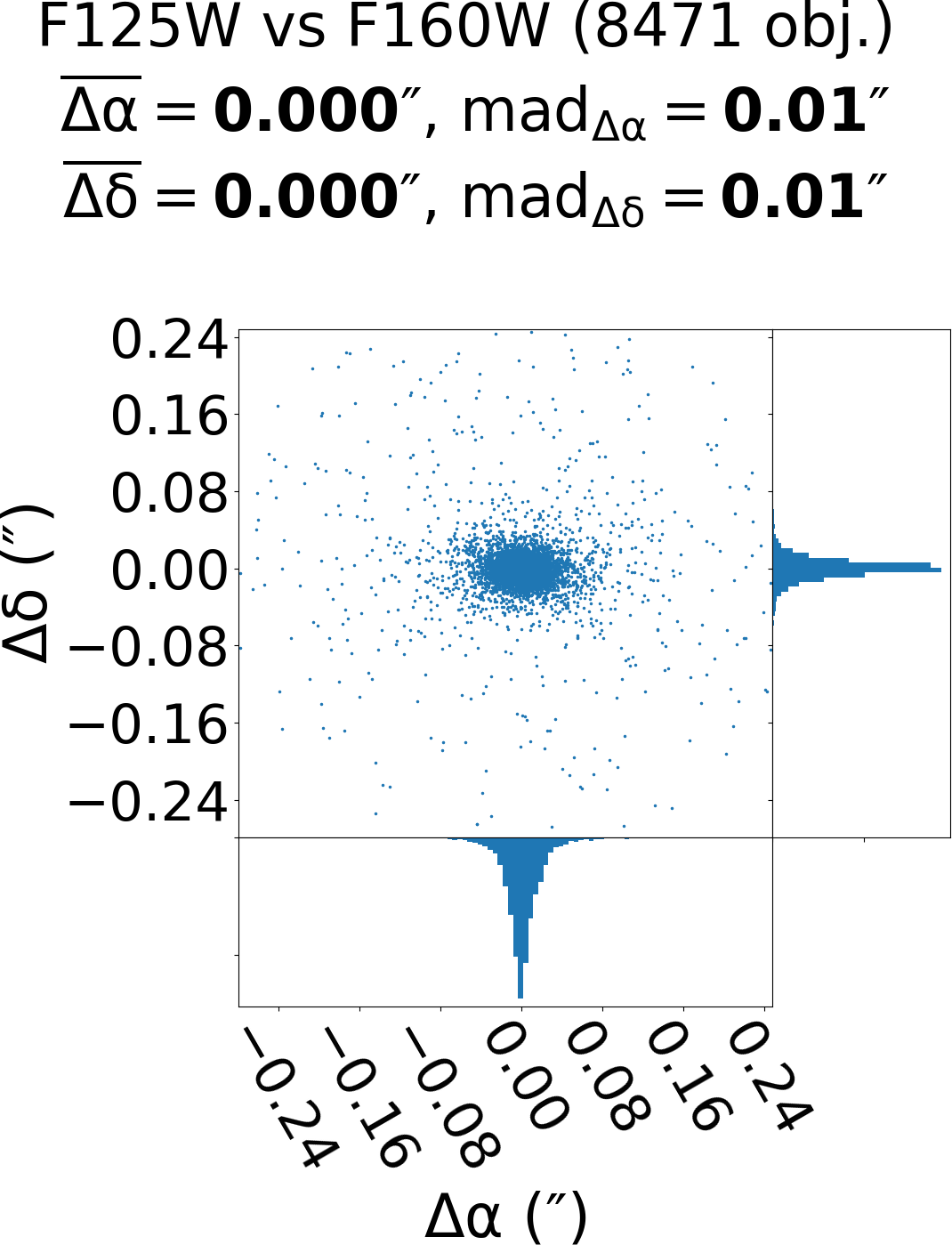}
\caption{Example of the validation tests made to fix the astrometric registration of the \textrm{HST} bands. Shown is the displacement $\Delta$RA and $\Delta$DEC between the sources detected in the CEERS F125W image and those detected in the CANDELS EGS F160W image, re-aligned to the reference frame of the CEERS F150W image, before (left panel) and after (right panel) applying the correction described in Sect. \ref{align}.}
 \label{fig:astrom}
\end{figure}

We assessed the quality of the astrometric registration by cross-matching the source coordinates  \texttt{ALPHAWIN\_J2000} and \texttt{DELTAWIN\_J2000} obtained by running \textsc{SExtractor} on all the bands in each field before re-projecting the images to the same common grid. In most cases we found small $\Delta$RA and $\Delta$DEC offsets (of the order of a few mas) between the native astrometry of the \textrm{JWST} mosaics (which are aligned to Gaia) and the \textrm{HST} bands. However, since in some cases the offsets were larger and not negligible, we applied a custom algorithm to correct all of them. First, we computed the offset between the \textrm{HST} WFC3 F160W band and the \textrm{JWST} NIRCam F150W band, and re-aligned the F160W image (we applied a rigid shift to the whole image). Then we corrected the offsets of the other \textrm{HST} bands by cross-matching the coordinates of the sources with those extracted from the re-aligned F160W band (see Fig. \ref{fig:astrom} and Table B.1). Finally, we re-projected all the images on the same NIRCam pixel grid.

\subsection{RMS scaling} \label{rmsscale}

Because the photometric errors were computed by means of the RMS maps (see Sect. \ref{photometry}), we checked that the latter were indeed representative of the true uncertainties of the measurement, using an improved version of the technique described in M22. In short, the RMS image of each band was subdivided into two to four complementary sub-regions of comparable exposure time, by means of an automatic algorithm applied to the weight maps. Then, in each of these sub-region, 300 artificial point sources (WebbPSF simulated point spread functions) were injected at random positions, excluding areas assigned to real sources by means of \textsc{SExtractor} segmentation maps. The dispersion of their fluxes \citep[measured within an aperture of 0.1'' using the software \textsc{a-phot},][]{Merlin2019} was compared to their nominal errors to obtain a re-scaling factor for that region of the RMS map. Finally, the original map (which includes Poissonian photon noise)  was then re-scaled by the median value of such factors. The typical values for SW bands are below $\sim$1.5, while they can get as high as $\sim$2.0 for some LW bands and fields.

\begin{figure*}
\sidecaption
\includegraphics[width=12cm]{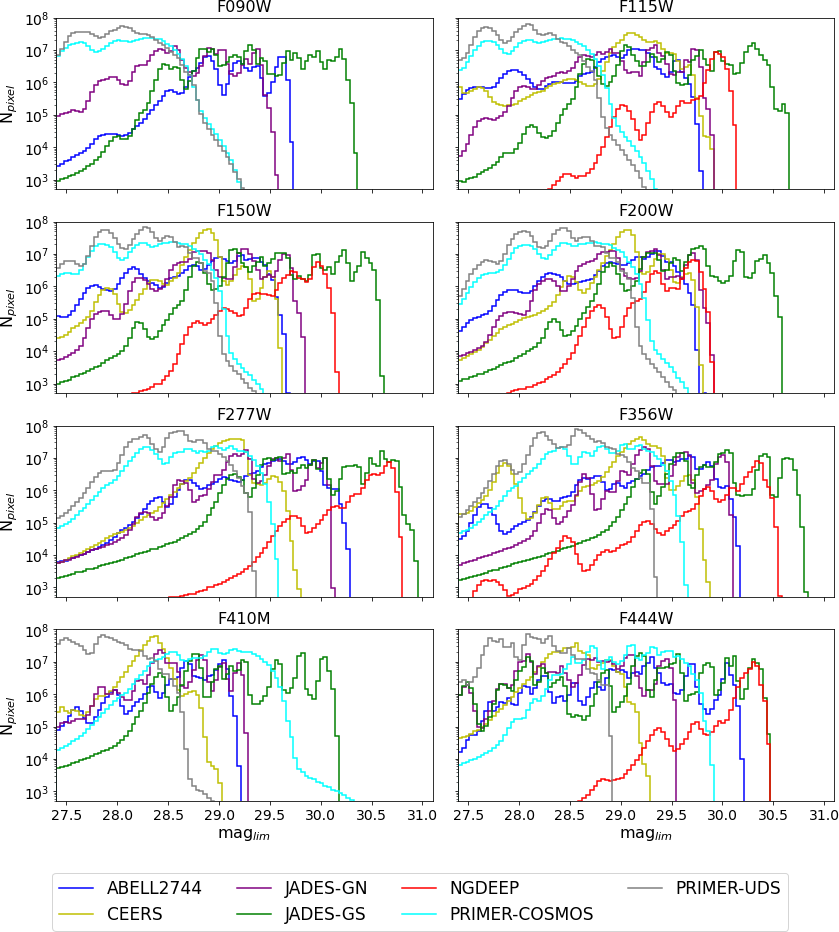} 
\caption{Histograms of the pixel distributions of limiting magnitudes (total at 5$\sigma$ in 0.2" diameter apertures), computed as described in Sect. \ref{rmsscale}, for all bands and fields.} \label{depth}
\end{figure*}

The histograms in Fig. \ref{depth} show the distribution of the limiting magnitudes (total at 5$\sigma$ in apertures of 0.2") for all the NIRCam bands, as obtained from the re-scaled RMS map pixel values by means of the formula: 
\begin{equation}
\mbox{depth}_{5\sigma,i}=-2.5\times \mbox{log}(5\times \sqrt{A}\times \mbox{RMS}_i / fr_{0.2}) + \mbox{ZP}, 
\end{equation}
\noindent where $i$ is a pixel index, ZP is the zero-point of the image (23.9 in our case), $A=\pi (0.5 \times 0.2/\mbox{ps})^2$ is the area of the circular aperture of 0.2" diameter (ps is the pixel scale), and $fr_{0.2}$ is the fraction of the flux of a point source enclosed in the aperture (see Table \ref{tab1}).

\begin{table}
\caption{\textsc{SExtractor} parameters used for the detection.} \label{sexpar}
\centering
\begin{tabular}{ll}    
\hline\hline
Parameter & Value \\ 
\hline
\texttt{DETECT\_MINAREA} & 5 \\ 
\texttt{DETECT\_THRESH} & 0.65/0.55/0.85\tablefootmark{a}  \\ % NB CEERS v1: 0.6
\texttt{ANALYSIS\_THRESH} & 0.65/0.55/0.85\tablefootmark{a}  \\ 
\texttt{DEBLEND\_THRESH} & 32 \\ 
\texttt{DEBLEND\_MINCONT} & 0.0003  \\ 
\texttt{MEMORY\_OBJSTACK} & 50,000\\ 
\texttt{MEMORY\_PIXSTACK} & 10,000,000\\ 
\texttt{MEMORY\_BUFSIZE} & 4096 \\ 
\hline
\end{tabular}
\tablefoot{
\tablefoottext{a}{ABELL2744: 0.55; PRIMER-COSMOS: 0.85; other fields: 0.65.}
}
\end{table}

\begin{table}
\caption{Number of detections in the six fields.} \label{detect}
\centering
\begin{tabular}{ll}    
\hline\hline
Field & Detections \\ \hline
ABELL2744 & 42,491 \\ 
CEERS & 82,547 \\ 
JADES-GN & 58,385 \\ 
JADES-GS & 73,638 \\ 
NGDEEP & 14,752 \\
PRIMER-COSMOS & 123,094 \\ 
PRIMER-UDS & 136,266 \\ 
Total & 531,173 \\ \hline
\end{tabular}
\end{table}

\subsection{Detection} \label{detection}

\begin{figure*}
\sidecaption
%\centering
%\includegraphics[width=0.85\textwidth]{figs/depths_001.png}
%\includegraphics[width=0.85\textwidth]
\includegraphics[width=12cm]{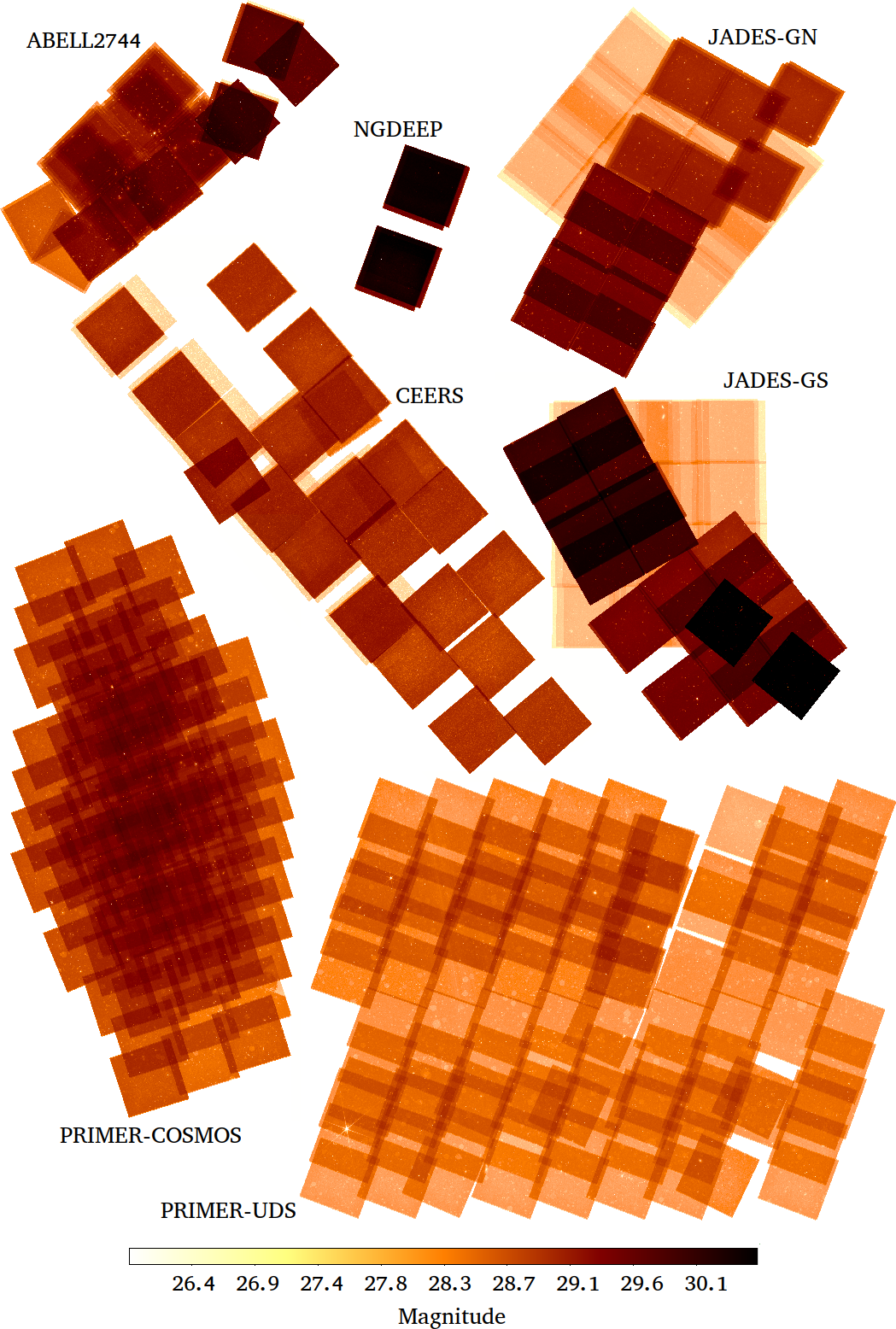}
\caption{Limiting magnitudes (total at 5$\sigma$ in 0.2" diameter apertures) of the detection F356W+F444W stack mosaics.}\label{depth001}
\end{figure*}

Because the catalogues have  mainly been designed for the study of the high-redshift Universe, major effort was put into optimising the measurements of faint extended objects, rather than those of bright local galaxies or stars. The strategy is similar to that adopted in M22 and P23: we chose to carry out our detection in the infrared, smoothing the scientific image with a Gaussian convolution filter with full width at half maximum (FWHM) of 0.14" (close to the F356W and F444W FWHMs; see below), and applying a detection threshold corresponding to S/N$\sim$2, to pick up as many faint high-redshift sources as possible. However, there is a substantial difference with respect to previous work: rather than just using F444W as the detection band, we created a weighted stack of F356W and F444W, allowing us to single out sources that peak at 3.5 to 4 $\mu$m, and exploiting the F356W mosaics, which are often as deep as (or deeper than) the F444W ones (see Fig. \ref{depth}). The resulting depths of the detection stacks are shown in Fig. \ref{depth001}.

We then ran \textsc{SExtractor} v2.8.6, in the customised version used for the CANDELS campaign. Most of the parameters were left at the default values, with the exception of those given in Table \ref{sexpar}. We tried to use the same values for all fields, but we found that \texttt{DETECT\_THRESH} and \texttt{ANALYSIS\_THRESH} needed adjustment in some cases, to ensure optimal results. In particular, we set them to 0.85 for PRIMER-COSMOS, and to 0.55 for ABELL2744; these values yielded the best trade-off between purity and completeness, allowing for the detection of faint sources, while avoiding too many spurious objects from being included on the list. 

The final detection catalogues contain a grand total of 531,173 sources, as reported in Table \ref{detect}. In the catalogues we include a unique object identifier number, the equatorial position (right ascension and declination, in degrees), and basic morphological information. The latter includes the area of the segmented cluster of pixels  and the half light radius from \textsc{SExtractor}, while the semi-major axis, ellipticity, and position angle of the elliptical isophote were obtained by means of \textsc{a-phot} (see Appendix A for more details).

Figure \ref{comple} shows the point-source detection completeness for the six fields, determined by injecting fake PSF-shaped objects in empty regions of the F356W+F444W stack (using the detection segmentation map as a mask), running \textsc{SExtractor} with the same parameters used in the actual detection process, and checking the fraction of them being actually detected. The different depths of the detection images is evident, with NGDEEP being the deepest with 90\% at AB$\simeq$30 and 50\% at AB$\simeq$30.5 and PRIMER-UDS the shallowest, with 90\% at AB$\simeq$28.5 and $50$\% at AB$\simeq$29. We note the complicated pattern of the JADES-GS field, which comprises deep and shallow regions (90\% completeness at AB$\simeq$28.5, but 50\% at AB$\simeq$30).

\begin{figure}
\includegraphics[width=0.49\textwidth]{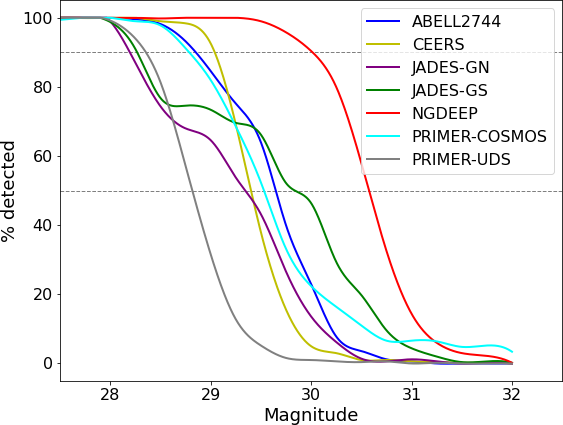}
\caption{Detection completeness:\ Fraction of fake point sources of a given magnitude injected in the detection F356W+F444W stacks and detected with \textsc{SExtractor} runs performed with the parameters used in real runs.} \label{comple}
\end{figure}

\subsection{Photometry} \label{photometry}

As in our previous efforts, in our public catalogues we provide total fluxes. The main methods used for this work are the same ones adopted in M22 and P23. In short, we compute colours in fixed circular apertures on PSF-matched images; as in P23, we smoothed all bands to the F444W resolution, except for the \textrm{HST} WFC3 bands which, having a broader FWHM, were smoothed to the F160W resolution. Assuming no colour gradients outside the apertures, we then scale to total fluxes, multiplying the colour term by the flux within \citet{Kron1980} elliptical apertures in the F356W+F444W stack as measured by the software \textsc{a-phot} \citep{Merlin2019}. So, the total flux in each band is given by the formula $f_{tot,band} = c_{ap,det} \times f_{ap,band}$, where $c_{ap,det} = f_{tot,det} / f_{ap,det}$. As pointed out in M22, our \textsc{a-phot} Kron-like aperture on average tends to gather more light than the standard \textsc{SExtractor} \texttt{MAG\_AUTO} (see Fig. 4 in M22), so we did not apply any further aperture correction to the total fluxes.  
Errors are estimated by summing in quadrature the relevant pixels from the RMS maps and applying the same formula, $e_{tot,band} = c_{ap,det} \times e_{ap,band}$; this choice is motivated by the fact that most scientific applications (e.g. SED-fitting or colour-selections) are essentially based on colours, so the propagation of the total flux error would overestimate the relevant uncertainty. 
We computed the fluxes within nine apertures (with diameters 0.2", 0.28", 0.33", 0.50", 0.66", 0.70", 1.32", 2.65", and 5.30"), and built just as many catalogues of total fluxes. The values of the apertures correspond to integer multiples of the F444W FWHM 0.165", except for the two smallest ones (a fixed 0.1" radius aperture and the value corresponding to the WebbPSF\footnote{\texttt{https://www.stsci.edu/jwst/science-planning/\\proposal-planning-toolbox/psf-simulation-tool}} F444W FWHM, 0.14") and the sixth one \citep[corresponding to the larger aperture in the UNCOVER catalogue by][]{Weaver2024}. We point out that we did not correct the fluxes for the cluster magnification effect in the ABELL2744 field.

For each field, we provide an `optimal' catalogue, in which the total flux of each source is obtained from the colours computed in a preferred aperture chosen on the basis of the object segmentation area. We used \textsc{SExtractor} \texttt{ISOAREA\_IMAGE} as a proxy for it, and selected as the preferred aperture the one immediately larger than the value $\sqrt{\texttt{ISOAREA\_IMAGE}/\pi}$ (this implies that the diameter of the optimal aperture is close to the radius of the circularized detected area). The value of this preferred aperture is also reported in the catalogue.

\subsubsection{PSF models}

\begin{figure}
\includegraphics[width=0.49\textwidth]{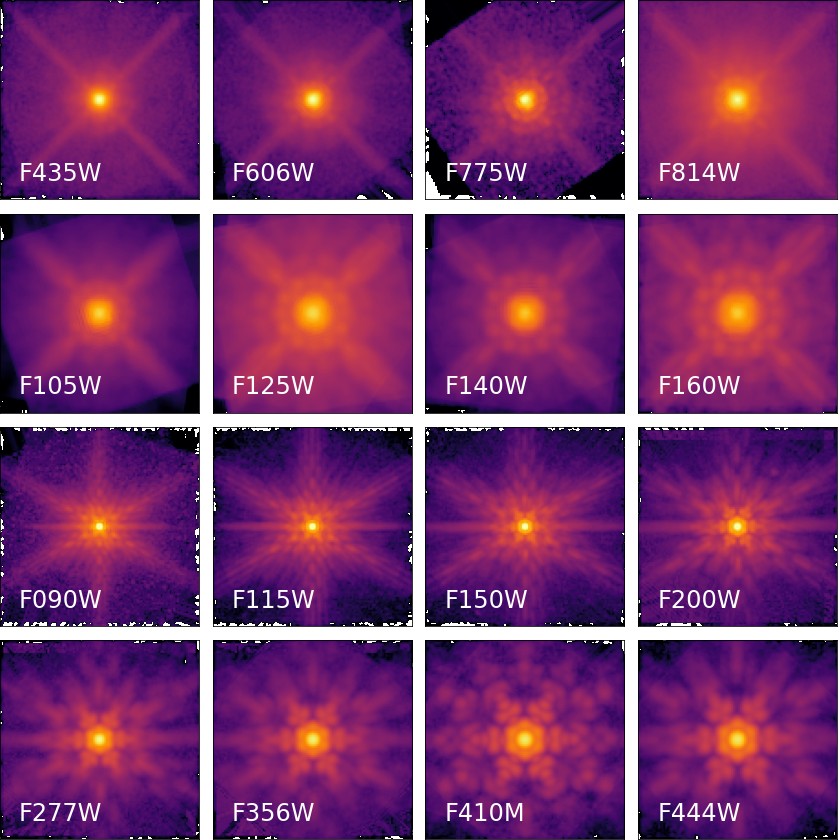}
\caption{PSF ``{\"u}ber''-models for all the bands in the catalogues. See text for details.}  \label{psfs}
\end{figure}

A major difference with respect to our previous efforts is the usage of empirical PSFs, rather than models simulated using the WebbPSF application. For each band, we created empirical models stacking isolated, high-S/N stars, singled out visually after an automatic pre-selection performed using catalogues created with ad-hoc \textsc{SExtactor} runs on each band. After trying various options, including using stars from each field to create the PSFs for that field only, we found that the best results in terms of growth curves and photometric versus spectroscopic redshifts comparison (see Sect. \ref{photoz}) was to create an `{\"u}ber' model for each band, stacking all the available good stars from all fields after rotating each one by the average position angle (PA) of the observation and, finally, rotating back the obtained model to the PA of the considered field (we used the Python module \texttt{numpy.rotate} for this task). In  cases where the mosaics are made up of stacks of images from different epochs, we created models using the stars with the most common orientation in our selection (also considering those consisting of two superposed PSFs with different orientations). These models typically sample the largest area in the field. 
Figure \ref{psfs} shows the final PSF `{\"u}ber' models before the final rotations for their usage in the different fields.

\subsubsection{Background subtraction} \label{bkgsub}

\begin{figure*} 
\sidecaption
%\centering
%\includegraphics[width=0.95\textwidth]{figs/comp_bksub_allfields.png}
%\includegraphics[width=0.9\textwidth]
\includegraphics[width=12cm]{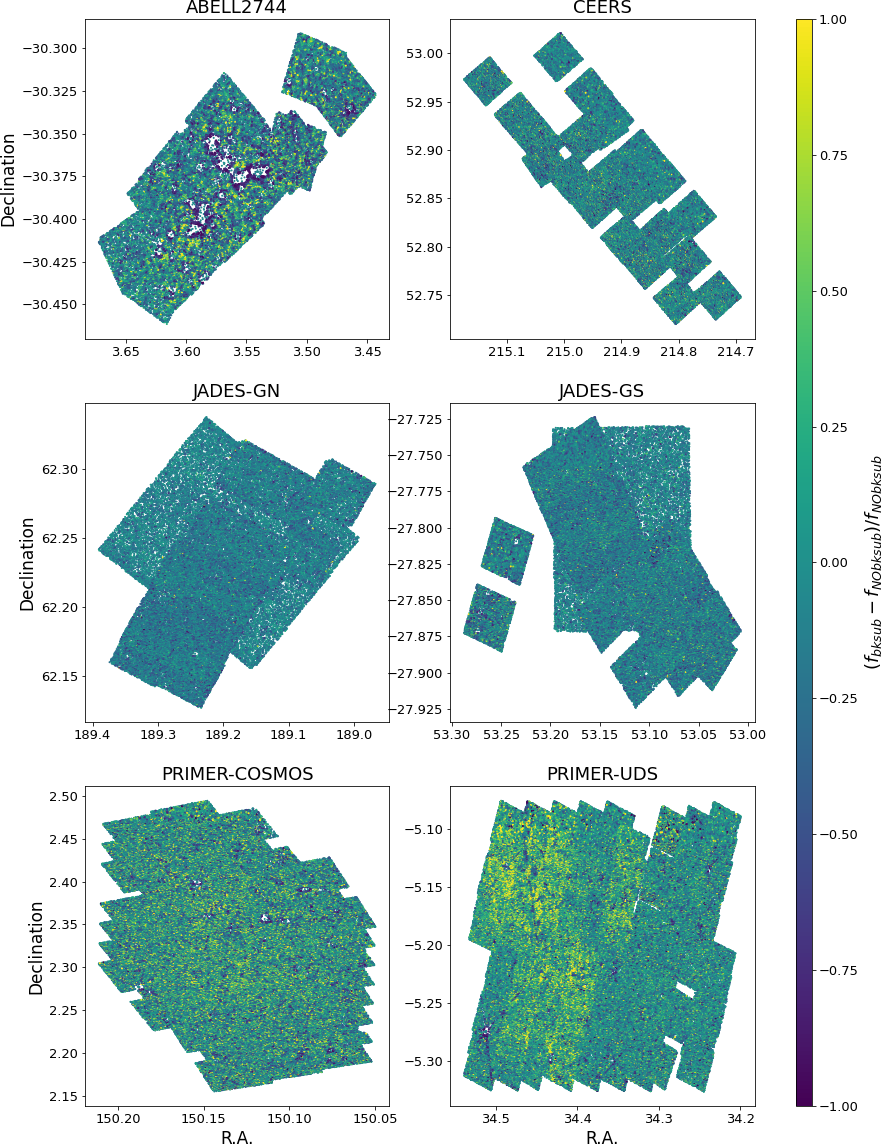}
\caption{Relative difference of measured fluxes in F444W with the \textsc{a-phot} background subtraction switched on and off.}\label{comp_bkg}
\end{figure*}

\textsc{a-phot} can perform `on-the-fly' local background subtraction, while measuring the fluxes. Including this feature is not necessarily the best choice in all cases, as it may yield sub-optimal estimates close to bright sources or in densely populated regions; however, it is typically reasonable to apply it. Therefore, we released both sets of catalogues with the local background subtraction option switched on and off, but we suggest  using the background-subtracted catalogues as the optimal choice. We checked that the difference is not dramatic for the vast majority of the sources, but it does have an impact in some cases. Figure \ref{comp_bkg} shows the difference in measured fluxes in six bands with and without \textsc{a-phot} background subtraction, in all fields. Major differences can be seen around the bright galaxies in the ABELL2744 cluster; we note how the \textsc{a-phot} background subtraction makes the objects closest to the cluster members fainter, but those slightly farther away brighter, compensating for over-subtraction in the image processing phase. Such differences are also seen in an extended area of the PRIMER-UDS field.

\subsubsection{Galactic extinction}

Finally, we corrected all the total fluxes for the effects of galactic extinction, taking advantage of the calculator provided by the NASA/IPAC Extragalactic Database (NED)\footnote{\texttt{https://ned.ipac.caltech.edu/extinction\_calculator}}, which gives the average dimming in a number of bands, at any given equatorial coordinates (we provide those corresponding to the centre of the FoVs of the fields). NIRCam filters are not included in the list, so for them, we interpolated between the available bands at the closest wavelengths. We used such values to compute the extinction-corrected fluxes and included the latter for the final catalogues.

\begin{figure*}
\includegraphics[width=0.95\textwidth]{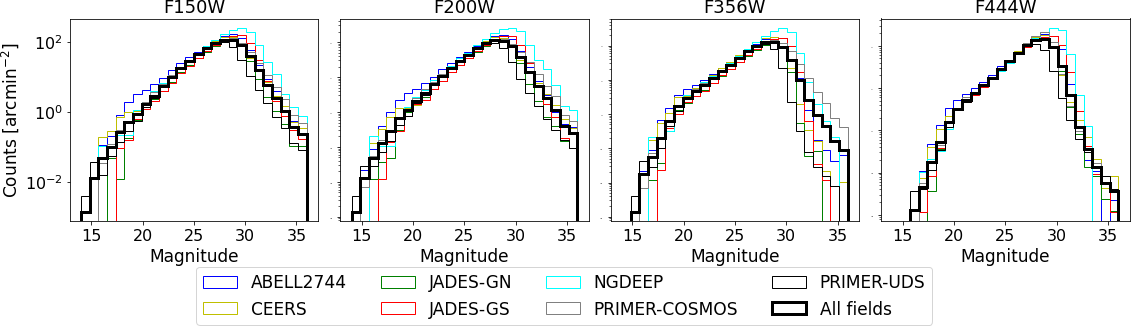}
\caption{Number counts in four NIRCam bands (total magnitudes from the optimal catalogues, normalised to field area). The thick black  line is the total of all fields.} \label{counts}
\end{figure*}

\subsubsection{Number counts}

Figure \ref{counts} shows the number counts of all fields in four bands using the magnitudes obtained using the total fluxes in the `optimal' catalogue and after the correction for galactic extinction. Comparing it with Fig. \ref{comple}, it can be seen that the counts in the F444W and F356 bands typically peak close to the 50\% completeness detection magnitude, which is consistent with previous studies \citep[see e.g.][their Fig. 4]{Guo2013}.

\begin{table}
\caption{Flags assigned in the catalogues.} \label{flags}
\centering
\begin{tabular}{p{1.5cm} p{6.5cm}}     %{ | l || l |}
\hline\hline
Flag & Description \\ \hline
+1-8 & Source is missing  \textrm{HST} coverage \\ 
+10-80 & Source is missing \textrm{JWST} coverage \\ 
+100 & Source is contaminated by close neighbours, or has bad pixels in detection \\
+200 & Source is blended with another in detection \\ 
+400 & Source is saturated in detection \\ 
+800 & Source is close to a border \\ 
+10,000 & Point-like \\ 
+20,000 & All \textrm{HST} fluxes are negative \\ 
+40,000 & Bad measurements in both F356W and F444W \\ 
+100,000 & Spurious detection \\ 
\hline
\end{tabular}
\tablefoot{The final flag assigned to each source is the sum of the individual flagging values.}
\end{table}

\subsection{Point-like sources, spurious detections, and flagging} \label{flag}

\begin{figure}
\includegraphics[width=0.49\textwidth]{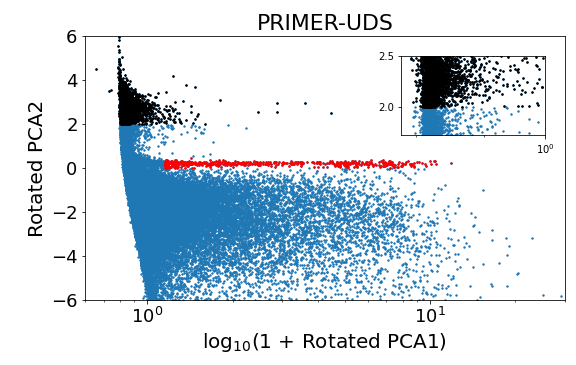}
\caption{Example of the PCA diagram used for the identification of point-like (red points) and spurious (black points) sources.} \label{PCAex}
\end{figure}

To identify and flag point-like sources and potentially spurious detections (PLS and PSD, respectively) we adopted a principal component analysis (PCA) technique, described in detail in Appendix C. In short, taking advantage of the quantities \texttt{MU\_MAX}, \texttt{MAG\_AUTO}, \texttt{MAGERR\_AUTO} and \texttt{FLUX\_RADIUS} estimated with \textsc{SExtractor} on the F356W+F444W detection stack, we obtained a two-dimensional projection space in which PLSs occupy an extremely well-defined and tight locus and PSDs gather in a relatively confined region. The first column of Fig. C.1 shows the resulting diagram for all the fields, with the inner subplot magnifying the region where spurious and regular sources overlap; an example for one field (PRIMER-UDS) is given in Fig. \ref{PCAex}. The exact formulae used to isolate PLS and PSD in the diagram are given in Appendix C and Table C.1. The second and third columns of Fig. C.1 show the position of the sources in two complementary diagnostic planes, namely: the \texttt{MU\_MAX} - \texttt{MAG\_AUTO} plane and the \texttt{FLUX\_RADIUS} versus S/N plane, where PLS and PSD also occupy well-defined regions. The application of the PCA approach allows for a more general identification, combining the two in a fluid way.

We then assigned a further flag to each source, based on the detection and photometric measurements. Table \ref{flags} describes the used values, the final flag being the sum of all the individual addenda. Power-of-two values multiplied by 100 were assigned on the basis of the detection measurements, as an output by the \textsc{a-phot} code, the maximum total value being 1500. Addenda below 100 indicate the number of \textrm{HST} and/or \textrm{JWST} bands missing because of the different observational coverage of the areas. We also included a special flag for the sources with all \textrm{HST} measurement having negative values (a few sources in a limited region of the ABELL2744 field, because of problematic background subtraction in the cluster core area). To give an example, a source flagged with the value 11016 identifies a point-like object, which is blended and saturated in the detection band, and has one \textrm{JWST} and six \textrm{HST} bands missing. Thus, a flag lower than 199 typically indicates a `regular' galaxy-like source which might have missing bands and/or be contaminated in its detection total flux, but it is not blended in detection, not saturated, not close to the borders of the image, nor identified as a star or a spurious detection. Clearly, these flags are a useful diagnostic tool, but we suggest  using them with caution, as they are the result of many automatic processes that cannot achieve full accuracy.
A certain number of spurious detections is unavoidably destined to remain in the catalogues. To further improve the purity, we visually checked all sources with photometric redshifts estimate above 10 (see Sect. \ref{photoz}) to exclude at least the most obvious errors in this important sub-space of the catalogues. In doing so, we found that most remaining cases were defects in the detection stack, either caused by missing coverage in the F356W or F444W images and/or by obvious reduction errors or noise features that had not been singled out in the diagnostic planes. The most affected fields were ABELL2744, CEERS, and PRIMER-UDS. 

It is reasonable to assume that similar spurious detections could exist at photo-$z$$<$10, although  making detections in the reddest bands favours high photometric redshift estimates for these kinds of fake sources. However, since it would be impossible to go through all of the catalogues to find them, we invite users to carefully inspect any objects selected for scientific purposes.

\section{ Photometry validation}
\label{sec:valid}

To validate the accuracy of our catalogues, we compared our `optimal' catalogues (see Sect. \ref{photometry}) to other available ones, either published or obtained via private communication with the proprietary teams. We considered both archival catalogues based on \textrm{HST} observations (CANDELS and \textsc{Astrodeep} releases), and recent \textrm{JWST} catalogues from various research groups: P23 and \citet{Weaver2024} for ABELL2744, \citet{Finkelstein2023a} for CEERS, S. Finkelstein's priv. comm. catalogue for NGDEEP, and the JADES public catalogues for the two JADES fields, v2.0 for GOODS-South and v1.0 for GOODS-North \citep[see][]{Rieke2023}\footnote{\texttt{https://archive.stsci.edu/hlsp/jades}}, which also include photometry from the JEMS programme \citep{Williams2023}.

\begin{figure*}
\centering
\includegraphics[width=0.9\textwidth]{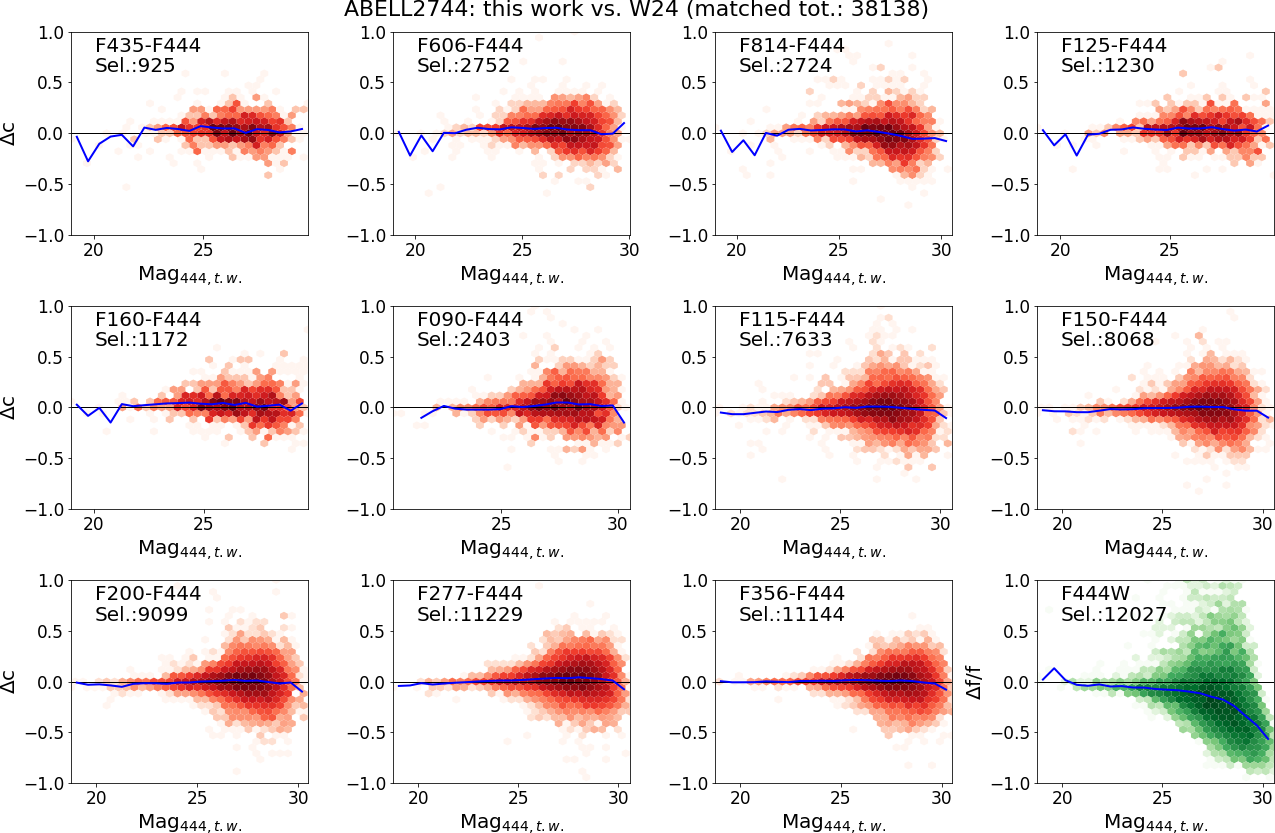} 
\caption{Example of a photometry validation plot, comparing the catalogues for the ABELL2744 field from this work and from \citet{Weaver2024}. Relative errors, $\Delta c$, are given, namely, the colours measured in this work minus those in the reference catalogues), versus the second band magnitude in this work catalogue (e.g. F444W for the F356-F444W colour). The number of cross-matched sources after excluding those with  S/N $<5$ in any of the two catalogues or flag $\geq200$ in this work is also given; the blue line is the median of the distribution. Similar plots for all fields are available in the appendix.}\label{compfluxex}
%\end{minipage}%
\end{figure*}

We cross-matched the catalogues using the equatorial coordinates of the detected sources with a searching radius of 0.35". For the comparison, we only considered sources that have S/N $>5$ in both catalogues and flagged $<200$ in this work (i.e. well-behaved galaxies, non-blended in the detection stack; see Sect. \ref{flag}).

Since many details about the detection procedure vary significantly because of the different techniques adopted in the various works, we chose to focus on colours rather than on total fluxes, as they convey a more robust diagnostic on the accuracy of the estimated SED of the galaxies. The results are shown in Figs. D.1 to D.6. The last panel for each field shows the relative difference in the total flux in F444W for \textrm{JWST} catalogues and in F160W for \textrm{HST} catalogues, for the sake of completeness. An example for one field (ABELL2744) is given in Fig. \ref{compfluxex}.

The agreement is generally good, in particular with \textrm{JWST} catalogues and especially in the NIRCam bands. The larger differences are found with respect to the P23 catalogue for ABEL2744 in the \textrm{HST} bluer bands, most likely because of the different PSF models and the local background subtraction introduced in this work. We also noticed a systematic trend with respect to the JADES catalogues in the \textrm{HST} bands, with our colours typically becoming redder towards the faint end of the distribution; given the good agreement of the F444W flux band estimates, this must be due to our fluxes being fainter in the \textrm{HST} bands.
Concerning the comparisons with the archival \textrm{HST} catalogues, we notice an evident declining trend at the faint end of most plots, which must be due to the deeper sensitivity of the \textrm{JWST} data. Faint objects are now measured with more accuracy, whereas in the old catalogue, they often happened to have spurious positive fluxes (higher than their nominal errors, thus they were not classified as upper limits) as a result of local noise fluctuations; especially in the less resolved redder bands ($Ks$ and IRAC), resulting in larger colours with respect to those measured in the high quality NIRCam images. In the cases of ABELL2744 versus the \textsc{Astrodeep} catalogue by \citet{Merlin2016b} and NGDEEP versus the CANDELS by \citet{Guo2013}, there are very few matched sources, so the comparison is less significant. 

We feel confident that most of the discrepancies can be explained considering the differences in the adopted processing techniques, particularly concerning the apertures used to measure the fluxes, the PSF models, the background subtraction algorithms, and the correction for galactic extinction. Similar discrepancies have been found among multi-wavelength catalogues in past efforts \citep[see e.g.][]{Stefanon2017}. A more detailed analysis is beyond the scope of the present work.

\section{Photometric redshifts} \label{photoz}

We estimated photometric redshifts on our `optimal' catalogues using the software packages \textsc{zphot} \citep{Fontana2000} and \textsc{EAzY} \citep{Brammer2008}.
For \textsc{zphot}, we adopted templates from \cite{Bruzual2003} and we assumed exponentially declining star formation histories (SFHs), with timescales, $\tau$, ranging from 0.1 to 15 Gyr. We included nebular emission lines according to \cite{Castellano2014} and \cite{Schaerer2009}. We considered metallicity values of 0.02, 0.2, 1, and 2.5 times Solar and the age was allowed to vary from 10 Myr to the age of the Universe at a given redshift. Finally, we adopted a \cite{Calzetti2000} extinction law with E(B-V) in the range of 0--1.1.

For \textsc{EAzY}, we carried out three runs: one with the pre-defined set of templates \texttt{eazy\_v1.3} and the other two with the two sets from the ones presented by \citet{Larson2023}. We used the FSPS + Set 1 + Set 3 or `LyaReduced' and FSPS Set 1 + Set 4 or `Lya' combinations, as suggested by the authors\footnote{\texttt{https://ceers.github.io/LarsonSEDTemplates}}). So we end up with four redshift estimates, which we list in our final catalogues.

As a final test of our photometry, we checked the accuracy of the photometric redshift estimates by considering the sub-sample of sources having spectroscopic information from the literature. We matched these spectroscopic targets with our catalogues adopting a conservative searching radius of 0.3". We considered the following spectroscopic samples: 
(i) NIRSpec: for ABELL2744, data from the programmes GLASS-JWST-ERS-1324 \citep{Treu2022}, UNCOVER-GO-2561 \citep{Bezanson2022} and GO-3073 \citep{Castellano2024}, using the \citet{Mascia2024} and \citet[][UNCOVER DR4]{Price2024} data releases, plus additional ones from our own on-going data analysis \citep{Napolitano2024}; for CEERS, data from the programme CEERS-ERS-1345 \citep{Finkelstein2023a}, retrieving them from the public DAWN JWST Archive\footnote{\texttt{https://dawn-cph.github.io/dja}} \citep[see details in][]{Heintz2024}; for the JADES fields, data from the programme JADES-GTO-1180 \citep{Eisenstein2023}, using the DR3 spectroscopic catalogues \citep{DEugenio2024} released by the team\footnote{\texttt{https://jades-survey.github.io/scientists/data.html}}, and FRESCO-GO-1895 \citep{Oesch2023,Meyer2024}. Additionally, we also included the collection of spectra from various programmes analysed and listed in \cite{RobertsBorsani2024}, and more data from the DAWN JWST Archive as available at the moment of writing this paper (July 2024); (ii) CANDELSz7 \citep{Pentericci2018b} data, for JADES-GS, NGDEEP, and PRIMER; (iii) VANDELS DR4 \citep{Pentericci2018,McLure2018,Garilli2021} data, for PRIMER-UDS, JADES-GS and NGDEEP; (iv) the collection of ground based spectroscopy obtained with different instruments by different projects (VIMOS, \citealt{Braglia2009}, and MUSE, \citealt{Mahler2018,Richard2021, Bergamini2023a, Bergamini2023}) at the VLT, AAOmega on the Anglo-Australian Telescope, \citealt{Owers2011}, and of the HST WFC3/IR grism through the HST GO programme GLASS \citep{Treu2015, Schmidt2014} for the ABELL27444 field; (v) the collection compiled by the CANDELS collaboration and described in \cite{Kodra2023} for  CEERS, JADES-GN PRIMER-UDS, PRIMER-COSMOS; (vi) the updated compilation of \cite{Merlin2021} for JADES-GS and NGDEEP; (vii) additional redshifts from \cite{Cowie2004, Reddy2006, Trump2009, vanderWel2016, Inami2017, Damjanov2018, Straatman2018, Scodeggio2018, Masters2019, Wisnioski2019, Urrutia2019, Ning2020, Jones2021, Pharo2022, Bacon2023} and the redshifts collected by \cite{Grazian2006, Wuyts2008, Xue2011}; (viii) the MUSE \citep{Schmidt2021,Rosani2020}, zCOSMOS \citep{Lilly2007}, DEIMOS \citep{Hasinger2018}, and VUDS \citep{Tasca2017} catalogues for PRIMER-COSMOS.
For our final comparisons, we only included robust redshift estimates on the basis of the various quality flags provided by the authors and took care to avoid repetitions.
Whenever a spectroscopic target was listed in more than one survey and the inferred redshift estimates (flagged as robust) differed by more than $0.05\times(1+z_{avg})$, where $z_{avg}$ is the average spectroscopic redshift, we removed the target from the sample; if the difference was lower than this value but larger than 0.1, we considered the redshift; however, we conservatively removed it from the statistics adopted to evaluate the accuracy of our photometric redshifts (see details below).

\begin{figure}[ht!]
\centering
\includegraphics[width=0.49\textwidth]{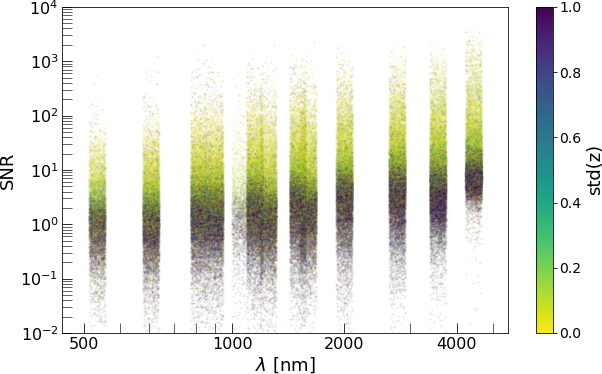} 
\caption{Comparative accuracy of photometric redshift estimates as a function of the S/N of the sources. The plot shows, for a random sample of 20,000 sources from the PRIMER-COSMOS field, the S/N at the wavelengths corresponding to the photometric bands included in the catalogue, colour-coding each point with the standard deviation of the three photo-$z$ estimates used to compute the median photo-$z$ (see text for details). Objects with S/N<10 in the red bands (and/or S/N<1 in the blue bands) tend to have discordant photo-$z$ estimates. See text for more details.} \label{phzsnr}
\end{figure}

\begin{figure}[ht!]
\centering
\includegraphics[width=0.49\textwidth]{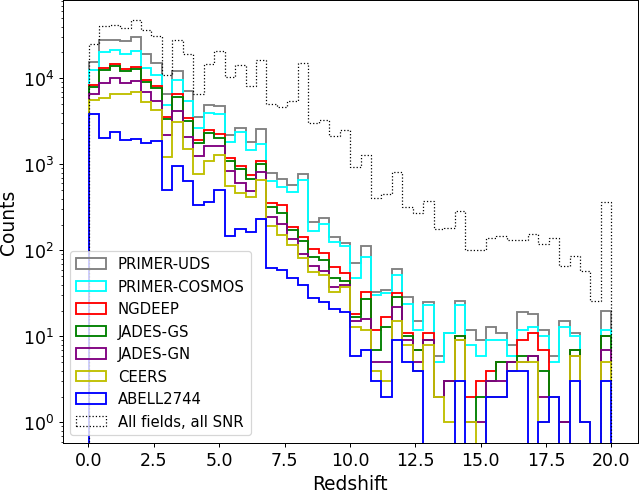}
\caption{Distribution of the median photometric redshifts obtained with the runs described in Sect. \ref{photoz} (raw counts per redshift bin, $\Delta z$=0.4). Stacked coloured histograms refer to individual fields, showing only sources with S/N$_{F356W+F444W}>10$ and flag $< 400$. The dotted black line shows the cumulative distribution for all fields, including all S/N.} 
\label{globalphotoz}
\end{figure}

Since the templates do not include an AGN component, to correctly evaluate the accuracy of the inferred photometric redshifts, we removed known AGN from the catalogues, regardless of their nuclear contribution to the galaxy SED. To this aim, we took advantage of the lists based on JWST data presented by \cite{Harikane2023b},  \cite{Goulding2023}, \cite{Maiolino2024a}, \cite{Kocevski2023}, \cite{Larson2023b}, \cite{RobertsBorsani2024}, \cite{Greene2024}, \cite{Barro2024}, and of X-ray catalogues \citep{Nandra2015,Xue2016,Luo2017,Kocevski2018}, by flagging as AGN all sources with 2-10 keV luminosity larger than $10^{42}$ erg/s or identified as AGN by the authors, and we removed spectroscopic targets flagged as AGN from the VANDELS, zCOSMOS, DEIMOS, VUDS, and COSMOS2020 \citep{Weaver2022} datasets.

Finally, we removed from the lists problematic sources using the flags described in Sect.~\ref{flag}. In particular, we excluded PLS and PSS, and sources that are saturated or at the boundary of the images; namely, we only kept sources with flags of $<$400. In addition, we removed sources lacking JWST photometry in more than three bands (i.e. having the second-to-last figure in the flag value larger than 3). Using these criteria, we were left with a total sample of 16,666 spectra.

Figure \ref{zspec} shows the global comparison between the spectroscopic redshifts and the median values of the photo-$z$ estimates. The latter were computed excluding the \textsc{EAzY} `LyA' run, to avoid over-weighting the results from the runs using the templates by Larson, which are very similar in the two cases. Similar plots for each field individually are shown in Appendix E; Table E.1 reports the full statistics for the four runs made with \textsc{zphot} and \textsc{EAzY}. We define $dz=|z_{phot}-z_{spec}|/(1+z_{spec})$, and $f_{outliers}$ as the percentage of objects with $dz<0.15$. We then compute mean, median, standard deviation, and NMAD (defined as 1.48$\times$median($|dz|$)) of non-outliers. The four runs yield comparable statistics.
The overall accuracy of the median estimates is good (NMAD 0.031, standard deviation 0.041 considering all fields together) and better than that from any individual  \textsc{zphot} or \textsc{EAzY} run. There is a non-negligible fraction of outliers (6.3\%) comparable to (albeit larger than) the one reached in recent efforts on multi-wavelength catalogues with a larger number of bands \citep[and more sohisticated approaches; see e.g.][]{Merlin2021}. Looking at Fig. E.1, we also note that the accuracy varies from field to field, with ABELL2744 and NGDEEP having the largest outlier fractions. This is most likely because of the effect of the cluster on photometric measurements for the former and of the low number of available spectroscopic redshifts for the latter.

\begin{figure}
\includegraphics[width=0.49\textwidth]{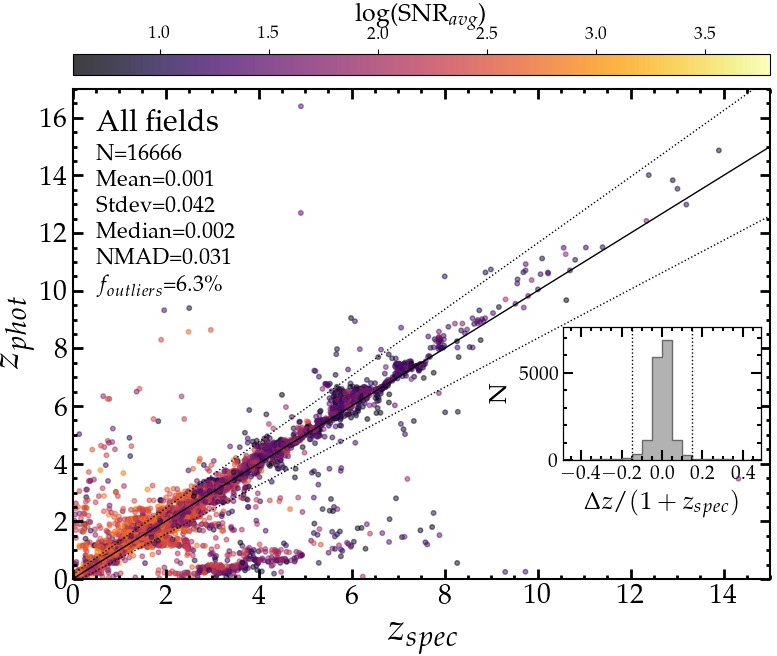}
\captionof{figure}{Global accuracy of photometric redshifts with respect to spectroscopic samples (all fields together). $z_{phot}$ values are the median of the three runs described in Sect. \ref{photoz}; N is the number of sources; $f_{outliers}$ is the percentage of sources with $|z_{phot}-z_{spec}|/(1+z_{spec})>0.15$. The other statistics are computed on non-outliers, with NMAD=$1.48\times \mbox{median}[|z_{phot}-z_{spec}|/(1+z_{spec})]$. Colour-coding is $<\mbox{S/N}>=(\mbox{S/N}_{\mbox{F444W}}+\mbox{S/N}_{\mbox{F356W}}+\mbox{S/N}_{\mbox{F277W}}+\mbox{S/N}_{\mbox{F200W}})/4$.}\label{zspec}
\end{figure}

We report that we also tested delayed exponentially declining SFHs with the \textsc{zphot} code. We verified that our choice of standard exponentially declining models performs better in terms of agreement with spectroscopic redshifts. 

We point out that the photometric redshift estimates are prone to substantial uncertainties for the sources at low S/N. Figure \ref{phzsnr} exemplifies this by showing the S/N as a function of the band wavelength for a random sub-sample of 20,000 sources in the PRIMER-COSMOS catalogue, colour-coding the points by the value of the standard deviation of the three photo-$z$ estimates used to compute the median value. Clearly, the discrepancy between the estimates grows with decreasing S/N. Since we detected in the reddest bands, sources close to the detection limit in the F356W+F444W stack typically become even fainter in bluer images, thus making their characterisation very hard. This is mainly because  different codes and/or templates prefer different solutions for poorly constrained photometry.

Then, as discussed in Sect. \ref{flag}, taking the \textsc{zphot} runs as reference (given its overall slightly better performance), we also visually inspected all of the sources with estimated redshift above 10 (which were initially 7123). By doing so, not only did we exclude a further population of spurious detections, but we identified some systematic cases where the redshift estimate was wrong due to unfortunate lack of relevant data. For example, in PRIMER-UDS some sources have been observed only with the LW NIRCam filters; thus, the missing information at wavelengths blue-ward of F277W caused an erroneous identification of the Lyman break and therefore of the redshift. We marked these and similar cases with a negative sign before the flag value in the photometric redshift catalogues. 
After these checks, excluding  sources flagged as spurious we were left with 3068 objects with photo-$z>10$. This number is certainly an over-estimation. Indeed, only 798 have a standard deviation of the estimates of the three codes lower than 0.5; again highlighting the challenging task of robustly constraining the photometric redshift for such faint, distant sources. Furthermore, some red low redshift interlopers must be present \citep[see][]{ArrabalHaro2023b,Harikane2024} and we deliberately only removed clearly spurious detections, not the uncertain or suspect ones. Thus, there are certainly still a number of fake sources polluting the sample. Comparing, for instance, with the estimated surface density of high-$z$ objects by \citet[][as seen in their Fig. 8]{Finkelstein2024}, we would expect $\sim$20 galaxies at $z>10$ with F277W$<$28.5 in CEERS; we found 59 in the \textsc{zphot} sample (a value close to the expected number after correcting for completeness), but just 9 considering a restricted sample of sources having a standard deviation of the estimates from the three codes that is lower than 0.5.

Having clarified this issue, we show in Fig. \ref{globalphotoz}  the final distributions of the estimated redshifts (median estimate). The coloured stacked histograms refer to the individual fields and show the distribution of sources with detection S/N$>$10 and not flagged as point-like, spurious, or with a bad redshift estimate because of lacking bands, as previously explained. The dotted thin black line shows the total distribution including all S/N.
The  issues discussed here lead to some evidently artificial features in the global photo-$z$ distribution, which largely disappear when considering only high S/N (e.g. S/N$_{F444W}>$10) sources. 
A thorough comparative analysis of the probability distribution functions from the four runs would be needed to disentangle the degeneracy for the faint objects, but this goes beyond the scope of this work. Therefore, we have restricted our study to the release of  the output of the runs. For similar reasons, we also postpone the release of the physical properties for the galaxies \citep[see e.g.][]{Markov2023}.

\section{Summary and conclusions}
\label{sec:concl}

We present and discuss a major release of photometric catalogues, collecting data from eight \textrm{JWST} observational programmes on six deep extra-galactic fields. We created a new reduction of the Abell 2744 composite mosaics, gathering the GLASS-JWST, UNCOVER, DDT2756, and GO3990 programmes. For the other fields, we used the reductions provided by the teams of the other programmes. The NIRCam data are complemented with archival \textrm{HST} images, which we used to obtain new measurements.

The catalogues, mainly conceived for high-redshift science, include a grand-total of 531,173 objects, of which 18,563 are tagged as spurious and 2,217 more have bad photometric measurements (flag$\geq$400; see Sect. \ref{flag}). Sources were detected with \textsc{SExtractor} on stacks of the F356W and F444W mosaics of each field. For all the detections, we provide positions in equatorial and pixel coordinates, basic morphological parameters, total fluxes, and corresponding uncertainties in 16 photometric bands computed by means of the software \textsc{a-phot}, a diagnostic flag, and four photometric redshift estimates obtained with \textsc{zphot} and \textsc{EAzY}.

We performed validation tests on the astrometry, the photometry and the photo-$z$ accuracy, comparing our catalogues against other releases and finding a general good agreement. We encourage the community to exploit these catalogues for any suitable scientific purpose.

\section{Data availability} \label{dataav}

The catalogues are available for download from the \textsc{Astrodeep} website\footnote{\texttt{https://astrodeep.eu/astrodeep-jwst-catalogs/}}. Updates and/or additional releases will be documented on the website. It is also possible to visualize and explore the catalogues on the Cosmological Surveys Rainbow Database\footnote{\texttt{http://arcoirix.cab.inta-csic.es/\\Rainbow\_navigator\_public/}} \citep{2005ApJ...630...82P,PerezGonzalez2008,Barro2019}. 

The appendix sections of this paper are available on Zenodo\footnote{\texttt{https://zenodo.org/records/13886578}}.

The `optimal' photometric catalogues (see Sect. \ref{photometry}) and the photo-$z$ catalogues for all fields are also available in electronic form at the CDS via anonymous ftp to cdsarc.u-strasbg.fr (130.79.128.5) or via http://cdsweb.u-strasbg.fr/cgi-bin/qcat?J/A+A/.

\begin{acknowledgements}

This work is based in part on observations made with the NASA/ESA/CSA \textit{James Webb} Space Telescope. The data were obtained from the Mikulski Archive for Space Telescopes at the Space Telescope Science Institute, which is operated by the Association of Universities for Research in Astronomy, Inc., under NASA contract NAS 5-03127 for JWST. These observations are associated with program JWST-ERS-1342.\\

This research is also based in part on observations made with the NASA/ESA Hubble Space Telescope obtained from the Space Telescope Science Institute, which is operated by the Association of Universities for Research in Astronomy, Inc., under NASA contract NAS 5–26555. These observations are associated with program HST-GO-17321.\\

This research is also supported in part by the Australian Research Council Centre of Excellence for All Sky Astrophysics in 3 Dimensions (ASTRO 3D), through project number CE170100013.\\

We also acknowledge support from the INAF Theory Grant 2022 ``Forward modeling of theoretical models for upcoming surveys'' (PI Merlin), Large Grant 2022 ``Extragalactic Surveys with JWST'' (PI Pentericci), Mini Grant 2022 ``The evolution of passive galaxies through cosmic time'' (PI Santini) and Mini Grant 2022 ``Reionization and Fundamental Cosmology with High-Redshift Galaxies'' (PI Castellano). \\

B.V. is supported by the European Union – NextGenerationEU RFF M4C2 1.1 PRIN 2022 project 2022ZSL4BL INSIGHT.\\

P.G.P.-G. acknowledges support from grant PID2022-139567NB-I00 funded by Spanish Ministerio de Ciencia, Innovación y Universidades MCIU/AEI/10.13039/501100011033, and the European Union through FEDER {\it Una manera de hacer Europa.}\\

%Support from NASA through grants JWST-GO-1342, and HST-GO-17231 is also gratefully acknowledged.\\

%D.M. acknowledges financial support from program HST-GO-17231,provided through a grant from the STScI under NASA contract NAS5-26555.

\end{acknowledgements}

\bibliographystyle{aa}
\bibliography{biblio.bib}{}

\begin{thebibliography}{145}
\expandafter\ifx\csname natexlab\endcsname\relax\def\natexlab#1{#1}\fi

\bibitem[{{Adamo} {et~al.}(2024){Adamo}, {Atek}, {Bagley}, {Ba{\~n}ados}, {Barrow}, {Berg}, {Bezanson}, {Brada{\v{c}}}, {Brammer}, {Carnall}, {Chisholm}, {Coe}, {Dayal}, {Eisenstein}, {Eldridge}, {Ferrara}, {Fujimoto}, {de Graaff}, {Habouzit}, {Hutchison}, {Kartaltepe}, {Kassin}, {Kriek}, {Labb{\'e}}, {Maiolino}, {Marques-Chaves}, {Maseda}, {Mason}, {Matthee}, {McQuinn}, {Meynet}, {Naidu}, {Oesch}, {Pentericci}, {P{\'e}rez-Gonz{\'a}lez}, {Rigby}, {Roberts-Borsani}, {Schaerer}, {Shapley}, {Stark}, {Stiavelli}, {Strom}, {Vanzella}, {Wang}, {Wilkins}, {Williams}, {Willott}, {Wylezalek}, \& {Nota}}]{Adamo2024}
{Adamo}, A., {Atek}, H., {Bagley}, M.~B., {et~al.} 2024, arXiv e-prints, arXiv:2405.21054

\bibitem[{{Arrabal Haro} {et~al.}(2023{\natexlab{a}}){Arrabal Haro}, {Dickinson}, {Finkelstein}, {Fujimoto}, {Fern{\'a}ndez}, {Kartaltepe}, {Jung}, {Cole}, {Burgarella}, {Chworowsky}, {Hutchison}, {Morales}, {Papovich}, {Simons}, {Amor{\'\i}n}, {Backhaus}, {Bagley}, {Bisigello}, {Calabr{\`o}}, {Castellano}, {Cleri}, {Dav{\'e}}, {Dekel}, {Ferguson}, {Fontana}, {Gawiser}, {Giavalisco}, {Harish}, {Hathi}, {Hirschmann}, {Holwerda}, {Huertas-Company}, {Koekemoer}, {Larson}, {Lucas}, {Mobasher}, {P{\'e}rez-Gonz{\'a}lez}, {Pirzkal}, {Rose}, {Santini}, {Trump}, {de la Vega}, {Wang}, {Weiner}, {Wilkins}, {Yang}, {Yung}, \& {Zavala}}]{ArrabalHaro2023}
{Arrabal Haro}, P., {Dickinson}, M., {Finkelstein}, S.~L., {et~al.} 2023{\natexlab{a}}, \apjl, 951, L22

\bibitem[{{Arrabal Haro} {et~al.}(2023{\natexlab{b}}){Arrabal Haro}, {Dickinson}, {Finkelstein}, {Kartaltepe}, {Donnan}, {Burgarella}, {Carnall}, {Cullen}, {Dunlop}, {Fern{\'a}ndez}, {Fujimoto}, {Jung}, {Krips}, {Larson}, {Papovich}, {P{\'e}rez-Gonz{\'a}lez}, {Amor{\'\i}n}, {Bagley}, {Buat}, {Casey}, {Chworowsky}, {Cohen}, {Ferguson}, {Giavalisco}, {Huertas-Company}, {Hutchison}, {Kocevski}, {Koekemoer}, {Lucas}, {McLeod}, {McLure}, {Pirzkal}, {Seill{\'e}}, {Trump}, {Weiner}, {Wilkins}, \& {Zavala}}]{ArrabalHaro2023b}
{Arrabal Haro}, P., {Dickinson}, M., {Finkelstein}, S.~L., {et~al.} 2023{\natexlab{b}}, \nat, 622, 707

\bibitem[{{Bacon} {et~al.}(2023){Bacon}, {Brinchmann}, {Conseil}, {Maseda}, {Nanayakkara}, {Wendt}, {Bacher}, {Mary}, {Weilbacher}, {Krajnovi{\'c}}, {Boogaard}, {Bouch{\'e}}, {Contini}, {Epinat}, {Feltre}, {Guo}, {Herenz}, {Kollatschny}, {Kusakabe}, {Leclercq}, {Michel-Dansac}, {Pello}, {Richard}, {Roth}, {Salvignol}, {Schaye}, {Steinmetz}, {Tresse}, {Urrutia}, {Verhamme}, {Vitte}, {Wisotzki}, \& {Zoutendijk}}]{Bacon2023}
{Bacon}, R., {Brinchmann}, J., {Conseil}, S., {et~al.} 2023, \aap, 670, A4

\bibitem[{{Bagley} {et~al.}(2023){Bagley}, {Finkelstein}, {Koekemoer}, {Ferguson}, {Arrabal Haro}, {Dickinson}, {Kartaltepe}, {Papovich}, {P{\'e}rez-Gonz{\'a}lez}, {Pirzkal}, {Somerville}, {Willmer}, {Yang}, {Yung}, {Fontana}, {Grazian}, {Grogin}, {Hirschmann}, {Kewley}, {Kirkpatrick}, {Kocevski}, {Lotz}, {Medrano}, {Morales}, {Pentericci}, {Ravindranath}, {Trump}, {Wilkins}, {Calabr{\`o}}, {Cooper}, {Costantin}, {de la Vega}, {Hilbert}, {Hutchison}, {Larson}, {Lucas}, {McGrath}, {Ryan}, {Wang}, \& {Wuyts}}]{Bagley2023}
{Bagley}, M.~B., {Finkelstein}, S.~L., {Koekemoer}, A.~M., {et~al.} 2023, \apjl, 946, L12

\bibitem[{{Bagley} {et~al.}(2024){Bagley}, {Pirzkal}, {Finkelstein}, {Papovich}, {Berg}, {Lotz}, {Leung}, {Ferguson}, {Koekemoer}, {Dickinson}, {Kartaltepe}, {Kocevski}, {Somerville}, {Yung}, {Backhaus}, {Casey}, {Castellano}, {Ch{\'a}vez Ortiz}, {Chworowsky}, {Cox}, {Dav{\'e}}, {Davis}, {Estrada-Carpenter}, {Fontana}, {Fujimoto}, {Gardner}, {Giavalisco}, {Grazian}, {Grogin}, {Hathi}, {Hutchison}, {Jaskot}, {Jung}, {Kewley}, {Kirkpatrick}, {Larson}, {Matharu}, {Natarajan}, {Pentericci}, {P{\'e}rez-Gonz{\'a}lez}, {Ravindranath}, {Rothberg}, {Ryan}, {Shen}, {Simons}, {Snyder}, {Trump}, \& {Wilkins}}]{Bagley2024}
{Bagley}, M.~B., {Pirzkal}, N., {Finkelstein}, S.~L., {et~al.} 2024, \apjl, 965, L6

\bibitem[{{Barro} {et~al.}(2019){Barro}, {P{\'e}rez-Gonz{\'a}lez}, {Cava}, {Brammer}, {Pandya}, {Eliche Moral}, {Esquej}, {Dom{\'\i}nguez-S{\'a}nchez}, {Alcalde Pampliega}, {Guo}, {Koekemoer}, {Trump}, {Ashby}, {Cardiel}, {Castellano}, {Conselice}, {Dickinson}, {Dolch}, {Donley}, {Espino Briones}, {Faber}, {Fazio}, {Ferguson}, {Finkelstein}, {Fontana}, {Galametz}, {Gardner}, {Gawiser}, {Giavalisco}, {Grazian}, {Grogin}, {Hathi}, {Hemmati}, {Hern{\'a}n-Caballero}, {Kocevski}, {Koo}, {Kodra}, {Lee}, {Lin}, {Lucas}, {Mobasher}, {McGrath}, {Nandra}, {Nayyeri}, {Newman}, {Pforr}, {Peth}, {Rafelski}, {Rodr{\'\i}guez-Munoz}, {Salvato}, {Stefanon}, {van der Wel}, {Willner}, {Wiklind}, \& {Wuyts}}]{Barro2019}
{Barro}, G., {P{\'e}rez-Gonz{\'a}lez}, P.~G., {Cava}, A., {et~al.} 2019, \apjs, 243, 22

\bibitem[{{Barro} {et~al.}(2024){Barro}, {P{\'e}rez-Gonz{\'a}lez}, {Kocevski}, {McGrath}, {Trump}, {Simons}, {Somerville}, {Yung}, {Arrabal Haro}, {Akins}, {Bagley}, {Cleri}, {Costantin}, {Davis}, {Dickinson}, {Finkelstein}, {Giavalisco}, {G{\'o}mez-Guijarro}, {Hathi}, {Hirschmann}, {Holwerda}, {Huertas-Company}, {Kartaltepe}, {Koekemoer}, {Lucas}, {Papovich}, {Pirzkal}, {Seill{\'e}}, {Tacchella}, {Wuyts}, {Wilkins}, {de la Vega}, {Yang}, \& {Zavala}}]{Barro2024}
{Barro}, G., {P{\'e}rez-Gonz{\'a}lez}, P.~G., {Kocevski}, D.~D., {et~al.} 2024, \apj, 963, 128

\bibitem[{{Bergamini} {et~al.}(2023{\natexlab{a}}){Bergamini}, {Acebron}, {Grillo}, {Rosati}, {Caminha}, {Mercurio}, {Vanzella}, {Angora}, {Brammer}, {Meneghetti}, \& {Nonino}}]{Bergamini2023a}
{Bergamini}, P., {Acebron}, A., {Grillo}, C., {et~al.} 2023{\natexlab{a}}, \aap, 670, A60

\bibitem[{{Bergamini} {et~al.}(2023{\natexlab{b}}){Bergamini}, {Acebron}, {Grillo}, {Rosati}, {Caminha}, {Mercurio}, {Vanzella}, {Mason}, {Treu}, {Angora}, {Brammer}, {Meneghetti}, {Nonino}, {Boyett}, {Brada{\v{c}}}, {Castellano}, {Fontana}, {Morishita}, {Paris}, {Prieto-Lyon}, {Roberts-Borsani}, {Roy}, {Santini}, {Vulcani}, {Wang}, \& {Yang}}]{Bergamini2023}
{Bergamini}, P., {Acebron}, A., {Grillo}, C., {et~al.} 2023{\natexlab{b}}, \apj, 952, 84

\bibitem[{{Bertin} \& {Arnouts}(1996)}]{Bertin1996}
{Bertin}, E. \& {Arnouts}, S. 1996, \aaps, 117, 393

\bibitem[{{Bezanson} {et~al.}(2022){Bezanson}, {Labbe}, {Whitaker}, {Leja}, {Price}, {Franx}, {Brammer}, {Marchesini}, {Zitrin}, {Wang}, {Weaver}, {Furtak}, {Atek}, {Coe}, {Cutler}, {Dayal}, {van Dokkum}, {Feldmann}, {Forster Schreiber}, {Fujimoto}, {Geha}, {Glazebrook}, {de Graaff}, {Greene}, {Juneau}, {Kassin}, {Kriek}, {Khullar}, {Maseda}, {Mowla}, {Muzzin}, {Nanayakkara}, {Nelson}, {Oesch}, {Pacifici}, {Pan}, {Papovich}, {Setton}, {Shapley}, {Smit}, {Stefanon}, {Taylor}, \& {Williams}}]{Bezanson2022}
{Bezanson}, R., {Labbe}, I., {Whitaker}, K.~E., {et~al.} 2022, submitted to ApJ, arXiv:2212.04026

\bibitem[{{Braglia} {et~al.}(2009){Braglia}, {Pierini}, {Biviano}, \& {B{\"o}hringer}}]{Braglia2009}
{Braglia}, F.~G., {Pierini}, D., {Biviano}, A., \& {B{\"o}hringer}, H. 2009, \aap, 500, 947

\bibitem[{{Brammer} {et~al.}(2008){Brammer}, {van Dokkum}, \& {Coppi}}]{Brammer2008}
{Brammer}, G.~B., {van Dokkum}, P.~G., \& {Coppi}, P. 2008, \apj, 686, 1503

\bibitem[{{Bruzual} \& {Charlot}(2003)}]{Bruzual2003}
{Bruzual}, G. \& {Charlot}, S. 2003, \mnras, 344, 1000

\bibitem[{{Calzetti} {et~al.}(2000){Calzetti}, {Armus}, {Bohlin}, {Kinney}, {Koornneef}, \& {Storchi-Bergmann}}]{Calzetti2000}
{Calzetti}, D., {Armus}, L., {Bohlin}, R.~C., {et~al.} 2000, \apj, 533, 682

\bibitem[{{Carnall} {et~al.}(2023){Carnall}, {McLeod}, {McLure}, {Dunlop}, {Begley}, {Cullen}, {Donnan}, {Hamadouche}, {Jewell}, {Jones}, {Pollock}, \& {Wild}}]{Carnall2023}
{Carnall}, A.~C., {McLeod}, D.~J., {McLure}, R.~J., {et~al.} 2023, \mnras, 520, 3974

\bibitem[{{Carniani} {et~al.}(2024){Carniani}, {Hainline}, {D'Eugenio}, {Eisenstein}, {Jakobsen}, {Witstok}, {Johnson}, {Chevallard}, {Maiolino}, {Helton}, {Willott}, {Robertson}, {Alberts}, {Arribas}, {Baker}, {Bhatawdekar}, {Boyett}, {Bunker}, {Cameron}, {Cargile}, {Charlot}, {Curti}, {Curtis-Lake}, {Egami}, {Giardino}, {Isaak}, {Ji}, {Jones}, {Kumari}, {Maseda}, {Parlanti}, {P{\'e}rez-Gonz{\'a}lez}, {Rawle}, {Rieke}, {Rieke}, {Del Pino}, {Saxena}, {Scholtz}, {Smit}, {Sun}, {Tacchella}, {{\"U}bler}, {Venturi}, {Williams}, \& {Willmer}}]{Carniani2024}
{Carniani}, S., {Hainline}, K., {D'Eugenio}, F., {et~al.} 2024, \nat, 633, 318

\bibitem[{{Castellano} {et~al.}(2023){Castellano}, {Fontana}, {Treu}, {Merlin}, {Santini}, {Bergamini}, {Grillo}, {Rosati}, {Acebron}, {Leethochawalit}, {Paris}, {Bonchi}, {Belfiori}, {Calabr{\`o}}, {Correnti}, {Nonino}, {Polenta}, {Trenti}, {Boyett}, {Brammer}, {Broadhurst}, {Caminha}, {Chen}, {Filippenko}, {Fortuni}, {Glazebrook}, {Mascia}, {Mason}, {Menci}, {Meneghetti}, {Mercurio}, {Metha}, {Morishita}, {Nanayakkara}, {Pentericci}, {Roberts-Borsani}, {Roy}, {Vanzella}, {Vulcani}, {Yang}, \& {Wang}}]{Castellano2023}
{Castellano}, M., {Fontana}, A., {Treu}, T., {et~al.} 2023, \apjl, 948, L14

\bibitem[{{Castellano} {et~al.}(2022){Castellano}, {Fontana}, {Treu}, {Santini}, {Merlin}, {Leethochawalit}, {Trenti}, {Vanzella}, {Mestric}, {Bonchi}, {Belfiori}, {Nonino}, {Paris}, {Polenta}, {Roberts-Borsani}, {Boyett}, {Brada{\v{c}}}, {Calabr{\`o}}, {Glazebrook}, {Grillo}, {Mascia}, {Mason}, {Mercurio}, {Morishita}, {Nanayakkara}, {Pentericci}, {Rosati}, {Vulcani}, {Wang}, \& {Yang}}]{Castellano2022}
{Castellano}, M., {Fontana}, A., {Treu}, T., {et~al.} 2022, \apjl, 938, L15

\bibitem[{{Castellano} {et~al.}(2024){Castellano}, {Napolitano}, {Fontana}, {Roberts-Borsani}, {Treu}, {Vanzella}, {Zavala}, {Arrabal Haro}, {Calabr{\`o}}, {Llerena}, {Mascia}, {Merlin}, {Paris}, {Pentericci}, {Santini}, {Bakx}, {Bergamini}, {Cupani}, {Dickinson}, {Filippenko}, {Glazebrook}, {Grillo}, {Kelly}, {Malkan}, {Mason}, {Morishita}, {Nanayakkara}, {Rosati}, {Sani}, {Wang}, \& {Yoon}}]{Castellano2024}
{Castellano}, M., {Napolitano}, L., {Fontana}, A., {et~al.} 2024, \apj, 972, 143

\bibitem[{{Castellano} {et~al.}(2014){Castellano}, {Sommariva}, {Fontana}, {Pentericci}, {Santini}, {Grazian}, {Amorin}, {Donley}, {Dunlop}, {Ferguson}, {Fiore}, {Galametz}, {Giallongo}, {Guo}, {Huang}, {Koekemoer}, {Maiolino}, {McLure}, {Paris}, {Schaerer}, {Troncoso}, \& {Vanzella}}]{Castellano2014}
{Castellano}, M., {Sommariva}, V., {Fontana}, A., {et~al.} 2014, \aap, 566, A19

\bibitem[{{Ciesla} {et~al.}(2024){Ciesla}, {Elbaz}, {Ilbert}, {Buat}, {Magnelli}, {Narayanan}, {Daddi}, {G{\'o}mez-Guijarro}, \& {Arango-Toro}}]{Ciesla2024}
{Ciesla}, L., {Elbaz}, D., {Ilbert}, O., {et~al.} 2024, \aap, 686, A128

\bibitem[{{Conselice} {et~al.}(2024){Conselice}, {Adams}, {Harvey}, {Austin}, {Ferreira}, {Ormerod}, {Duan}, {Trussler}, {Li}, {Juodzbalis}, {Westcott}, {Harris}, {Seeyave}, {Bluck}, {Windhorst}, {Bhatawdekar}, {Coe}, {Cohen}, {Cheng}, {Driver}, {Frye}, {Furtak}, {Grogin}, {Hathi}, {Holwerda}, {Jansen}, {Koekemoer}, {Marshall}, {Nonino}, {Robotham}, {Summers}, {Wilkins}, {Willmer}, {Yan}, \& {Zitrin}}]{Conselice2024}
{Conselice}, C.~J., {Adams}, N., {Harvey}, T., {et~al.} 2024, arXiv e-prints, arXiv:2407.14973

\bibitem[{{Cowie} {et~al.}(2004){Cowie}, {Barger}, {Hu}, {Capak}, \& {Songaila}}]{Cowie2004}
{Cowie}, L.~L., {Barger}, A.~J., {Hu}, E.~M., {Capak}, P., \& {Songaila}, A. 2004, \aj, 127, 3137

\bibitem[{{Damjanov} {et~al.}(2018){Damjanov}, {Zahid}, {Geller}, {Fabricant}, \& {Hwang}}]{Damjanov2018}
{Damjanov}, I., {Zahid}, H.~J., {Geller}, M.~J., {Fabricant}, D.~G., \& {Hwang}, H.~S. 2018, \apjs, 234, 21

\bibitem[{{Davis} {et~al.}(2023){Davis}, {Trump}, {Simons}, {Mcgrath}, {Wilkins}, {Arrabal Haro}, {Bagley}, {Dickinson}, {Fern{\'A}ndez}, {Amor{\'I}n}, {Backhaus}, {Cleri}, {Llerena}, {Brunker}, {Barro}, {Bisigello}, {Brooks}, {Costantin}, {De La Vega}, {Dekel}, {Finkelstein}, {Hathi}, {Hirschmann}, {Kartaltepe}, {Koekemoer}, {Lucas}, {Papovich}, {P{\'E}rez-Gonz{\'A}lez}, {Pirzkal}, {Rodighiero}, {Rose}, \& {Yung}}]{Davis2023}
{Davis}, K., {Trump}, J.~R., {Simons}, R.~C., {et~al.} 2023, submitted to ApJ, arXiv:2312.07799

\bibitem[{{Davis} {et~al.}(2007){Davis}, {Guhathakurta}, {Konidaris}, {Newman}, {Ashby}, {Biggs}, {Barmby}, {Bundy}, {Chapman}, {Coil}, {Conselice}, {Cooper}, {Croton}, {Eisenhardt}, {Ellis}, {Faber}, {Fang}, {Fazio}, {Georgakakis}, {Gerke}, {Goss}, {Gwyn}, {Harker}, {Hopkins}, {Huang}, {Ivison}, {Kassin}, {Kirby}, {Koekemoer}, {Koo}, {Laird}, {Le Floc'h}, {Lin}, {Lotz}, {Marshall}, {Martin}, {Metevier}, {Moustakas}, {Nandra}, {Noeske}, {Papovich}, {Phillips}, {Rich}, {Rieke}, {Rigopoulou}, {Salim}, {Schiminovich}, {Simard}, {Smail}, {Small}, {Weiner}, {Willmer}, {Willner}, {Wilson}, {Wright}, \& {Yan}}]{Davis2007}
{Davis}, M., {Guhathakurta}, P., {Konidaris}, N.~P., {et~al.} 2007, \apjl, 660, L1

\bibitem[{{Dekel} {et~al.}(2023){Dekel}, {Sarkar}, {Birnboim}, {Mandelker}, \& {Li}}]{Dekel2023}
{Dekel}, A., {Sarkar}, K.~C., {Birnboim}, Y., {Mandelker}, N., \& {Li}, Z. 2023, \mnras, 523, 3201

\bibitem[{{D'Eugenio} {et~al.}(2024){D'Eugenio}, {Cameron}, {Scholtz}, {Carniani}, {Willott}, {Curtis-Lake}, {Bunker}, {Parlanti}, {Maiolino}, {Willmer}, {Jakobsen}, {Robertson}, {Johnson}, {Tacchella}, {Cargile}, {Rawle}, {Arribas}, {Chevallard}, {Curti}, {Egami}, {Eisenstein}, {Kumari}, {Looser}, {Rieke}, {Rodr{\'\i}guez Del Pino}, {Saxena}, {{\"U}bler}, {Venturi}, {Witstok}, {Baker}, {Bhatawdekar}, {Bonaventura}, {Boyett}, {Charlot}, {Danhaive}, {Hainline}, {Hausen}, {Helton}, {Ji}, {Ji}, {Jones}, {Joud{\v{z}}balis}, {Maseda}, {P{\'e}rez-Gonz{\'a}lez}, {Perna}, {Pusk{\'a}s}, {Shivaei}, {Silcock}, {Simmonds}, {Smit}, {Sun}, {Villanueva}, {Williams}, \& {Zhu}}]{DEugenio2024}
{D'Eugenio}, F., {Cameron}, A.~J., {Scholtz}, J., {et~al.} 2024, submitted to ApJS, arXiv:2404.06531

\bibitem[{{Dressler} {et~al.}(2023){Dressler}, {Vulcani}, {Treu}, {Rieke}, {Burns}, {Calabr{\`o}}, {Bonchi}, {Castellano}, {Fontana}, {Leethochawalit}, {Mason}, {Merlin}, {Morishita}, {Paris}, {Bradac}, {Mercurio}, {Nanayakkara}, {Poggianti}, {Santini}, {Wang}, {Misselt}, {Stark}, \& {Willmer}}]{Dressler2023}
{Dressler}, A., {Vulcani}, B., {Treu}, T., {et~al.} 2023, \apjl, 947, L27

\bibitem[{{Eisenstein} {et~al.}(2023){Eisenstein}, {Willott}, {Alberts}, {Arribas}, {Bonaventura}, {Bunker}, {Cameron}, {Carniani}, {Charlot}, {Curtis-Lake}, {D'Eugenio}, {Endsley}, {Ferruit}, {Giardino}, {Hainline}, {Hausen}, {Jakobsen}, {Johnson}, {Maiolino}, {Rieke}, {Rieke}, {Rix}, {Robertson}, {Stark}, {Tacchella}, {Williams}, {Willmer}, {Baker}, {Baum}, {Bhatawdekar}, {Boyett}, {Chen}, {Chevallard}, {Circosta}, {Curti}, {Danhaive}, {DeCoursey}, {de Graaff}, {Dressler}, {Egami}, {Helton}, {Hviding}, {Ji}, {Jones}, {Kumari}, {L{\"u}tzgendorf}, {Laseter}, {Looser}, {Lyu}, {Maseda}, {Nelson}, {Parlanti}, {Perna}, {Pusk{\'a}s}, {Rawle}, {Rodr{\'\i}guez Del Pino}, {Sandles}, {Saxena}, {Scholtz}, {Sharpe}, {Shivaei}, {Silcock}, {Simmonds}, {Skarbinski}, {Smit}, {Stone}, {Suess}, {Sun}, {Tang}, {Topping}, {{\"U}bler}, {Villanueva}, {Wallace}, {Whitler}, {Witstok}, \& {Woodrum}}]{Eisenstein2023}
{Eisenstein}, D.~J., {Willott}, C., {Alberts}, S., {et~al.} 2023, submitted to ApJS, arXiv:2306.02465

\bibitem[{{Ferrara} {et~al.}(2023){Ferrara}, {Pallottini}, \& {Dayal}}]{Ferrara2023}
{Ferrara}, A., {Pallottini}, A., \& {Dayal}, P. 2023, \mnras, 522, 3986

\bibitem[{{Ferreira} {et~al.}(2022){Ferreira}, {Adams}, {Conselice}, {Sazonova}, {Austin}, {Caruana}, {Ferrari}, {Verma}, {Trussler}, {Broadhurst}, {Diego}, {Frye}, {Pascale}, {Wilkins}, {Windhorst}, \& {Zitrin}}]{Ferreira2022}
{Ferreira}, L., {Adams}, N., {Conselice}, C.~J., {et~al.} 2022, \apjl, 938, L2

\bibitem[{{Ferreira} {et~al.}(2023){Ferreira}, {Conselice}, {Sazonova}, {Ferrari}, {Caruana}, {Tohill}, {Lucatelli}, {Adams}, {Irodotou}, {Marshall}, {Roper}, {Lovell}, {Verma}, {Austin}, {Trussler}, \& {Wilkins}}]{Ferreira2023}
{Ferreira}, L., {Conselice}, C.~J., {Sazonova}, E., {et~al.} 2023, \apj, 955, 94

\bibitem[{{Finkelstein} {et~al.}(2022){Finkelstein}, {Bagley}, {Arrabal Haro}, {Dickinson}, {Ferguson}, {Kartaltepe}, {Papovich}, {Burgarella}, {Kocevski}, {Huertas-Company}, {Iyer}, {Koekemoer}, {Larson}, {P{\'e}rez-Gonz{\'a}lez}, {Rose}, {Tacchella}, {Wilkins}, {Chworowsky}, {Medrano}, {Morales}, {Somerville}, {Yung}, {Fontana}, {Giavalisco}, {Grazian}, {Grogin}, {Kewley}, {Kirkpatrick}, {Kurczynski}, {Lotz}, {Pentericci}, {Pirzkal}, {Ravindranath}, {Ryan}, {Trump}, {Yang}, {Almaini}, {Amor{\'\i}n}, {Annunziatella}, {Backhaus}, {Barro}, {Behroozi}, {Bell}, {Bhatawdekar}, {Bisigello}, {Bromm}, {Buat}, {Buitrago}, {Calabr{\`o}}, {Casey}, {Castellano}, {Ch{\'a}vez Ortiz}, {Ciesla}, {Cleri}, {Cohen}, {Cole}, {Cooke}, {Cooper}, {Cooray}, {Costantin}, {Cox}, {Croton}, {Daddi}, {Dav{\'e}}, {de La Vega}, {Dekel}, {Elbaz}, {Estrada-Carpenter}, {Faber}, {Fern{\'a}ndez}, {Finkelstein}, {Freundlich}, {Fujimoto}, {Garc{\'\i}a-Argum{\'a}nez}, {Gardner}, {Gawiser}, {G{\'o}mez-Guijarro}, {Guo}, {Hamblin}, {Hamilton},
  {Hathi}, {Holwerda}, {Hirschmann}, {Hutchison}, {Jaskot}, {Jha}, {Jogee}, {Juneau}, {Jung}, {Kassin}, {Le Bail}, {Leung}, {Lucas}, {Magnelli}, {Mantha}, {Matharu}, {McGrath}, {McIntosh}, {Merlin}, {Mobasher}, {Newman}, {Nicholls}, {Pandya}, {Rafelski}, {Ronayne}, {Santini}, {Seill{\'e}}, {Shah}, {Shen}, {Simons}, {Snyder}, {Stanway}, {Straughn}, {Teplitz}, {Vanderhoof}, {Vega-Ferrero}, {Wang}, {Weiner}, {Willmer}, {Wuyts}, {Zavala}, \& {Ceers Team}}]{Finkelstein2022}
{Finkelstein}, S.~L., {Bagley}, M.~B., {Arrabal Haro}, P., {et~al.} 2022, \apjl, 940, L55

\bibitem[{{Finkelstein} {et~al.}(2023){Finkelstein}, {Bagley}, {Ferguson}, {Wilkins}, {Kartaltepe}, {Papovich}, {Yung}, {Arrabal Haro}, {Behroozi}, {Dickinson}, {Kocevski}, {Koekemoer}, {Larson}, {Le Bail}, {Morales}, {P{\'e}rez-Gonz{\'a}lez}, {Burgarella}, {Dav{\'e}}, {Hirschmann}, {Somerville}, {Wuyts}, {Bromm}, {Casey}, {Fontana}, {Fujimoto}, {Gardner}, {Giavalisco}, {Grazian}, {Grogin}, {Hathi}, {Hutchison}, {Jha}, {Jogee}, {Kewley}, {Kirkpatrick}, {Long}, {Lotz}, {Pentericci}, {Pierel}, {Pirzkal}, {Ravindranath}, {Ryan}, {Trump}, {Yang}, {Bhatawdekar}, {Bisigello}, {Buat}, {Calabr{\`o}}, {Castellano}, {Cleri}, {Cooper}, {Croton}, {Daddi}, {Dekel}, {Elbaz}, {Franco}, {Gawiser}, {Holwerda}, {Huertas-Company}, {Jaskot}, {Leung}, {Lucas}, {Mobasher}, {Pandya}, {Tacchella}, {Weiner}, \& {Zavala}}]{Finkelstein2023a}
{Finkelstein}, S.~L., {Bagley}, M.~B., {Ferguson}, H.~C., {et~al.} 2023, \apjl, 946, L13

\bibitem[{{Finkelstein} {et~al.}(2024){Finkelstein}, {Leung}, {Bagley}, {Dickinson}, {Ferguson}, {Papovich}, {Akins}, {Arrabal Haro}, {Dav{\'e}}, {Dekel}, {Kartaltepe}, {Kocevski}, {Koekemoer}, {Pirzkal}, {Somerville}, {Yung}, {Amor{\'\i}n}, {Backhaus}, {Behroozi}, {Bisigello}, {Bromm}, {Casey}, {Ch{\'a}vez Ortiz}, {Cheng}, {Chworowsky}, {Cleri}, {Cooper}, {Davis}, {de la Vega}, {Elbaz}, {Franco}, {Fontana}, {Fujimoto}, {Giavalisco}, {Grogin}, {Holwerda}, {Huertas-Company}, {Hirschmann}, {Iyer}, {Jogee}, {Jung}, {Larson}, {Lucas}, {Mobasher}, {Morales}, {Morley}, {Mukherjee}, {P{\'e}rez-Gonz{\'a}lez}, {Ravindranath}, {Rodighiero}, {Rowland}, {Tacchella}, {Taylor}, {Trump}, \& {Wilkins}}]{Finkelstein2024}
{Finkelstein}, S.~L., {Leung}, G. C.~K., {Bagley}, M.~B., {et~al.} 2024, \apjl, 969, L2

\bibitem[{{Fontana} {et~al.}(2000){Fontana}, {D'Odorico}, {Poli}, {Giallongo}, {Arnouts}, {Cristiani}, {Moorwood}, \& {Saracco}}]{Fontana2000}
{Fontana}, A., {D'Odorico}, S., {Poli}, F., {et~al.} 2000, \aj, 120, 2206

\bibitem[{{Galametz} {et~al.}(2013){Galametz}, {Grazian}, {Fontana}, {Ferguson}, {Ashby}, {Barro}, {Castellano}, {Dahlen}, {Donley}, {Faber}, {Grogin}, {Guo}, {Huang}, {Kocevski}, {Koekemoer}, {Lee}, {McGrath}, {Peth}, {Willner}, {Almaini}, {Cooper}, {Cooray}, {Conselice}, {Dickinson}, {Dunlop}, {Fazio}, {Foucaud}, {Gardner}, {Giavalisco}, {Hathi}, {Hartley}, {Koo}, {Lai}, {de Mello}, {McLure}, {Lucas}, {Paris}, {Pentericci}, {Santini}, {Simpson}, {Sommariva}, {Targett}, {Weiner}, {Wuyts}, \& {the CANDELS Team}}]{Galametz2013}
{Galametz}, A., {Grazian}, A., {Fontana}, A., {et~al.} 2013, \apjs, 206, 10

\bibitem[{{Gardner} {et~al.}(2023){Gardner}, {Mather}, {Abbott}, {Abell}, {Abernathy}, {Abney}, {Abraham}, {Abraham}, {Abul-Huda}, {Acton}, {Adams}, {Adams}, {Adler}, {Adriaensen}, {Aguilar}, {Ahmed}, {Ahmed}, {Ahmed}, {Albat}, {Albert}, {Alberts}, {Aldridge}, {Allen}, {Allen}, {Altenburg}, {Altunc}, {Alvarez}, {{\'A}lvarez-M{\'a}rquez}, {Alves de Oliveira}, {Ambrose}, {Anandakrishnan}, {Andersen}, {Anderson}, {Anderson}, {Anderson}, {Anderson}, {Aprea}, {Archer}, {Arenberg}, {Argyriou}, {Arribas}, {Artigau}, {Arvai}, {Atcheson}, {Atkinson}, {Averbukh}, {Aymergen}, {Bacinski}, {Baggett}, {Bagnasco}, {Baker}, {Balzano}, {Banks}, {Baran}, {Barker}, {Barrett}, {Barringer}, {Barto}, {Bast}, {Baudoz}, {Baum}, {Beatty}, {Beaulieu}, {Bechtold}, {Beck}, {Beddard}, {Beichman}, {Bellagama}, {Bely}, {Berger}, {Bergeron}, {Bernier}, {Bertch}, {Beskow}, {Betz}, {Biagetti}, {Birkmann}, {Bjorklund}, {Blackwood}, {Blazek}, {Blossfeld}, {Bluth}, {Boccaletti}, {Boegner}, {Bohlin}, {Boia}, {B{\"o}ker}, {Bonaventura}, {Bond},
  {Bosley}, {Boucarut}, {Bouchet}, {Bouwman}, {Bower}, {Bowers}, {Bowers}, {Boyce}, {Boyer}, {Boyer}, {Boyer}, {Boyer}, {Bradley}, {Brady}, {Brandl}, {Brannen}, {Breda}, {Bremmer}, {Brennan}, {Bresnahan}, {Bright}, {Broiles}, {Bromenschenkel}, {Brooks}, {Brooks}, {Brown}, {Brown}, {Brown}, {Bruce}, {Bryson}, {Bujanda}, {Bullock}, {Bunker}, {Bureo}, {Burt}, {Bush}, {Bushouse}, {Bussman}, {Cabaud}, {Cale}, {Calhoon}, {Calvani}, {Canipe}, {Caputo}, {Cara}, {Carey}, {Case}, {Cesari}, {Cetorelli}, {Chance}, {Chandler}, {Chaney}, {Chapman}, {Charlot}, {Chayer}, {Cheezum}, {Chen}, {Chen}, {Cherinka}, {Chichester}, {Chilton}, {Chittiraibalan}, {Clampin}, {Clark}, {Clark}, {Clark}, {Claybrooks}, {Cleveland}, {Cohen}, {Cohen}, {Col{\'o}n}, {Coleman}, {Colina}, {Comber}, {Comeau}, {Comer}, {Conde Reis}, {Connolly}, {Conroy}, {Contos}, {Contreras}, {Cook}, {Cooper}, {Cooper}, {Correia}, {Correnti}, {Cossou}, {Costanza}, {Coulais}, {Cox}, {Coyle}, {Cracraft}, {Crew}, {Curtis}, {Cusveller}, {Da Costa Maciel}, {Dailey},
  {Daugeron}, {Davidson}, {Davies}, {Davis}, {Davis}, {Day}, {de Chambure}, {de Jong}, {De Marchi}, {Dean}, {Decker}, {Delisa}, {Dell}, {Dellagatta}, {Dembinska}, {Demosthenes}, {Dencheva}, {Deneu}, {DePriest}, {Deschenes}, {Dethienne}, {Detre}, {Diaz}, {Dicken}, {DiFelice}, {Dillman}, {Disharoon}, {Dixon}, {Doggett}, {Dominguez}, {Donaldson}, {Doria-Warner}, {Santos}, {Doty}, {Douglas}, {Doyon}, {Dressler}, {Driggers}, {Driggers}, {Dunn}, {DuPrie}, {Dupuis}, {Durning}, {Dutta}, {Earl}, {Eccleston}, {Ecobichon}, {Egami}, {Ehrenwinkler}, {Eisenhamer}, {Eisenhower}, {Eisenstein}, {El Hamel}, {Elie}, {Elliott}, {Elliott}, {Engesser}, {Espinoza}, {Etienne}, {Etxaluze}, {Evans}, {Fabreguettes}, {Falcolini}, {Falini}, {Fatig}, {Feeney}, {Feinberg}, {Fels}, {Ferdous}, {Ferguson}, {Ferrarese}, {Ferreira}, {Ferruit}, {Ferry}, {Filippazzo}, {Firre}, {Fix}, {Flagey}, {Flanagan}, {Fleming}, {Florian}, {Flynn}, {Foiadelli}, {Fontaine}, {Fontanella}, {Forshay}, {Fortner}, {Fox}, {Framarini}, {Francisco}, {Franck}, {Franx},
  {Franz}, {Friedman}, {Friend}, {Frost}, {Fu}, {Fullerton}, {Gaillard}, {Galkin}, {Gallagher}, {Galyer}, {Garc{\'\i}a Mar{\'\i}n}, {Gardner}, {Garland}, {Garrett}, {Gasman}, {G{\'a}sp{\'a}r}, {Gastaud}, {Gaudreau}, {Gauthier}, {Geers}, {Geithner}, {Gennaro}, {Gerber}, {Gereau}, {Giampaoli}, {Giardino}, {Gibbons}, {Gilbert}, {Gilman}, {Girard}, {Giuliano}, {Gkountis}, {Glasse}, {Glassmire}, {Glauser}, {Glazer}, {Goldberg}, {Golimowski}, {Gonzaga}, {Gordon}, {Gordon}, {Goudfrooij}, {Gough}, {Graham}, {Grau}, {Green}, {Greene}, {Greene}, {Greenfield}, {Greenhouse}, {Greve}, {Greville}, {Grimaldi}, {Groe}, {Groebner}, {Grumm}, {Grundy}, {G{\"u}del}, {Guillard}, {Guldalian}, {Gunn}, {Gurule}, {Gutman}, {Guy}, {Guyot}, {Hack}, {Haderlein}, {Hagan}, {Hagedorn}, {Hainline}, {Haley}, {Hami}, {Hamilton}, {Hammann}, {Hammel}, {Hanley}, {Hansen}, {Hardy}, {Harnisch}, {Harr}, {Harris}, {Hart}, {Hartig}, {Hasan}, {Hashim}, {Hashimoto}, {Haskins}, {Hawkins}, {Hayden}, {Hayden}, {Healy}, {Hecht}, {Heeg}, {Hejal}, {Helm},
  {Hengemihle}, {Henning}, {Henry}, {Henry}, {Henshaw}, {Hernandez}, {Herrington}, {Heske}, {Hesman}, {Hickey}, {Hilbert}, {Hines}, {Hinz}, {Hirsch}, {Hitcho}, {Hodapp}, {Hodge}, {Hoffman}, {Holfeltz}, {Holler}, {Hoppa}, {Horner}, {Howard}, {Howard}, {Huber}, {Hunkeler}, {Hunter}, {Hunter}, {Hurd}, {Hurst}, {Hutchings}, {Hylan}, {Ignat}, {Illingworth}, {Irish}, {Isaacs}, {Jackson}, {Jaffe}, {Jahic}, {Jahromi}, {Jakobsen}, {James}, {James}, {James}, {Jamieson}, {Jandra}, {Jayawardhana}, {Jedrzejewski}, {Jeffers}, {Jensen}, {Joanne}, {Johns}, {Johnson}, {Johnson}, {Johnson}, {Johnson}, {Johnson}, {Johnson}, {Johnstone}, {Jollet}, {Jones}, {Jones}, {Jones}, {Jones}, {Jones}, {Jordan}, {Jordan}, {Jue}, {Jurkowski}, {Justis}, {Justtanont}, {Kaleida}, {Kalirai}, {Kalmanson}, {Kaltenegger}, {Kammerer}, {Kan}, {Kanarek}, {Kao}, {Karakla}, {Karl}, {Kassin}, {Kauffman}, {Kavanagh}, {Kelley}, {Kelly}, {Kendrew}, {Kennedy}, {Kenny}, {Keski-Kuha}, {Keyes}, {Khan}, {Kidwell}, {Kimble}, {King}, {King}, {Kinzel}, {Kirk},
  {Kirkpatrick}, {Klaassen}, {Klingemann}, {Klintworth}, {Knapp}, {Knight}, {Knollenberg}, {Knutsen}, {Koehler}, {Koekemoer}, {Kofler}, {Kontson}, {Kovacs}, {Kozhurina-Platais}, {Krause}, {Kriss}, {Krist}, {Kristoffersen}, {Krogel}, {Krueger}, {Kulp}, {Kumari}, {Kwan}, {Kyprianou}, {Labador}, {Labiano}, {Lafreni{\`e}re}, {Lagage}, {Laidler}, {Laine}, {Laird}, {Lajoie}, {Lallo}, {Lam}, {LaMassa}, {Lambros}, {Lampenfield}, {Lander}, {Langston}, {Larson}, {Larson}, {LaVerghetta}, {Law}, {Lawrence}, {Lee}, {Lee}, {Lee}, {Leisenring}, {Leveille}, {Levenson}, {Levi}, {Levine}, {Lewis}, {Lewis}, {Lewis}, {Libralato}, {Lidon}, {Liebrecht}, {Lightsey}, {Lilly}, {Lim}, {Lim}, {Ling}, {Link}, {Link}, {Lipinski}, {Liu}, {Lo}, {Lobmeyer}, {Logue}, {Long}, {Long}, {Long}, {Long}, {L{\'o}pez-Caniego}, {Lotz}, {Love-Pruitt}, {Lubskiy}, {Luers}, {Luetgens}, {Luevano}, {Lui}, {Lund}, {Lundquist}, {Lunine}, {L{\"u}tzgendorf}, {Lynch}, {MacDonald}, {MacDonald}, {Macias}, {Macklis}, {Maghami}, {Maharaja}, {Maiolino},
  {Makrygiannis}, {Malla}, {Malumuth}, {Manjavacas}, {Marini}, {Marrione}, {Marston}, {Martel}, {Martin}, {Martin}, {Martinez}, {Maschmann}, {Masci}, {Masetti}, {Maszkiewicz}, {Matthews}, {Matuskey}, {McBrayer}, {McCarthy}, {McCaughrean}, {McClare}, {McClare}, {McCloskey}, {McClurg}, {McCoy}, {McElwain}, {McGregor}, {McGuffey}, {McKay}, {McKenzie}, {McLean}, {McMaster}, {McNeil}, {De Meester}, {Mehalick}, {Meixner}, {Mel{\'e}ndez}, {Menzel}, {Menzel}, {Merz}, {Mesterharm}, {Meyer}, {Meyett}, {Meza}, {Midwinter}, {Milam}, {Miller}, {Miller}, {Miskey}, {Misselt}, {Mitchell}, {Mohan}, {Montoya}, {Moran}, {Morishita}, {Moro-Mart{\'\i}n}, {Morrison}, {Morrison}, {Morse}, {Moschos}, {Moseley}, {Mosier}, {Mosner}, {Mountain}, {Muckenthaler}, {Mueller}, {Mueller}, {Muhiem}, {M{\"u}hlmann}, {Mullally}, {Mullen}, {Munger}, {Murphy}, {Murray}, {Muzerolle}, {Mycroft}, {Myers}, {Myers}, {Myers}, {Myers}, {Myrick}, {Nagle}, {Nayak}, {Naylor}, {Neff}, {Nelan}, {Nella}, {Nguyen}, {Nguyen}, {Nickson}, {Nidhiry}, {Niedner},
  {Nieto-Santisteban}, {Nikolov}, {Nishisaka}, {Noriega-Crespo}, {Nota}, {O'Mara}, {Oboryshko}, {O'Brien}, {Ochs}, {Offenberg}, {Ogle}, {Ohl}, {Olmsted}, {Osborne}, {O'Shaughnessy}, {{\"O}stlin}, {O'Sullivan}, {Otor}, {Ottens}, {Ouellette}, {Outlaw}, {Owens}, {Pacifici}, {Page}, {Paranilam}, {Park}, {Parrish}, {Paschal}, {Patapis}, {Patel}, {Patrick}, {Pattishall}, {Paul}, {Paul}, {Pauly}, {Pavlovsky}, {Pe{\~n}a-Guerrero}, {Pedder}, {Peek}, {Pelham}, {Penanen}, {Perriello}, {Perrin}, {Perrine}, {Perrygo}, {Peslier}, {Petach}, {Peterson}, {Pfarr}, {Pierson}, {Pietraszkiewicz}, {Pilchen}, {Pipher}, {Pirzkal}, {Pitman}, {Player}, {Plesha}, {Plitzke}, {Pohner}, {Poletis}, {Pollizzi}, {Polster}, {Pontius}, {Pontoppidan}, {Porges}, {Potter}, {Prescott}, {Proffitt}, {Pueyo}, {Quispe Neira}, {Radich}, {Rager}, {Rameau}, {Ramey}, {Ramos Alarcon}, {Rampini}, {Rapp}, {Rashford}, {Rauscher}, {Ravindranath}, {Rawle}, {Rawlings}, {Ray}, {Regan}, {Rehm}, {Rehm}, {Reid}, {Reis}, {Renk}, {Reoch}, {Ressler}, {Rest},
  {Reynolds}, {Richon}, {Richon}, {Ridgaway}, {Riedel}, {Rieke}, {Rieke}, {Rifelli}, {Rigby}, {Riggs}, {Ringel}, {Ritchie}, {Rix}, {Robberto}, {Robinson}, {Robinson}, {Robinson}, {Rock}, {Rodriguez}, {Rodr{\'\i}guez del Pino}, {Roellig}, {Rohrbach}, {Roman}, {Romelfanger}, {Romo}, {Rosales}, {Rose}, {Roteliuk}, {Roth}, {Rothwell}, {Rouzaud}, {Rowe}, {Rowlands}, {Roy}, {Royer}, {Rui}, {Rumler}, {Rumpl}, {Russ}, {Ryan}, {Ryan}, {Saad}, {Sabata}, {Sabatino}, {Sabbi}, {Sabelhaus}, {Sabia}, {Sahu}, {Saif}, {Salvignol}, {Samara-Ratna}, {Samuelson}, {Sanders}, {Sappington}, {Sargent}, {Sauer}, {Savadkin}, {Sawicki}, {Schappell}, {Scheffer}, {Scheithauer}, {Scherer}, {Schiff}, {Schlawin}, {Schmeitzky}, {Schmitz}, {Schmude}, {Schneider}, {Schreiber}, {Schroeven-Deceuninck}, {Schultz}, {Schwab}, {Schwartz}, {Scoccimarro}, {Scott}, {Scott}, {Seaton}, {Seely}, {Seery}, {Seidleck}, {Sembach}, {Shanahan}, {Shaughnessy}, {Shaw}, {Shay}, {Sheehan}, {Sheth}, {Shih}, {Shivaei}, {Siegel}, {Sienkiewicz}, {Simmons}, {Simon},
  {Sirianni}, {Sivaramakrishnan}, {Slade}, {Sloan}, {Slocum}, {Slowinski}, {Smith}, {Smith}, {Smith}, {Smith}, {Smith}, {Smith}, {Smolik}, {Soderblom}, {Sohn}, {Sokol}, {Sonneborn}, {Sontag}, {Sooy}, {Soummer}, {Southwood}, {Spain}, {Sparmo}, {Speer}, {Spencer}, {Sprofera}, {Stallcup}, {Stanley}, {Stansberry}, {Stark}, {Starr}, {Stassi}, {Steck}, {Steeley}, {Stephens}, {Stephenson}, {Stewart}, {Stiavelli}, {}, {Strada}, {Straughn}, {Streetman}, {Strickland}, {Strobele}, {Stuhlinger}, {Stys}, {Such}, {Sukhatme}, {Sullivan}, {Sullivan}, {Sumner}, {Sun}, {Sunnquist}, {Swade}, {Swam}, {Swenton}, {Swoish}, {Tam Litten}, {Tamas}, {Tao}, {Taylor}, {Taylor}, {te Plate}, {Van Tea}, {Teague}, {Telfer}, {Temim}, {Texter}, {Thatte}, {Thompson}, {Thompson}, {Thomson}, {Thronson}, {Tierney}, {Tikkanen}, {Tinnin}, {Tippet}, {Todd}, {Tran}, {Trauger}, {Trejo}, {Vinh Truong}, {Tsukamoto}, {Tufail}, {Tumlinson}, {Tustain}, {Tyra}, {Ubeda}, {Underwood}, {Uzzo}, {Vaclavik}, {Valenduc}, {Valenti}, {Van Campen}, {van de Wetering},
  {Van Der Marel}, {van Haarlem}, {Vandenbussche}, {van Dishoeck}, {Vanterpool}, {Vernoy}, {Vila Costas}, {Volk}, {Voorzaat}, {Voyton}, {Vydra}, {Waddy}, {Waelkens}, {Wahlgren}, {Walker}, {Wander}, {Warfield}, {Warner}, {Wasiak}, {Wasiak}, {Wehner}, {Weiler}, {Weilert}, {Weiss}, {Wells}, {Welty}, {Wheate}, {Wheeler}, {White}, {Whitehouse}, {Whiteleather}, {Whitman}, {Williams}, {Willmer}, {Willott}, {Willoughby}, {Wilson}, {Wilson}, {Wilson}, {Windhorst}, {Wislowski}, {Wolfe}, {Wolfe}, {Wolff}, {Wondel}, {Woo}, {Woods}, {Worden}, {Workman}, {Wright}, {Wu}, {Wu}, {Wun}, {Wymer}, {Yadetie}, {Yan}, {Yang}, {Yates}, {Yeager}, {Yerger}, {Young}, {Young}, {Yu}, {Yu}, {Zak}, {Zeidler}, {Zepp}, {Zhou}, {Zincke}, {Zonak}, \& {Zondag}}]{Gardner2023}
{Gardner}, J.~P., {Mather}, J.~C., {Abbott}, R., {et~al.} 2023, \pasp, 135, 068001

\bibitem[{{Gardner} {et~al.}(2006){Gardner}, {Mather}, {Clampin}, {Doyon}, {Greenhouse}, {Hammel}, {Hutchings}, {Jakobsen}, {Lilly}, {Long}, {Lunine}, {McCaughrean}, {Mountain}, {Nella}, {Rieke}, {Rieke}, {Rix}, {Smith}, {Sonneborn}, {Stiavelli}, {Stockman}, {Windhorst}, \& {Wright}}]{Gardner2006}
{Gardner}, J.~P., {Mather}, J.~C., {Clampin}, M., {et~al.} 2006, \ssr, 123, 485

\bibitem[{{Garilli} {et~al.}(2021){Garilli}, {McLure}, {Pentericci}, {Franzetti}, {Gargiulo}, {Carnall}, {Cucciati}, {Iovino}, {Amorin}, {Bolzonella}, {Bongiorno}, {Castellano}, {Cimatti}, {Cirasuolo}, {Cullen}, {Dunlop}, {Elbaz}, {Finkelstein}, {Fontana}, {Fontanot}, {Fumana}, {Guaita}, {Hartley}, {Jarvis}, {Juneau}, {Maccagni}, {McLeod}, {Nandra}, {Pompei}, {Pozzetti}, {Scodeggio}, {Talia}, {Calabr{\`o}}, {Cresci}, {Fynbo}, {Hathi}, {Hibon}, {Koekemoer}, {Magliocchetti}, {Salvato}, {Vietri}, {Zamorani}, {Almaini}, {Balestra}, {Bardelli}, {Begley}, {Brammer}, {Bell}, {Bowler}, {Brusa}, {Buitrago}, {Caputi}, {Cassata}, {Charlot}, {Citro}, {Cristiani}, {Curtis-Lake}, {Dickinson}, {Fazio}, {Ferguson}, {Fiore}, {Franco}, {Georgakakis}, {Giavalisco}, {Grazian}, {Hamadouche}, {Jung}, {Kim}, {Khusanova}, {Le F{\`e}vre}, {Longhetti}, {Lotz}, {Mannucci}, {Maltby}, {Matsuoka}, {Mendez-Hernandez}, {Mendez-Abreu}, {Mignoli}, {Moresco}, {Nonino}, {Pannella}, {Papovich}, {Popesso}, {Roberts-Borsani}, {Rosario},
  {Saldana-Lopez}, {Santini}, {Saxena}, {Schaerer}, {Schreiber}, {Stark}, {Tasca}, {Thomas}, {Vanzella}, {Wild}, {Williams}, \& {Zucca}}]{Garilli2021}
{Garilli}, B., {McLure}, R., {Pentericci}, L., {et~al.} 2021, \aap, 647, A150

\bibitem[{{Giacconi} {et~al.}(2002){Giacconi}, {Zirm}, {Wang}, {Rosati}, {Nonino}, {Tozzi}, {Gilli}, {Mainieri}, {Hasinger}, {Kewley}, {Bergeron}, {Borgani}, {Gilmozzi}, {Grogin}, {Koekemoer}, {Schreier}, {Zheng}, \& {Norman}}]{Giacconi2002}
{Giacconi}, R., {Zirm}, A., {Wang}, J., {et~al.} 2002, \apjs, 139, 369

\bibitem[{{Giavalisco} {et~al.}(2004){Giavalisco}, {Ferguson}, {Koekemoer}, {Dickinson}, {Alexander}, {Bauer}, {Bergeron}, {Biagetti}, {Brandt}, {Casertano}, {Cesarsky}, {Chatzichristou}, {Conselice}, {Cristiani}, {Da Costa}, {Dahlen}, {de Mello}, {Eisenhardt}, {Erben}, {Fall}, {Fassnacht}, {Fosbury}, {Fruchter}, {Gardner}, {Grogin}, {Hook}, {Hornschemeier}, {Idzi}, {Jogee}, {Kretchmer}, {Laidler}, {Lee}, {Livio}, {Lucas}, {Madau}, {Mobasher}, {Moustakas}, {Nonino}, {Padovani}, {Papovich}, {Park}, {Ravindranath}, {Renzini}, {Richardson}, {Riess}, {Rosati}, {Schirmer}, {Schreier}, {Somerville}, {Spinrad}, {Stern}, {Stiavelli}, {Strolger}, {Urry}, {Vandame}, {Williams}, \& {Wolf}}]{Giavalisco2004}
{Giavalisco}, M., {Ferguson}, H.~C., {Koekemoer}, A.~M., {et~al.} 2004, \apjl, 600, L93

\bibitem[{{Glazebrook} {et~al.}(2023){Glazebrook}, {Nanayakkara}, {Jacobs}, {Leethochawalit}, {Calabr{\`o}}, {Bonchi}, {Castellano}, {Fontana}, {Mason}, {Merlin}, {Morishita}, {Paris}, {Trenti}, {Treu}, {Santini}, {Wang}, {Boyett}, {Bradac}, {Brammer}, {Jones}, {Marchesini}, {Nonino}, \& {Vulcani}}]{Glazebrook2023}
{Glazebrook}, K., {Nanayakkara}, T., {Jacobs}, C., {et~al.} 2023, \apjl, 947, L25

\bibitem[{{Goulding} {et~al.}(2023){Goulding}, {Greene}, {Setton}, {Labbe}, {Bezanson}, {Miller}, {Atek}, {Bogd{\'a}n}, {Brammer}, {Chemerynska}, {Cutler}, {Dayal}, {Fudamoto}, {Fujimoto}, {Furtak}, {Kokorev}, {Khullar}, {Leja}, {Marchesini}, {Natarajan}, {Nelson}, {Oesch}, {Pan}, {Papovich}, {Price}, {van Dokkum}, {Wang}, {Weaver}, {Whitaker}, \& {Zitrin}}]{Goulding2023}
{Goulding}, A.~D., {Greene}, J.~E., {Setton}, D.~J., {et~al.} 2023, \apjl, 955, L24

\bibitem[{{Grazian} {et~al.}(2006){Grazian}, {Fontana}, {de Santis}, {Nonino}, {Salimbeni}, {Giallongo}, {Cristiani}, {Gallozzi}, \& {Vanzella}}]{Grazian2006}
{Grazian}, A., {Fontana}, A., {de Santis}, C., {et~al.} 2006, \aap, 449, 951

\bibitem[{{Greene} {et~al.}(2024){Greene}, {Labbe}, {Goulding}, {Furtak}, {Chemerynska}, {Kokorev}, {Dayal}, {Volonteri}, {Williams}, {Wang}, {Setton}, {Burgasser}, {Bezanson}, {Atek}, {Brammer}, {Cutler}, {Feldmann}, {Fujimoto}, {Glazebrook}, {de Graaff}, {Khullar}, {Leja}, {Marchesini}, {Maseda}, {Matthee}, {Miller}, {Naidu}, {Nanayakkara}, {Oesch}, {Pan}, {Papovich}, {Price}, {van Dokkum}, {Weaver}, {Whitaker}, \& {Zitrin}}]{Greene2024}
{Greene}, J.~E., {Labbe}, I., {Goulding}, A.~D., {et~al.} 2024, \apj, 964, 39

\bibitem[{{Grogin} {et~al.}(2011){Grogin}, {Kocevski}, {Faber}, {Ferguson}, {Koekemoer}, {Riess}, {Acquaviva}, {Alexander}, {Almaini}, {Ashby}, {Barden}, {Bell}, {Bournaud}, {Brown}, {Caputi}, {Casertano}, {Cassata}, {Castellano}, {Challis}, {Chary}, {Cheung}, {Cirasuolo}, {Conselice}, {Roshan Cooray}, {Croton}, {Daddi}, {Dahlen}, {Dav{\'e}}, {de Mello}, {Dekel}, {Dickinson}, {Dolch}, {Donley}, {Dunlop}, {Dutton}, {Elbaz}, {Fazio}, {Filippenko}, {Finkelstein}, {Fontana}, {Gardner}, {Garnavich}, {Gawiser}, {Giavalisco}, {Grazian}, {Guo}, {Hathi}, {H{\"a}ussler}, {Hopkins}, {Huang}, {Huang}, {Jha}, {Kartaltepe}, {Kirshner}, {Koo}, {Lai}, {Lee}, {Li}, {Lotz}, {Lucas}, {Madau}, {McCarthy}, {McGrath}, {McIntosh}, {McLure}, {Mobasher}, {Moustakas}, {Mozena}, {Nandra}, {Newman}, {Niemi}, {Noeske}, {Papovich}, {Pentericci}, {Pope}, {Primack}, {Rajan}, {Ravindranath}, {Reddy}, {Renzini}, {Rix}, {Robaina}, {Rodney}, {Rosario}, {Rosati}, {Salimbeni}, {Scarlata}, {Siana}, {Simard}, {Smidt}, {Somerville}, {Spinrad},
  {Straughn}, {Strolger}, {Telford}, {Teplitz}, {Trump}, {van der Wel}, {Villforth}, {Wechsler}, {Weiner}, {Wiklind}, {Wild}, {Wilson}, {Wuyts}, {Yan}, \& {Yun}}]{Grogin2011}
{Grogin}, N.~A., {Kocevski}, D.~D., {Faber}, S.~M., {et~al.} 2011, 197, 35

\bibitem[{{Guo} {et~al.}(2013){Guo}, {Ferguson}, {Giavalisco}, {Barro}, {Willner}, {Ashby}, {Dahlen}, {Donley}, {Faber}, {Fontana}, {Galametz}, {Grazian}, {Huang}, {Kocevski}, {Koekemoer}, {Koo}, {McGrath}, {Peth}, {Salvato}, {Wuyts}, {Castellano}, {Cooray}, {Dickinson}, {Dunlop}, {Fazio}, {Gardner}, {Gawiser}, {Grogin}, {Hathi}, {Hsu}, {Lee}, {Lucas}, {Mobasher}, {Nandra}, {Newman}, \& {van der Wel}}]{Guo2013}
{Guo}, Y., {Ferguson}, H.~C., {Giavalisco}, M., {et~al.} 2013, 207, 24

\bibitem[{{Harikane} {et~al.}(2024){Harikane}, {Nakajima}, {Ouchi}, {Umeda}, {Isobe}, {Ono}, {Xu}, \& {Zhang}}]{Harikane2024}
{Harikane}, Y., {Nakajima}, K., {Ouchi}, M., {et~al.} 2024, \apj, 960, 56

\bibitem[{{Harikane} {et~al.}(2023){Harikane}, {Zhang}, {Nakajima}, {Ouchi}, {Isobe}, {Ono}, {Hatano}, {Xu}, \& {Umeda}}]{Harikane2023b}
{Harikane}, Y., {Zhang}, Y., {Nakajima}, K., {et~al.} 2023, \apj, 959, 39

\bibitem[{{Harvey} {et~al.}(2024){Harvey}, {Conselice}, {Adams}, {Austin}, {Juodzbalis}, {Trussler}, {Li}, {Ormerod}, {Ferreira}, {Duan}, {Westcott}, {Harris}, {Bhatawdekar}, {Coe}, {Cohen}, {Caruana}, {Cheng}, {Driver}, {Frye}, {Furtak}, {Grogin}, {Hathi}, {Holwerda}, {Jansen}, {Koekemoer}, {Lovell}, {Marshall}, {Nonino}, {Smail}, {Vijayan}, {Wilkins}, {Windhorst}, {Willmer}, {Yan}, \& {Zitrin}}]{Harvey2024}
{Harvey}, T., {Conselice}, C., {Adams}, N.~J., {et~al.} 2024, submitted to ApJ, arXiv:2403.03908

\bibitem[{{Hasinger} {et~al.}(2018){Hasinger}, {Capak}, {Salvato}, {Barger}, {Cowie}, {Faisst}, {Hemmati}, {Kakazu}, {Kartaltepe}, {Masters}, {Mobasher}, {Nayyeri}, {Sanders}, {Scoville}, {Suh}, {Steinhardt}, \& {Yang}}]{Hasinger2018}
{Hasinger}, G., {Capak}, P., {Salvato}, M., {et~al.} 2018, \apj, 858, 77

\bibitem[{{Heintz} {et~al.}(2024){Heintz}, {Brammer}, {Watson}, {Oesch}, {Keating}, {Hayes}, {Abdurro'uf}, {Arellano-C{\'o}rdova}, {Carnall}, {Christiansen}, {Cullen}, {Dav{\'e}}, {Dayal}, {Ferrara}, {Finlator}, {Fynbo}, {Flury}, {Gelli}, {Gillman}, {Gottumukkala}, {Gould}, {Greve}, {Hardin}, {Y. -Y Hsiao}, {Hutter}, {Jakobsson}, {Killi}, {Khosravaninezhad}, {Laursen}, {Lee}, {Magdis}, {Matthee}, {Naidu}, {Narayanan}, {Pollock}, {Prescott}, {Rusakov}, {Shuntov}, {Sneppen}, {Smit}, {Tanvir}, {Terp}, {Toft}, {Valentino}, {Vijayan}, {Weaver}, {Wise}, \& {Witstok}}]{Heintz2024}
{Heintz}, K.~E., {Brammer}, G.~B., {Watson}, D., {et~al.} 2024, submitted to \aap, arXiv:2404.02211

\bibitem[{{Illingworth} {et~al.}(2016){Illingworth}, {Magee}, {Bouwens}, {Oesch}, {Labbe}, {van Dokkum}, {Whitaker}, {Holden}, {Franx}, \& {Gonzalez}}]{Illingworth2016}
{Illingworth}, G., {Magee}, D., {Bouwens}, R., {et~al.} 2016, arXiv e-prints, arXiv:1606.00841

\bibitem[{{Inami} {et~al.}(2017){Inami}, {Bacon}, {Brinchmann}, {Richard}, {Contini}, {Conseil}, {Hamer}, {Akhlaghi}, {Bouch{\'e}}, {Cl{\'e}ment}, {Desprez}, {Drake}, {Hashimoto}, {Leclercq}, {Maseda}, {Michel-Dansac}, {Paalvast}, {Tresse}, {Ventou}, {Kollatschny}, {Boogaard}, {Finley}, {Marino}, {Schaye}, \& {Wisotzki}}]{Inami2017}
{Inami}, H., {Bacon}, R., {Brinchmann}, J., {et~al.} 2017, \aap, 608, A2

\bibitem[{{Jacobs} {et~al.}(2023){Jacobs}, {Glazebrook}, {Calabr{\`o}}, {Treu}, {Nannayakkara}, {Jones}, {Merlin}, {Abraham}, {Stevens}, {Vulcani}, {Yang}, {Bonchi}, {Boyett}, {Brada{\v{c}}}, {Castellano}, {Fontana}, {Marchesini}, {Malkan}, {Mason}, {Morishita}, {Paris}, {Santini}, {Trenti}, \& {Wang}}]{Jacobs2023}
{Jacobs}, C., {Glazebrook}, K., {Calabr{\`o}}, A., {et~al.} 2023, \apjl, 948, L13

\bibitem[{{Jones} {et~al.}(2021){Jones}, {Rosenthal}, {Barger}, \& {Cowie}}]{Jones2021}
{Jones}, L.~H., {Rosenthal}, M.~J., {Barger}, A.~J., \& {Cowie}, L.~L. 2021, \apj, 916, 46

\bibitem[{{Kartaltepe} {et~al.}(2023){Kartaltepe}, {Rose}, {Vanderhoof}, {McGrath}, {Costantin}, {Cox}, {Yung}, {Kocevski}, {Wuyts}, {Ferguson}, {Bagley}, {Finkelstein}, {Amor{\'\i}n}, {Andrews}, {Arrabal Haro}, {Backhaus}, {Behroozi}, {Bisigello}, {Calabr{\`o}}, {Casey}, {Coogan}, {Cooper}, {Croton}, {de la Vega}, {Dickinson}, {Fontana}, {Franco}, {Grazian}, {Grogin}, {Hathi}, {Holwerda}, {Huertas-Company}, {Iyer}, {Jogee}, {Jung}, {Kewley}, {Kirkpatrick}, {Koekemoer}, {Liu}, {Lotz}, {Lucas}, {Newman}, {Pacifici}, {Pandya}, {Papovich}, {Pentericci}, {P{\'e}rez-Gonz{\'a}lez}, {Petersen}, {Pirzkal}, {Rafelski}, {Ravindranath}, {Simons}, {Snyder}, {Somerville}, {Stanway}, {Straughn}, {Tacchella}, {Trump}, {Vega-Ferrero}, {Wilkins}, {Yang}, \& {Zavala}}]{Kartaltepe2023}
{Kartaltepe}, J.~S., {Rose}, C., {Vanderhoof}, B.~N., {et~al.} 2023, \apjl, 946, L15

\bibitem[{{Kirkpatrick} {et~al.}(2023){Kirkpatrick}, {Yang}, {Le Bail}, {Troiani}, {Bell}, {Cleri}, {Elbaz}, {Finkelstein}, {Hathi}, {Hirschmann}, {Holwerda}, {Kocevski}, {Lucas}, {McKinney}, {Papovich}, {P{\'e}rez-Gonz{\'a}lez}, {de la Vega}, {Bagley}, {Daddi}, {Dickinson}, {Ferguson}, {Fontana}, {Grazian}, {Grogin}, {Arrabal Haro}, {Kartaltepe}, {Kewley}, {Koekemoer}, {Lotz}, {Pentericci}, {Pirzkal}, {Ravindranath}, {Somerville}, {Trump}, {Wilkins}, \& {Yung}}]{Kirkpatrick2023}
{Kirkpatrick}, A., {Yang}, G., {Le Bail}, A., {et~al.} 2023, \apjl, 959, L7

\bibitem[{{Kocevski} {et~al.}(2018){Kocevski}, {Hasinger}, {Brightman}, {Nandra}, {Georgakakis}, {Cappelluti}, {Civano}, {Li}, {Li}, {Aird}, {Alexander}, {Almaini}, {Brusa}, {Buchner}, {Comastri}, {Conselice}, {Dickinson}, {Finoguenov}, {Gilli}, {Koekemoer}, {Miyaji}, {Mullaney}, {Papovich}, {Rosario}, {Salvato}, {Silverman}, {Somerville}, \& {Ueda}}]{Kocevski2018}
{Kocevski}, D.~D., {Hasinger}, G., {Brightman}, M., {et~al.} 2018, \apjs, 236, 48

\bibitem[{{Kocevski} {et~al.}(2023){Kocevski}, {Onoue}, {Inayoshi}, {Trump}, {Arrabal Haro}, {Grazian}, {Dickinson}, {Finkelstein}, {Kartaltepe}, {Hirschmann}, {Aird}, {Holwerda}, {Fujimoto}, {Juneau}, {Amor{\'\i}n}, {Backhaus}, {Bagley}, {Barro}, {Bell}, {Bisigello}, {Calabr{\`o}}, {Cleri}, {Cooper}, {Ding}, {Grogin}, {Ho}, {Hutchison}, {Inoue}, {Jiang}, {Jones}, {Koekemoer}, {Li}, {Li}, {McGrath}, {Molina}, {Papovich}, {P{\'e}rez-Gonz{\'a}lez}, {Pirzkal}, {Wilkins}, {Yang}, \& {Yung}}]{Kocevski2023}
{Kocevski}, D.~D., {Onoue}, M., {Inayoshi}, K., {et~al.} 2023, \apjl, 954, L4

\bibitem[{{Kodra} {et~al.}(2023){Kodra}, {Andrews}, {Newman}, {Finkelstein}, {Fontana}, {Hathi}, {Salvato}, {Wiklind}, {Wuyts}, {Broussard}, {Chartab}, {Conselice}, {Cooper}, {Dekel}, {Dickinson}, {Ferguson}, {Gawiser}, {Grogin}, {Iyer}, {Kartaltepe}, {Kassin}, {Koekemoer}, {Koo}, {Lucas}, {Mantha}, {McIntosh}, {Mobasher}, {Pacifici}, {P{\'e}rez-Gonz{\'a}lez}, \& {Santini}}]{Kodra2023}
{Kodra}, D., {Andrews}, B.~H., {Newman}, J.~A., {et~al.} 2023, \apj, 942, 36

\bibitem[{{Koekemoer} {et~al.}(2011){Koekemoer}, {Faber}, {Ferguson}, {Grogin}, {Kocevski}, {Koo}, {Lai}, {Lotz}, {Lucas}, {McGrath}, {Ogaz}, {Rajan}, {Riess}, {Rodney}, {Strolger}, {Casertano}, {Castellano}, {Dahlen}, {Dickinson}, {Dolch}, {Fontana}, {Giavalisco}, {Grazian}, {Guo}, {Hathi}, {Huang}, {van der Wel}, {Yan}, {Acquaviva}, {Alexander}, {Almaini}, {Ashby}, {Barden}, {Bell}, {Bournaud}, {Brown}, {Caputi}, {Cassata}, {Challis}, {Chary}, {Cheung}, {Cirasuolo}, {Conselice}, {Roshan Cooray}, {Croton}, {Daddi}, {Dav{\'e}}, {de Mello}, {de Ravel}, {Dekel}, {Donley}, {Dunlop}, {Dutton}, {Elbaz}, {Fazio}, {Filippenko}, {Finkelstein}, {Frazer}, {Gardner}, {Garnavich}, {Gawiser}, {Gruetzbauch}, {Hartley}, {H{\"a}ussler}, {Herrington}, {Hopkins}, {Huang}, {Jha}, {Johnson}, {Kartaltepe}, {Khostovan}, {Kirshner}, {Lani}, {Lee}, {Li}, {Madau}, {McCarthy}, {McIntosh}, {McLure}, {McPartland}, {Mobasher}, {Moreira}, {Mortlock}, {Moustakas}, {Mozena}, {Nandra}, {Newman}, {Nielsen}, {Niemi}, {Noeske}, {Papovich},
  {Pentericci}, {Pope}, {Primack}, {Ravindranath}, {Reddy}, {Renzini}, {Rix}, {Robaina}, {Rosario}, {Rosati}, {Salimbeni}, {Scarlata}, {Siana}, {Simard}, {Smidt}, {Snyder}, {Somerville}, {Spinrad}, {Straughn}, {Telford}, {Teplitz}, {Trump}, {Vargas}, {Villforth}, {Wagner}, {Wandro}, {Wechsler}, {Weiner}, {Wiklind}, {Wild}, {Wilson}, {Wuyts}, \& {Yun}}]{Koekemoer2011}
{Koekemoer}, A.~M., {Faber}, S.~M., {Ferguson}, H.~C., {et~al.} 2011, \apjs, 197, 36

\bibitem[{{Kokorev} {et~al.}(2024){Kokorev}, {Caputi}, {Greene}, {Dayal}, {Trebitsch}, {Cutler}, {Fujimoto}, {Labb{\'e}}, {Miller}, {Iani}, {Navarro-Carrera}, \& {Rinaldi}}]{Kokorev2024}
{Kokorev}, V., {Caputi}, K.~I., {Greene}, J.~E., {et~al.} 2024, \apj, 968, 38

\bibitem[{{Kron}(1980)}]{Kron1980}
{Kron}, R.~G. 1980, \apjs, 43, 305

\bibitem[{{Larson} {et~al.}(2023{\natexlab{a}}){Larson}, {Finkelstein}, {Kocevski}, {Hutchison}, {Trump}, {Arrabal Haro}, {Bromm}, {Cleri}, {Dickinson}, {Fujimoto}, {Kartaltepe}, {Koekemoer}, {Papovich}, {Pirzkal}, {Tacchella}, {Zavala}, {Bagley}, {Behroozi}, {Champagne}, {Cole}, {Jung}, {Morales}, {Yang}, {Zhang}, {Zitrin}, {Amor{\'\i}n}, {Burgarella}, {Casey}, {Ch{\'a}vez Ortiz}, {Cox}, {Chworowsky}, {Fontana}, {Gawiser}, {Grazian}, {Grogin}, {Harish}, {Hathi}, {Hirschmann}, {Holwerda}, {Juneau}, {Leung}, {Lucas}, {McGrath}, {P{\'e}rez-Gonz{\'a}lez}, {Rigby}, {Seill{\'e}}, {Simons}, {de La Vega}, {Weiner}, {Wilkins}, {Yung}, \& {Ceers Team}}]{Larson2023b}
{Larson}, R.~L., {Finkelstein}, S.~L., {Kocevski}, D.~D., {et~al.} 2023{\natexlab{a}}, \apjl, 953, L29

\bibitem[{{Larson} {et~al.}(2023{\natexlab{b}}){Larson}, {Hutchison}, {Bagley}, {Finkelstein}, {Yung}, {Somerville}, {Hirschmann}, {Brammer}, {Holwerda}, {Papovich}, {Morales}, \& {Wilkins}}]{Larson2023}
{Larson}, R.~L., {Hutchison}, T.~A., {Bagley}, M., {et~al.} 2023{\natexlab{b}}, \apj, 958, 141

\bibitem[{{Lawrence} {et~al.}(2007){Lawrence}, {Warren}, {Almaini}, {Edge}, {Hambly}, {Jameson}, {Lucas}, {Casali}, {Adamson}, {Dye}, {Emerson}, {Foucaud}, {Hewett}, {Hirst}, {Hodgkin}, {Irwin}, {Lodieu}, {McMahon}, {Simpson}, {Smail}, {Mortlock}, \& {Folger}}]{Lawrence2007}
{Lawrence}, A., {Warren}, S.~J., {Almaini}, O., {et~al.} 2007, \mnras, 379, 1599

\bibitem[{{Leung} {et~al.}(2023){Leung}, {Bagley}, {Finkelstein}, {Ferguson}, {Koekemoer}, {P{\'e}rez-Gonz{\'a}lez}, {Morales}, {Kocevski}, {Yang}, {Somerville}, {Wilkins}, {Yung}, {Fujimoto}, {Larson}, {Papovich}, {Pirzkal}, {Berg}, {Lotz}, {Castellano}, {Ch{\'a}vez Ortiz}, {Cheng}, {Dickinson}, {Giavalisco}, {Hathi}, {Hutchison}, {Jung}, {Kartaltepe}, {Natarajan}, \& {Rothberg}}]{Leung2023}
{Leung}, G. C.~K., {Bagley}, M.~B., {Finkelstein}, S.~L., {et~al.} 2023, \apjl, 954, L46

\bibitem[{{Lilly} {et~al.}(2007){Lilly}, {Le F{\`e}vre}, {Renzini}, {Zamorani}, {Scodeggio}, {Contini}, {Carollo}, {Hasinger}, {Kneib}, {Iovino}, {Le Brun}, {Maier}, {Mainieri}, {Mignoli}, {Silverman}, {Tasca}, {Bolzonella}, {Bongiorno}, {Bottini}, {Capak}, {Caputi}, {Cimatti}, {Cucciati}, {Daddi}, {Feldmann}, {Franzetti}, {Garilli}, {Guzzo}, {Ilbert}, {Kampczyk}, {Kovac}, {Lamareille}, {Leauthaud}, {Le Borgne}, {McCracken}, {Marinoni}, {Pello}, {Ricciardelli}, {Scarlata}, {Vergani}, {Sanders}, {Schinnerer}, {Scoville}, {Taniguchi}, {Arnouts}, {Aussel}, {Bardelli}, {Brusa}, {Cappi}, {Ciliegi}, {Finoguenov}, {Foucaud}, {Franceschini}, {Halliday}, {Impey}, {Knobel}, {Koekemoer}, {Kurk}, {Maccagni}, {Maddox}, {Marano}, {Marconi}, {Meneux}, {Mobasher}, {Moreau}, {Peacock}, {Porciani}, {Pozzetti}, {Scaramella}, {Schiminovich}, {Shopbell}, {Smail}, {Thompson}, {Tresse}, {Vettolani}, {Zanichelli}, \& {Zucca}}]{Lilly2007}
{Lilly}, S.~J., {Le F{\`e}vre}, O., {Renzini}, A., {et~al.} 2007, \apjs, 172, 70

\bibitem[{{Looser} {et~al.}(2023){Looser}, {D'Eugenio}, {Maiolino}, {Tacchella}, {Curti}, {Arribas}, {Baker}, {Baum}, {Bonaventura}, {Boyett}, {Bunker}, {Carniani}, {Charlot}, {Chevallard}, {Curtis-Lake}, {Danhaive}, {Eisenstein}, {de Graaff}, {Hainline}, {Ji}, {Johnson}, {Kumari}, {Nelson}, {Parlanti}, {Rix}, {Robertson}, {Rodr{\'\i}guez Del Pino}, {Sandles}, {Scholtz}, {Smit}, {Stark}, {{\"U}bler}, {Williams}, {Willott}, \& {Witstok}}]{Looser2023}
{Looser}, T.~J., {D'Eugenio}, F., {Maiolino}, R., {et~al.} 2023, arXiv e-prints, arXiv:2306.02470

\bibitem[{{Lotz} {et~al.}(2017){Lotz}, {Koekemoer}, {Coe}, {Grogin}, {Capak}, {Mack}, {Anderson}, {Avila}, {Barker}, {Borncamp}, {Brammer}, {Durbin}, {Gunning}, {Hilbert}, {Jenkner}, {Khandrika}, {Levay}, {Lucas}, {MacKenty}, {Ogaz}, {Porterfield}, {Reid}, {Robberto}, {Royle}, {Smith}, {Storrie-Lombardi}, {Sunnquist}, {Surace}, {Taylor}, {Williams}, {Bullock}, {Dickinson}, {Finkelstein}, {Natarajan}, {Richard}, {Robertson}, {Tumlinson}, {Zitrin}, {Flanagan}, {Sembach}, {Soifer}, \& {Mountain}}]{Lotz2017}
{Lotz}, J.~M., {Koekemoer}, A., {Coe}, D., {et~al.} 2017, \apj, 837, 97

\bibitem[{{Luo} {et~al.}(2017){Luo}, {Brandt}, {Xue}, {Lehmer}, {Alexander}, {Bauer}, {Vito}, {Yang}, {Basu-Zych}, {Comastri}, {Gilli}, {Gu}, {Hornschemeier}, {Koekemoer}, {Liu}, {Mainieri}, {Paolillo}, {Ranalli}, {Rosati}, {Schneider}, {Shemmer}, {Smail}, {Sun}, {Tozzi}, {Vignali}, \& {Wang}}]{Luo2017}
{Luo}, B., {Brandt}, W.~N., {Xue}, Y.~Q., {et~al.} 2017, \apjs, 228, 2

\bibitem[{{Mahler} {et~al.}(2018){Mahler}, {Richard}, {Cl{\'e}ment}, {Lagattuta}, {Schmidt}, {Patr{\'\i}cio}, {Soucail}, {Bacon}, {Pello}, {Bouwens}, {Maseda}, {Martinez}, {Carollo}, {Inami}, {Leclercq}, \& {Wisotzki}}]{Mahler2018}
{Mahler}, G., {Richard}, J., {Cl{\'e}ment}, B., {et~al.} 2018, \mnras, 473, 663

\bibitem[{{Maiolino} {et~al.}(2024){Maiolino}, {Scholtz}, {Curtis-Lake}, {Carniani}, {Baker}, {de Graaff}, {Tacchella}, {{\"U}bler}, {D'Eugenio}, {Witstok}, {Curti}, {Arribas}, {Bunker}, {Charlot}, {Chevallard}, {Eisenstein}, {Egami}, {Ji}, {Jones}, {Lyu}, {Rawle}, {Robertson}, {Rujopakarn}, {Perna}, {Sun}, {Venturi}, {Williams}, \& {Willott}}]{Maiolino2024a}
{Maiolino}, R., {Scholtz}, J., {Curtis-Lake}, E., {et~al.} 2024, submitted to \aap, arXiv:2308.01230

\bibitem[{{Markov} {et~al.}(2023){Markov}, {Gallerani}, {Pallottini}, {Sommovigo}, {Carniani}, {Ferrara}, {Parlanti}, \& {Di Mascia}}]{Markov2023}
{Markov}, V., {Gallerani}, S., {Pallottini}, A., {et~al.} 2023, \aap, 679, A12

\bibitem[{{Mascia} {et~al.}(2024){Mascia}, {Pentericci}, {Calabr{\`o}}, {Santini}, {Napolitano}, {Arrabal Haro}, {Castellano}, {Dickinson}, {Ocvirk}, {Lewis}, {Amor{\'\i}n}, {Bagley}, {Bhatawdekar}, {Cleri}, {Costantin}, {Dekel}, {Finkelstein}, {Fontana}, {Giavalisco}, {Grogin}, {Hathi}, {Hirschmann}, {Holwerda}, {Jung}, {Kartaltepe}, {Koekemoer}, {Lucas}, {Papovich}, {P{\'e}rez-Gonz{\'a}lez}, {Pirzkal}, {Trump}, {Wilkins}, \& {Yung}}]{Mascia2024}
{Mascia}, S., {Pentericci}, L., {Calabr{\`o}}, A., {et~al.} 2024, \aap, 685, A3

\bibitem[{{Mason} {et~al.}(2023){Mason}, {Trenti}, \& {Treu}}]{Mason2023}
{Mason}, C.~A., {Trenti}, M., \& {Treu}, T. 2023, \mnras, 521, 497

\bibitem[{{Masters} {et~al.}(2019){Masters}, {Stern}, {Cohen}, {Capak}, {Stanford}, {Hernitschek}, {Galametz}, {Davidzon}, {Rhodes}, {Sanders}, {Mobasher}, {Castander}, {Pruett}, \& {Fotopoulou}}]{Masters2019}
{Masters}, D.~C., {Stern}, D.~K., {Cohen}, J.~G., {et~al.} 2019, \apj, 877, 81

\bibitem[{{McLeod} {et~al.}(2024){McLeod}, {Donnan}, {McLure}, {Dunlop}, {Magee}, {Begley}, {Carnall}, {Cullen}, {Ellis}, {Hamadouche}, \& {Stanton}}]{McLeod2024}
{McLeod}, D.~J., {Donnan}, C.~T., {McLure}, R.~J., {et~al.} 2024, \mnras, 527, 5004

\bibitem[{{McLure} {et~al.}(2018){McLure}, {Pentericci}, {Cimatti}, {Dunlop}, {Elbaz}, {Fontana}, {Nandra}, {Amorin}, {Bolzonella}, {Bongiorno}, {Carnall}, {Castellano}, {Cirasuolo}, {Cucciati}, {Cullen}, {De Barros}, {Finkelstein}, {Fontanot}, {Franzetti}, {Fumana}, {Gargiulo}, {Garilli}, {Guaita}, {Hartley}, {Iovino}, {Jarvis}, {Juneau}, {Karman}, {Maccagni}, {Marchi}, {M{\'a}rmol-Queralt{\'o}}, {Pompei}, {Pozzetti}, {Scodeggio}, {Sommariva}, {Talia}, {Almaini}, {Balestra}, {Bardelli}, {Bell}, {Bourne}, {Bowler}, {Brusa}, {Buitrago}, {Caputi}, {Cassata}, {Charlot}, {Citro}, {Cresci}, {Cristiani}, {Curtis-Lake}, {Dickinson}, {Fazio}, {Ferguson}, {Fiore}, {Franco}, {Fynbo}, {Galametz}, {Georgakakis}, {Giavalisco}, {Grazian}, {Hathi}, {Jung}, {Kim}, {Koekemoer}, {Khusanova}, {Le F{\`e}vre}, {Lotz}, {Mannucci}, {Maltby}, {Matsuoka}, {McLeod}, {Mendez-Hernandez}, {Mendez-Abreu}, {Mignoli}, {Moresco}, {Mortlock}, {Nonino}, {Pannella}, {Papovich}, {Popesso}, {Rosario}, {Salvato}, {Santini}, {Schaerer},
  {Schreiber}, {Stark}, {Tasca}, {Thomas}, {Treu}, {Vanzella}, {Wild}, {Williams}, {Zamorani}, \& {Zucca}}]{McLure2018}
{McLure}, R.~J., {Pentericci}, L., {Cimatti}, A., {et~al.} 2018, \mnras, 479, 25

\bibitem[{{Merlin} {et~al.}(2016{\natexlab{a}}){Merlin}, {Amor{\'\i}n}, {Castellano}, {Fontana}, {Buitrago}, {Dunlop}, {Elbaz}, {Boucaud}, {Bourne}, {Boutsia}, {Brammer}, {Bruce}, {Capak}, {Cappelluti}, {Ciesla}, {Comastri}, {Cullen}, {Derriere}, {Faber}, {Ferguson}, {Giallongo}, {Grazian}, {Lotz}, {Micha{\l}owski}, {Paris}, {Pentericci}, {Pilo}, {Santini}, {Schreiber}, {Shu}, \& {Wang}}]{Merlin2016b}
{Merlin}, E., {Amor{\'\i}n}, R., {Castellano}, M., {et~al.} 2016{\natexlab{a}}, \aap, 590, A30

\bibitem[{{Merlin} {et~al.}(2022){Merlin}, {Bonchi}, {Paris}, {Belfiori}, {Fontana}, {Castellano}, {Nonino}, {Polenta}, {Santini}, {Yang}, {Glazebrook}, {Treu}, {Roberts-Borsani}, {Trenti}, {Birrer}, {Brammer}, {Grillo}, {Calabr{\`o}}, {Marchesini}, {Mason}, {Mercurio}, {Morishita}, {Strait}, {Boyett}, {Leethochawalit}, {Nanayakkara}, {Vulcani}, {Bradac}, \& {Wang}}]{Merlin2022}
{Merlin}, E., {Bonchi}, A., {Paris}, D., {et~al.} 2022, \apjl, 938, L14

\bibitem[{{Merlin} {et~al.}(2016{\natexlab{b}}){Merlin}, {Bourne}, {Castellano}, {Ferguson}, {Wang}, {Derriere}, {Dunlop}, {Elbaz}, \& {Fontana}}]{Merlin2016a}
{Merlin}, E., {Bourne}, N., {Castellano}, M., {et~al.} 2016{\natexlab{b}}, \aap, 595, A97

\bibitem[{{Merlin} {et~al.}(2021){Merlin}, {Castellano}, {Santini}, {Cipolletta}, {Boutsia}, {Schreiber}, {Buitrago}, {Fontana}, {Elbaz}, {Dunlop}, {Grazian}, {McLure}, {McLeod}, {Nonino}, {Milvang-Jensen}, {Derriere}, {Hathi}, {Pentericci}, {Fortuni}, \& {Calabr{\`o}}}]{Merlin2021}
{Merlin}, E., {Castellano}, M., {Santini}, P., {et~al.} 2021, \aap, 649, A22

\bibitem[{{Merlin} {et~al.}(2019){Merlin}, {Fortuni}, {Torelli}, {Santini}, {Castellano}, {Fontana}, {Grazian}, {Pentericci}, {Pilo}, \& {Schmidt}}]{Merlin2019}
{Merlin}, E., {Fortuni}, F., {Torelli}, M., {et~al.} 2019, \mnras, 490, 3309

\bibitem[{{Meyer} {et~al.}(2024){Meyer}, {Oesch}, {Giovinazzo}, {Weibel}, {Brammer}, {Matthee}, {Naidu}, {Bouwens}, {Chisholm}, {Covelo-Paz}, {Fudamoto}, {Maseda}, {Nelson}, {Shivaei}, {Xiao}, {Herard-Demanche}, {Illingworth}, {Kerutt}, {Kramarenko}, {Labbe}, {Leonova}, {Magee}, {Matharu}, {Prieto Lyon}, {Reddy}, {Schaerer}, {Shapley}, {Stefanon}, {Wozniak}, \& {Wuyts}}]{Meyer2024}
{Meyer}, R.~A., {Oesch}, P.~A., {Giovinazzo}, E., {et~al.} 2024, submitted to MNRAS, arXiv:2405.05111

\bibitem[{{Naidu} {et~al.}(2022){Naidu}, {Oesch}, {van Dokkum}, {Nelson}, {Suess}, {Brammer}, {Whitaker}, {Illingworth}, {Bouwens}, {Tacchella}, {Matthee}, {Allen}, {Bezanson}, {Conroy}, {Labbe}, {Leja}, {Leonova}, {Magee}, {Price}, {Setton}, {Strait}, {Stefanon}, {Toft}, {Weaver}, \& {Weibel}}]{Naidu2022}
{Naidu}, R.~P., {Oesch}, P.~A., {van Dokkum}, P., {et~al.} 2022, \apjl, 940, L14

\bibitem[{{Nanayakkara} {et~al.}(2024){Nanayakkara}, {Glazebrook}, {Jacobs}, {Kawinwanichakij}, {Schreiber}, {Brammer}, {Esdaile}, {Kacprzak}, {Labbe}, {Lagos}, {Marchesini}, {Marsan}, {Oesch}, {Papovich}, {Remus}, \& {Tran}}]{Nanayakkara2024}
{Nanayakkara}, T., {Glazebrook}, K., {Jacobs}, C., {et~al.} 2024, Scientific Reports, 14, 3724

\bibitem[{{Nandra} {et~al.}(2015){Nandra}, {Laird}, {Aird}, {Salvato}, {Georgakakis}, {Barro}, {Perez-Gonzalez}, {Barmby}, {Chary}, {Coil}, {Cooper}, {Davis}, {Dickinson}, {Faber}, {Fazio}, {Guhathakurta}, {Gwyn}, {Hsu}, {Huang}, {Ivison}, {Koo}, {Newman}, {Rangel}, {Yamada}, \& {Willmer}}]{Nandra2015}
{Nandra}, K., {Laird}, E.~S., {Aird}, J.~A., {et~al.} 2015, \apjs, 220, 10

\bibitem[{{Napolitano} {et~al.}(2024){Napolitano}, {Castellano}, {Pentericci}, {Arrabal Haro}, {Fontana}, \& {Treu}}]{Napolitano2024}
{Napolitano}, L., {Castellano}, M., {Pentericci}, L., {et~al.} 2024, submitted to \aap

\bibitem[{{Nayyeri} {et~al.}(2017){Nayyeri}, {Hemmati}, {Mobasher}, {Ferguson}, {Cooray}, {Barro}, {Faber}, {Dickinson}, {Koekemoer}, {Peth}, {Salvato}, {Ashby}, {Darvish}, {Donley}, {Durbin}, {Finkelstein}, {Fontana}, {Grogin}, {Gruetzbauch}, {Huang}, {Khostovan}, {Kocevski}, {Kodra}, {Lee}, {Newman}, {Pacifici}, {Pforr}, {Stefanon}, {Wiklind}, {Willner}, {Wuyts}, {Castellano}, {Conselice}, {Dolch}, {Dunlop}, {Galametz}, {Hathi}, {Lucas}, \& {Yan}}]{Nayyeri2017}
{Nayyeri}, H., {Hemmati}, S., {Mobasher}, B., {et~al.} 2017, \apjs, 228, 7

\bibitem[{{Ning} {et~al.}(2020){Ning}, {Jiang}, {Zheng}, {Wu}, {Bian}, {Egami}, {Fan}, {Ho}, {Shen}, {Wang}, \& {Wu}}]{Ning2020}
{Ning}, Y., {Jiang}, L., {Zheng}, Z.-Y., {et~al.} 2020, \apj, 903, 4

\bibitem[{{Oesch} {et~al.}(2023){Oesch}, {Brammer}, {Naidu}, {Bouwens}, {Chisholm}, {Illingworth}, {Matthee}, {Nelson}, {Qin}, {Reddy}, {Shapley}, {Shivaei}, {van Dokkum}, {Weibel}, {Whitaker}, {Wuyts}, {Covelo-Paz}, {Endsley}, {Fudamoto}, {Giovinazzo}, {Herard-Demanche}, {Kerutt}, {Kramarenko}, {Labbe}, {Leonova}, {Lin}, {Magee}, {Marchesini}, {Maseda}, {Mason}, {Matharu}, {Meyer}, {Neufeld}, {Prieto Lyon}, {Schaerer}, {Sharma}, {Shuntov}, {Smit}, {Stefanon}, {Wyithe}, \& {Xiao}}]{Oesch2023}
{Oesch}, P.~A., {Brammer}, G., {Naidu}, R.~P., {et~al.} 2023, \mnras, 525, 2864

\bibitem[{{Oke} \& {Gunn}(1983)}]{Oke83}
{Oke}, J.~B. \& {Gunn}, J.~E. 1983, \apj, 266, 713

\bibitem[{{Owers} {et~al.}(2011){Owers}, {Randall}, {Nulsen}, {Couch}, {David}, \& {Kempner}}]{Owers2011}
{Owers}, M.~S., {Randall}, S.~W., {Nulsen}, P. E.~J., {et~al.} 2011, \apj, 728, 27

\bibitem[{{Padmanabhan} \& {Loeb}(2023)}]{Padmanabhan2023}
{Padmanabhan}, H. \& {Loeb}, A. 2023, \apjl, 953, L4

\bibitem[{{Paris} {et~al.}(2023){Paris}, {Merlin}, {Fontana}, {Bonchi}, {Brammer}, {Correnti}, {Treu}, {Boyett}, {Calabr{\`o}}, {Castellano}, {Chen}, {Yang}, {Glazebrook}, {Kelly}, {Koekemoer}, {Leethochawalit}, {Mascia}, {Mason}, {Morishita}, {Nonino}, {Pentericci}, {Polenta}, {Roberts-Borsani}, {Santini}, {Trenti}, {Vanzella}, {Vulcani}, {Windhorst}, {Nanayakkara}, \& {Wang}}]{Paris2023}
{Paris}, D., {Merlin}, E., {Fontana}, A., {et~al.} 2023, \apj, 952, 20

\bibitem[{{Pentericci} {et~al.}(2018{\natexlab{a}}){Pentericci}, {McLure}, {Garilli}, {Cucciati}, {Franzetti}, {Iovino}, {Amorin}, {Bolzonella}, {Bongiorno}, {Carnall}, {Castellano}, {Cimatti}, {Cirasuolo}, {Cullen}, {De Barros}, {Dunlop}, {Elbaz}, {Finkelstein}, {Fontana}, {Fontanot}, {Fumana}, {Gargiulo}, {Guaita}, {Hartley}, {Jarvis}, {Juneau}, {Karman}, {Maccagni}, {Marchi}, {Marmol-Queralto}, {Nandra}, {Pompei}, {Pozzetti}, {Scodeggio}, {Sommariva}, {Talia}, {Almaini}, {Balestra}, {Bardelli}, {Bell}, {Bourne}, {Bowler}, {Brusa}, {Buitrago}, {Caputi}, {Cassata}, {Charlot}, {Citro}, {Cresci}, {Cristiani}, {Curtis-Lake}, {Dickinson}, {Fazio}, {Ferguson}, {Fiore}, {Franco}, {Fynbo}, {Galametz}, {Georgakakis}, {Giavalisco}, {Grazian}, {Hathi}, {Jung}, {Kim}, {Koekemoer}, {Khusanova}, {Le F{\`e}vre}, {Lotz}, {Mannucci}, {Maltby}, {Matsuoka}, {McLeod}, {Mendez-Hernandez}, {Mendez-Abreu}, {Mignoli}, {Moresco}, {Mortlock}, {Nonino}, {Pannella}, {Papovich}, {Popesso}, {Rosario}, {Salvato}, {Santini}, {Schaerer},
  {Schreiber}, {Stark}, {Tasca}, {Thomas}, {Treu}, {Vanzella}, {Wild}, {Williams}, {Zamorani}, \& {Zucca}}]{Pentericci2018}
{Pentericci}, L., {McLure}, R.~J., {Garilli}, B., {et~al.} 2018{\natexlab{a}}, \aap, 616, A174

\bibitem[{{Pentericci} {et~al.}(2018{\natexlab{b}}){Pentericci}, {Vanzella}, {Castellano}, {Fontana}, {De Barros}, {Grazian}, {Marchi}, {Bradac}, {Conselice}, {Cristiani}, {Dickinson}, {Finkelstein}, {Giallongo}, {Guaita}, {Koekemoer}, {Maiolino}, {Santini}, \& {Tilvi}}]{Pentericci2018b}
{Pentericci}, L., {Vanzella}, E., {Castellano}, M., {et~al.} 2018{\natexlab{b}}, \aap, 619, A147

\bibitem[{{P{\'e}rez-Gonz{\'a}lez} {et~al.}(2023{\natexlab{a}}){P{\'e}rez-Gonz{\'a}lez}, {Barro}, {Annunziatella}, {Costantin}, {Garc{\'\i}a-Argum{\'a}nez}, {McGrath}, {M{\'e}rida}, {Zavala}, {Arrabal Haro}, {Bagley}, {Backhaus}, {Behroozi}, {Bell}, {Bisigello}, {Buat}, {Calabr{\`o}}, {Casey}, {Cleri}, {Coogan}, {Cooper}, {Cooray}, {Dekel}, {Dickinson}, {Elbaz}, {Ferguson}, {Finkelstein}, {Fontana}, {Franco}, {Gardner}, {Giavalisco}, {G{\'o}mez-Guijarro}, {Grazian}, {Grogin}, {Guo}, {Huertas-Company}, {Jogee}, {Kartaltepe}, {Kewley}, {Kirkpatrick}, {Kocevski}, {Koekemoer}, {Long}, {Lotz}, {Lucas}, {Papovich}, {Pirzkal}, {Ravindranath}, {Somerville}, {Tacchella}, {Trump}, {Wang}, {Wilkins}, {Wuyts}, {Yang}, \& {Yung}}]{PerezGonzalez2023}
{P{\'e}rez-Gonz{\'a}lez}, P.~G., {Barro}, G., {Annunziatella}, M., {et~al.} 2023{\natexlab{a}}, \apjl, 946, L16

\bibitem[{{P{\'e}rez-Gonz{\'a}lez} {et~al.}(2024){P{\'e}rez-Gonz{\'a}lez}, {Barro}, {Rieke}, {Lyu}, {Rieke}, {Alberts}, {Williams}, {Hainline}, {Sun}, {Pusk{\'a}s}, {Annunziatella}, {Baker}, {Bunker}, {Egami}, {Ji}, {Johnson}, {Robertson}, {Rodr{\'\i}guez Del Pino}, {Rujopakarn}, {Shivaei}, {Tacchella}, {Willmer}, \& {Willott}}]{PerezGonzalez2024}
{P{\'e}rez-Gonz{\'a}lez}, P.~G., {Barro}, G., {Rieke}, G.~H., {et~al.} 2024, \apj, 968, 4

\bibitem[{{P{\'e}rez-Gonz{\'a}lez} {et~al.}(2023{\natexlab{b}}){P{\'e}rez-Gonz{\'a}lez}, {Costantin}, {Langeroodi}, {Rinaldi}, {Annunziatella}, {Ilbert}, {Colina}, {N{\o}rgaard-Nielsen}, {Greve}, {{\"O}stlin}, {Wright}, {Alonso-Herrero}, {{\'A}lvarez-M{\'a}rquez}, {Caputi}, {Eckart}, {Le F{\`e}vre}, {Labiano}, {Garc{\'\i}a-Mar{\'\i}n}, {Hjorth}, {Kendrew}, {Pye}, {Tikkanen}, {van der Werf}, {Walter}, {Ward}, {Bik}, {Boogaard}, {Bosman}, {G{\'o}mez}, {Gillman}, {Iani}, {Jermann}, {Melinder}, {Meyer}, {Moutard}, {van Dishoek}, {Henning}, {Lagage}, {Guedel}, {Peissker}, {Ray}, {Vandenbussche}, {Garc{\'\i}a-Argum{\'a}nez}, \& {Mar{\'\i}a M{\'e}rida}}]{PerezGonzalez2023b}
{P{\'e}rez-Gonz{\'a}lez}, P.~G., {Costantin}, L., {Langeroodi}, D., {et~al.} 2023{\natexlab{b}}, \apjl, 951, L1

\bibitem[{{P{\'e}rez-Gonz{\'a}lez} {et~al.}(2005){P{\'e}rez-Gonz{\'a}lez}, {Rieke}, {Egami}, {Alonso-Herrero}, {Dole}, {Papovich}, {Blaylock}, {Jones}, {Rieke}, {Rigby}, {Barmby}, {Fazio}, {Huang}, \& {Martin}}]{2005ApJ...630...82P}
{P{\'e}rez-Gonz{\'a}lez}, P.~G., {Rieke}, G.~H., {Egami}, E., {et~al.} 2005, \apj, 630, 82

\bibitem[{{P{\'e}rez-Gonz{\'a}lez} {et~al.}(2008){P{\'e}rez-Gonz{\'a}lez}, {Rieke}, {Villar}, {Barro}, {Blaylock}, {Egami}, {Gallego}, {Gil de Paz}, {Pascual}, {Zamorano}, \& {Donley}}]{PerezGonzalez2008}
{P{\'e}rez-Gonz{\'a}lez}, P.~G., {Rieke}, G.~H., {Villar}, V., {et~al.} 2008, 675, 234

\bibitem[{{Pharo} {et~al.}(2022){Pharo}, {Guo}, {Calvo}, {Carleton}, {Faber}, {Guhathakurta}, {Kassin}, {Koo}, {Lonergan}, {Teppala}, {Wang}, {Yesuf}, {Bian}, {Dav{\'e}}, {Forbes}, {Keres}, {Perez-Gonzalez}, {Martin}, {Puleo}, {Williams}, \& {Winningham}}]{Pharo2022}
{Pharo}, J., {Guo}, Y., {Calvo}, G.~B., {et~al.} 2022, \apjs, 261, 12

\bibitem[{{Price} {et~al.}(2024){Price}, {Bezanson}, {Labbe}, {Furtak}, {de Graaff}, {Greene}, {Kokorev}, {Setton}, {Suess}, {Brammer}, {Cutler}, {Leja}, {Pan}, {Wang}, {Weaver}, {Whitaker}, {Atek}, {Burgasser}, {Chemerynska}, {Dayal}, {Feldmann}, {F{\"o}rster Schreiber}, {Fudamoto}, {Fujimoto}, {Glazebrook}, {Goulding}, {Khullar}, {Kriek}, {Marchesini}, {Maseda}, {Miller}, {Muzzin}, {Nanayakkara}, {Nelson}, {Oesch}, {Shipley}, {Smit}, {Taylor}, {van Dokkum}, {Williams}, \& {Zitrin}}]{Price2024}
{Price}, S.~H., {Bezanson}, R., {Labbe}, I., {et~al.} 2024, submitted to ApJ, arXiv:2408.03920

\bibitem[{{Reddy} {et~al.}(2006){Reddy}, {Steidel}, {Erb}, {Shapley}, \& {Pettini}}]{Reddy2006}
{Reddy}, N.~A., {Steidel}, C.~C., {Erb}, D.~K., {Shapley}, A.~E., \& {Pettini}, M. 2006, \apj, 653, 1004

\bibitem[{{Richard} {et~al.}(2021){Richard}, {Claeyssens}, {Lagattuta}, {Guaita}, {Bauer}, {Pello}, {Carton}, {Bacon}, {Soucail}, {Lyon}, {Kneib}, {Mahler}, {Cl{\'e}ment}, {Mercier}, {Variu}, {Tamone}, {Ebeling}, {Schmidt}, {Nanayakkara}, {Maseda}, {Weilbacher}, {Bouch{\'e}}, {Bouwens}, {Wisotzki}, {de la Vieuville}, {Martinez}, \& {Patr{\'\i}cio}}]{Richard2021}
{Richard}, J., {Claeyssens}, A., {Lagattuta}, D., {et~al.} 2021, \aap, 646, A83

\bibitem[{{Rieke} {et~al.}(2023){Rieke}, {Robertson}, {Tacchella}, {Hainline}, {Johnson}, {Hausen}, {Ji}, {Willmer}, {Eisenstein}, {Pusk{\'a}s}, {Alberts}, {Arribas}, {Baker}, {Baum}, {Bhatawdekar}, {Bonaventura}, {Boyett}, {Bunker}, {Cameron}, {Carniani}, {Charlot}, {Chevallard}, {Chen}, {Curti}, {Curtis-Lake}, {Danhaive}, {DeCoursey}, {Dressler}, {Egami}, {Endsley}, {Helton}, {Hviding}, {Kumari}, {Looser}, {Lyu}, {Maiolino}, {Maseda}, {Nelson}, {Rieke}, {Rix}, {Sandles}, {Saxena}, {Sharpe}, {Shivaei}, {Skarbinski}, {Smit}, {Stark}, {Stone}, {Suess}, {Sun}, {Topping}, {{\"U}bler}, {Villanueva}, {Wallace}, {Williams}, {Willott}, {Whitler}, {Witstok}, \& {Woodrum}}]{Rieke2023}
{Rieke}, M.~J., {Robertson}, B., {Tacchella}, S., {et~al.} 2023, \apjs, 269, 16

\bibitem[{{Roberts-Borsani} {et~al.}(2024){Roberts-Borsani}, {Treu}, {Shapley}, {Fontana}, {Pentericci}, {Castellano}, {Morishita}, {Bergamini}, \& {Rosati}}]{RobertsBorsani2024}
{Roberts-Borsani}, G., {Treu}, T., {Shapley}, A., {et~al.} 2024, submitted to ApJ, arXiv:2403.07103

\bibitem[{{Rodighiero} {et~al.}(2024){Rodighiero}, {Enia}, {Bisigello}, {Girardi}, {Gandolfi}, {Kohandel}, {Pallottini}, {Badinelli}, {Grazian}, {Ferrara}, {Vulcani}, {Bianchetti}, {Marasco}, {Sinigaglia}, {Castellano}, {Santini}, \& {Cassata}}]{Rodighiero2024}
{Rodighiero}, G., {Enia}, A., {Bisigello}, L., {et~al.} 2024, submitted to \aap, arXiv:2405.04572

\bibitem[{{Rosani} {et~al.}(2020){Rosani}, {Caminha}, {Caputi}, \& {Deshmukh}}]{Rosani2020}
{Rosani}, G., {Caminha}, G.~B., {Caputi}, K.~I., \& {Deshmukh}, S. 2020, \aap, 633, A159

\bibitem[{{Santini} {et~al.}(2023){Santini}, {Fontana}, {Castellano}, {Leethochawalit}, {Trenti}, {Treu}, {Belfiori}, {Birrer}, {Bonchi}, {Merlin}, {Mason}, {Morishita}, {Nonino}, {Paris}, {Polenta}, {Rosati}, {Yang}, {Boyett}, {Bradac}, {Calabr{\`o}}, {Dressler}, {Glazebrook}, {Marchesini}, {Mascia}, {Nanayakkara}, {Pentericci}, {Roberts-Borsani}, {Scarlata}, {Vulcani}, \& {Wang}}]{Santini2023}
{Santini}, P., {Fontana}, A., {Castellano}, M., {et~al.} 2023, \apjl, 942, L27

\bibitem[{{Schaerer} \& {de Barros}(2009)}]{Schaerer2009}
{Schaerer}, D. \& {de Barros}, S. 2009, \aap, 502, 423

\bibitem[{{Schmidt} {et~al.}(2021){Schmidt}, {Kerutt}, {Wisotzki}, {Urrutia}, {Feltre}, {Maseda}, {Nanayakkara}, {Bacon}, {Boogaard}, {Conseil}, {Contini}, {Herenz}, {Kollatschny}, {Krumpe}, {Leclercq}, {Mahler}, {Matthee}, {Mauerhofer}, {Richard}, \& {Schaye}}]{Schmidt2021}
{Schmidt}, K.~B., {Kerutt}, J., {Wisotzki}, L., {et~al.} 2021, \aap, 654, A80

\bibitem[{{Schmidt} {et~al.}(2014){Schmidt}, {Treu}, {Brammer}, {Brada{\v{c}}}, {Wang}, {Dijkstra}, {Dressler}, {Fontana}, {Gavazzi}, {Henry}, {Hoag}, {Jones}, {Kelly}, {Malkan}, {Mason}, {Pentericci}, {Poggianti}, {Stiavelli}, {Trenti}, {von der Linden}, \& {Vulcani}}]{Schmidt2014}
{Schmidt}, K.~B., {Treu}, T., {Brammer}, G.~B., {et~al.} 2014, \apjl, 782, L36

\bibitem[{{Scodeggio} {et~al.}(2018){Scodeggio}, {Guzzo}, {Garilli}, {Granett}, {Bolzonella}, {de la Torre}, {Abbas}, {Adami}, {Arnouts}, {Bottini}, {Cappi}, {Coupon}, {Cucciati}, {Davidzon}, {Franzetti}, {Fritz}, {Iovino}, {Krywult}, {Le Brun}, {Le F{\`e}vre}, {Maccagni}, {Ma{\l}ek}, {Marchetti}, {Marulli}, {Polletta}, {Pollo}, {Tasca}, {Tojeiro}, {Vergani}, {Zanichelli}, {Bel}, {Branchini}, {De Lucia}, {Ilbert}, {McCracken}, {Moutard}, {Peacock}, {Zamorani}, {Burden}, {Fumana}, {Jullo}, {Marinoni}, {Mellier}, {Moscardini}, \& {Percival}}]{Scodeggio2018}
{Scodeggio}, M., {Guzzo}, L., {Garilli}, B., {et~al.} 2018, \aap, 609, A84

\bibitem[{{Stefanon} {et~al.}(2017){Stefanon}, {Yan}, {Mobasher}, {Barro}, {Donley}, {Fontana}, {Hemmati}, {Koekemoer}, {Lee}, {Lee}, {Nayyeri}, {Peth}, {Pforr}, {Salvato}, {Wiklind}, {Wuyts}, {Ashby}, {Castellano}, {Conselice}, {Cooper}, {Cooray}, {Dolch}, {Ferguson}, {Galametz}, {Giavalisco}, {Guo}, {Willner}, {Dickinson}, {Faber}, {Fazio}, {Gardner}, {Gawiser}, {Grazian}, {Grogin}, {Kocevski}, {Koo}, {Lee}, {Lucas}, {McGrath}, {Nandra}, {Newman}, \& {van der Wel}}]{Stefanon2017}
{Stefanon}, M., {Yan}, H., {Mobasher}, B., {et~al.} 2017, \apjs, 229, 32

\bibitem[{{Straatman} {et~al.}(2018){Straatman}, {van der Wel}, {Bezanson}, {Pacifici}, {Gallazzi}, {Wu}, {Noeske}, {Bari{\v{s}}i{\'c}}, {Bell}, {Brammer}, {Calhau}, {Chauke}, {Franx}, {van Houdt}, {Labb{\'e}}, {Maseda}, {Mu{\~n}oz-Mateos}, {Muzzin}, {van de Sande}, {Sobral}, \& {Spilker}}]{Straatman2018}
{Straatman}, C. M.~S., {van der Wel}, A., {Bezanson}, R., {et~al.} 2018, \apjs, 239, 27

\bibitem[{{Tasca} {et~al.}(2017){Tasca}, {Le F{\`e}vre}, {Ribeiro}, {Thomas}, {Moreau}, {Cassata}, {Garilli}, {Le Brun}, {Lemaux}, {Maccagni}, {Pentericci}, {Schaerer}, {Vanzella}, {Zamorani}, {Zucca}, {Amorin}, {Bardelli}, {Cassar{\`a}}, {Castellano}, {Cimatti}, {Cucciati}, {Durkalec}, {Fontana}, {Giavalisco}, {Grazian}, {Hathi}, {Ilbert}, {Paltani}, {Pforr}, {Scodeggio}, {Sommariva}, {Talia}, {Tresse}, {Vergani}, {Capak}, {Charlot}, {Contini}, {de la Torre}, {Dunlop}, {Fotopoulou}, {Guaita}, {Koekemoer}, {L{\'o}pez-Sanjuan}, {Mellier}, {Salvato}, {Scoville}, {Taniguchi}, \& {Wang}}]{Tasca2017}
{Tasca}, L.~A.~M., {Le F{\`e}vre}, O., {Ribeiro}, B., {et~al.} 2017, \aap, 600, A110

\bibitem[{{Treu} {et~al.}(2023){Treu}, {Calabr{\`o}}, {Castellano}, {Leethochawalit}, {Merlin}, {Fontana}, {Yang}, {Morishita}, {Trenti}, {Dressler}, {Mason}, {Paris}, {Pentericci}, {Roberts-Borsani}, {Vulcani}, {Boyett}, {Bradac}, {Glazebrook}, {Jones}, {Marchesini}, {Mascia}, {Nanayakkara}, {Santini}, {Strait}, {Vanzella}, \& {Wang}}]{Treu2023}
{Treu}, T., {Calabr{\`o}}, A., {Castellano}, M., {et~al.} 2023, \apjl, 942, L28

\bibitem[{{Treu} {et~al.}(2022){Treu}, {Roberts-Borsani}, {Bradac}, {Brammer}, {Fontana}, {Henry}, {Mason}, {Morishita}, {Pentericci}, {Wang}, {Acebron}, {Bagley}, {Bergamini}, {Belfiori}, {Bonchi}, {Boyett}, {Boutsia}, {Calabr{\'o}}, {Caminha}, {Castellano}, {Dressler}, {Glazebrook}, {Grillo}, {Jacobs}, {Jones}, {Kelly}, {Leethochawalit}, {Malkan}, {Marchesini}, {Mascia}, {Mercurio}, {Merlin}, {Nanayakkara}, {Nonino}, {Paris}, {Poggianti}, {Rosati}, {Santini}, {Scarlata}, {Shipley}, {Strait}, {Trenti}, {Tubthong}, {Vanzella}, {Vulcani}, \& {Yang}}]{Treu2022}
{Treu}, T., {Roberts-Borsani}, G., {Bradac}, M., {et~al.} 2022, \apj, 935, 110

\bibitem[{{Treu} {et~al.}(2015){Treu}, {Schmidt}, {Brammer}, {Vulcani}, {Wang}, {Brada{\v{c}}}, {Dijkstra}, {Dressler}, {Fontana}, {Gavazzi}, {Henry}, {Hoag}, {Huang}, {Jones}, {Kelly}, {Malkan}, {Mason}, {Pentericci}, {Poggianti}, {Stiavelli}, {Trenti}, \& {von der Linden}}]{Treu2015}
{Treu}, T., {Schmidt}, K.~B., {Brammer}, G.~B., {et~al.} 2015, \apj, 812, 114

\bibitem[{{Trinca} {et~al.}(2024){Trinca}, {Schneider}, {Valiante}, {Graziani}, {Ferrotti}, {Omukai}, \& {Chon}}]{Trinca2024}
{Trinca}, A., {Schneider}, R., {Valiante}, R., {et~al.} 2024, \mnras, 529, 3563

\bibitem[{{Trump} {et~al.}(2009){Trump}, {Impey}, {Elvis}, {McCarthy}, {Huchra}, {Brusa}, {Salvato}, {Capak}, {Cappelluti}, {Civano}, {Comastri}, {Gabor}, {Hao}, {Hasinger}, {Jahnke}, {Kelly}, {Lilly}, {Schinnerer}, {Scoville}, \& {Smol{\v{c}}i{\'c}}}]{Trump2009}
{Trump}, J.~R., {Impey}, C.~D., {Elvis}, M., {et~al.} 2009, \apj, 696, 1195

\bibitem[{{Urrutia} {et~al.}(2019){Urrutia}, {Wisotzki}, {Kerutt}, {Schmidt}, {Herenz}, {Klar}, {Saust}, {Werhahn}, {Diener}, {Caruana}, {Krajnovi{\'c}}, {Bacon}, {Boogaard}, {Brinchmann}, {Enke}, {Maseda}, {Nanayakkara}, {Richard}, {Steinmetz}, \& {Weilbacher}}]{Urrutia2019}
{Urrutia}, T., {Wisotzki}, L., {Kerutt}, J., {et~al.} 2019, \aap, 624, A141

\bibitem[{{van der Wel} {et~al.}(2016){van der Wel}, {Noeske}, {Bezanson}, {Pacifici}, {Gallazzi}, {Franx}, {Mu{\~n}oz-Mateos}, {Bell}, {Brammer}, {Charlot}, {Chauk{\'e}}, {Labb{\'e}}, {Maseda}, {Muzzin}, {Rix}, {Sobral}, {van de Sande}, {van Dokkum}, {Wild}, \& {Wolf}}]{vanderWel2016}
{van der Wel}, A., {Noeske}, K., {Bezanson}, R., {et~al.} 2016, \apjs, 223, 29

\bibitem[{{Wang} {et~al.}(2024{\natexlab{a}}){Wang}, {Leja}, {de Graaff}, {Brammer}, {Weibel}, {van Dokkum}, {Baggen}, {Suess}, {Greene}, {Bezanson}, {Cleri}, {Hirschmann}, {Labb{\'e}}, {Matthee}, {McConachie}, {Naidu}, {Nelson}, {Oesch}, {Setton}, \& {Williams}}]{Wang2024}
{Wang}, B., {Leja}, J., {de Graaff}, A., {et~al.} 2024{\natexlab{a}}, \apjl, 969, L13

\bibitem[{{Wang} {et~al.}(2024{\natexlab{b}}){Wang}, {Teplitz}, {Sun}, {Rafelski}, {Grogin}, {Prichard}, {Sunnquist}, {Alavi}, {Windhorst}, {Koekemoer}, {Ashcraft}, {Bagley}, {Baronchelli}, {Barro}, {Blanche}, {Brammer}, {Broussard}, {Carleton}, {Chartab}, {Cheng}, {Codoreanu}, {Cohen}, {Colbert}, {Conselice}, {Dai}, {Darvish}, {Dav{\'e}}, {DeGroot}, {De Mello}, {Dickinson}, {Emami}, {Ferguson}, {Ferreira}, {Finkelstein}, {Finkelstein}, {Gardner}, {Gawiser}, {Gburek}, {Giavalisco}, {Grazian}, {Gronwall}, {Guo}, {Arrabal Haro}, {Hathi}, {Hayes}, {Hemmati}, {Howell}, {Iyer}, {Jansen}, {Ji}, {Kaviraj}, {Kurczynski}, {Lazar}, {Lucas}, {MacKenty}, {Mehta}, {Mantha}, {Martin}, {Martin}, {McCabe}, {Mobasher}, {Nedkova}, {O'Connell}, {Olsen}, {Otteson}, {Ravindranath}, {Redshaw}, {Robertson}, {Rutkowski}, {Sattari}, {Scarlata}, {Siana}, {Smith}, {Soto}, {Vanzella}, {Yung}, \& {Zabelle}}]{WangXin2024}
{Wang}, X., {Teplitz}, H.~I., {Sun}, L., {et~al.} 2024{\natexlab{b}}, Research Notes of the American Astronomical Society, 8, 26

\bibitem[{{Ward} {et~al.}(2024){Ward}, {de la Vega}, {Mobasher}, {McGrath}, {Iyer}, {Calabr{\`o}}, {Costantin}, {Dickinson}, {Holwerda}, {Huertas-Company}, {Hirschmann}, {Lucas}, {Pandya}, {Wilkins}, {Yung}, {Arrabal Haro}, {Bagley}, {Finkelstein}, {Kartaltepe}, {Koekemoer}, {Papovich}, \& {Pirzkal}}]{Ward2024}
{Ward}, E., {de la Vega}, A., {Mobasher}, B., {et~al.} 2024, \apj, 962, 176

\bibitem[{{Weaver} {et~al.}(2024){Weaver}, {Cutler}, {Pan}, {Whitaker}, {Labb{\'e}}, {Price}, {Bezanson}, {Brammer}, {Marchesini}, {Leja}, {Wang}, {Furtak}, {Zitrin}, {Atek}, {Chemerynska}, {Coe}, {Dayal}, {van Dokkum}, {Feldmann}, {F{\"o}rster Schreiber}, {Franx}, {Fujimoto}, {Fudamoto}, {Glazebrook}, {de Graaff}, {Greene}, {Juneau}, {Kassin}, {Kriek}, {Khullar}, {Maseda}, {Mowla}, {Muzzin}, {Nanayakkara}, {Nelson}, {Oesch}, {Pacifici}, {Papovich}, {Setton}, {Shapley}, {Shipley}, {Smit}, {Stefanon}, {Taylor}, {Weibel}, \& {Williams}}]{Weaver2024}
{Weaver}, J.~R., {Cutler}, S.~E., {Pan}, R., {et~al.} 2024, \apjs, 270, 7

\bibitem[{{Weaver} {et~al.}(2022){Weaver}, {Kauffmann}, {Ilbert}, {McCracken}, {Moneti}, {Toft}, {Brammer}, {Shuntov}, {Davidzon}, {Hsieh}, {Laigle}, {Anastasiou}, {Jespersen}, {Vinther}, {Capak}, {Casey}, {McPartland}, {Milvang-Jensen}, {Mobasher}, {Sanders}, {Zalesky}, {Arnouts}, {Aussel}, {Dunlop}, {Faisst}, {Franx}, {Furtak}, {Fynbo}, {Gould}, {Greve}, {Gwyn}, {Kartaltepe}, {Kashino}, {Koekemoer}, {Kokorev}, {Le F{\`e}vre}, {Lilly}, {Masters}, {Magdis}, {Mehta}, {Peng}, {Riechers}, {Salvato}, {Sawicki}, {Scarlata}, {Scoville}, {Shirley}, {Silverman}, {Sneppen}, {Smolc̆i{\'c}}, {Steinhardt}, {Stern}, {Tanaka}, {Taniguchi}, {Teplitz}, {Vaccari}, {Wang}, \& {Zamorani}}]{Weaver2022}
{Weaver}, J.~R., {Kauffmann}, O.~B., {Ilbert}, O., {et~al.} 2022, \apjs, 258, 11

\bibitem[{{Weibel} {et~al.}(2024){Weibel}, {Oesch}, {Barrufet}, {Gottumukkala}, {Ellis}, {Santini}, {Weaver}, {Allen}, {Bouwens}, {Bowler}, {Brammer}, {Carnall}, {Cullen}, {Dayal}, {Dickinson}, {Donnan}, {Dunlop}, {Giavalisco}, {Grogin}, {Illingworth}, {Koekemoer}, {Labbe}, {Marchesini}, {McLeod}, {McLure}, {Naidu}, {P{\'e}rez-Gonz{\'a}lez}, {Shuntov}, {Stefanon}, {Toft}, \& {Xiao}}]{Weibel2024}
{Weibel}, A., {Oesch}, P.~A., {Barrufet}, L., {et~al.} 2024, \mnras, 533, 1808

\bibitem[{{Williams} {et~al.}(2024){Williams}, {Alberts}, {Ji}, {Hainline}, {Lyu}, {Rieke}, {Endsley}, {Suess}, {Sun}, {Johnson}, {Florian}, {Shivaei}, {Rujopakarn}, {Baker}, {Bhatawdekar}, {Boyett}, {Bunker}, {Cameron}, {Carniani}, {Charlot}, {Curtis-Lake}, {DeCoursey}, {de Graaff}, {Egami}, {Eisenstein}, {Gibson}, {Hausen}, {Helton}, {Maiolino}, {Maseda}, {Nelson}, {P{\'e}rez-Gonz{\'a}lez}, {Rieke}, {Robertson}, {Saxena}, {Tacchella}, {Willmer}, \& {Willott}}]{Williams2024}
{Williams}, C.~C., {Alberts}, S., {Ji}, Z., {et~al.} 2024, \apj, 968, 34

\bibitem[{{Williams} {et~al.}(2023){Williams}, {Tacchella}, {Maseda}, {Robertson}, {Johnson}, {Willott}, {Eisenstein}, {Willmer}, {Ji}, {Hainline}, {Helton}, {Alberts}, {Baum}, {Bhatawdekar}, {Boyett}, {Bunker}, {Carniani}, {Charlot}, {Chevallard}, {Curtis-Lake}, {de Graaff}, {Egami}, {Franx}, {Kumari}, {Maiolino}, {Nelson}, {Rieke}, {Sandles}, {Shivaei}, {Simmonds}, {Smit}, {Suess}, {Sun}, {{\"U}bler}, \& {Witstok}}]{Williams2023}
{Williams}, C.~C., {Tacchella}, S., {Maseda}, M.~V., {et~al.} 2023, \apjs, 268, 64

\bibitem[{{Wisnioski} {et~al.}(2019){Wisnioski}, {F{\"o}rster Schreiber}, {Fossati}, {Mendel}, {Wilman}, {Genzel}, {Bender}, {Wuyts}, {Davies}, {{\"U}bler}, {Bandara}, {Beifiori}, {Belli}, {Brammer}, {Chan}, {Davies}, {Fabricius}, {Galametz}, {Lang}, {Lutz}, {Nelson}, {Momcheva}, {Price}, {Rosario}, {Saglia}, {Seitz}, {Shimizu}, {Tacconi}, {Tadaki}, {van Dokkum}, \& {Wuyts}}]{Wisnioski2019}
{Wisnioski}, E., {F{\"o}rster Schreiber}, N.~M., {Fossati}, M., {et~al.} 2019, \apj, 886, 124

\bibitem[{{Wright} {et~al.}(2024){Wright}, {Whitaker}, {Weaver}, {Cutler}, {Wang}, {Carnall}, {Suess}, {Bezanson}, {Nelson}, {Miller}, {Ito}, \& {Valentino}}]{Wright2024}
{Wright}, L., {Whitaker}, K.~E., {Weaver}, J.~R., {et~al.} 2024, \apjl, 964, L10

\bibitem[{{Wuyts} {et~al.}(2008){Wuyts}, {Labb{\'e}}, {F{\"o}rster Schreiber}, {Franx}, {Rudnick}, {Brammer}, \& {van Dokkum}}]{Wuyts2008}
{Wuyts}, S., {Labb{\'e}}, I., {F{\"o}rster Schreiber}, N.~M., {et~al.} 2008, \apj, 682, 985

\bibitem[{{Xue} {et~al.}(2016){Xue}, {Luo}, {Brandt}, {Alexander}, {Bauer}, {Lehmer}, \& {Yang}}]{Xue2016}
{Xue}, Y.~Q., {Luo}, B., {Brandt}, W.~N., {et~al.} 2016, \apjs, 224, 15

\bibitem[{{Xue} {et~al.}(2011){Xue}, {Luo}, {Brandt}, {Bauer}, {Lehmer}, {Broos}, {Schneider}, {Alexander}, {Brusa}, {Comastri}, {Fabian}, {Gilli}, {Hasinger}, {Hornschemeier}, {Koekemoer}, {Liu}, {Mainieri}, {Paolillo}, {Rafferty}, {Rosati}, {Shemmer}, {Silverman}, {Smail}, {Tozzi}, \& {Vignali}}]{Xue2011}
{Xue}, Y.~Q., {Luo}, B., {Brandt}, W.~N., {et~al.} 2011, \apjs, 195, 10

\bibitem[{{Yung} {et~al.}(2024){Yung}, {Somerville}, {Finkelstein}, {Wilkins}, \& {Gardner}}]{Yung2024}
{Yung}, L.~Y.~A., {Somerville}, R.~S., {Finkelstein}, S.~L., {Wilkins}, S.~M., \& {Gardner}, J.~P. 2024, \mnras, 527, 5929

\end{thebibliography}

%\newpage
\begin{appendix}

\section{Catalogue format} \label{format}

We release several catalogues for each field, among which one (named ``optap''-catalogue) is obtained assigning to each source the colours computed in its ``optimal'' aperture (see Section 3.4). The catalogues are published in \texttt{fits} format, and come with a \texttt{README} file that explains the meaning and format of the columns. All catalogues first list a set of columns indicating the unique identifiers of the detected objects, their position in equatorial coordinates and in pixels, and some values from the \textsc{SExtractor} detection run on the F356W+F444W stack mosaics: \texttt{ISOAREA\_IMAGE}, \texttt{CLASS\_STAR}, \texttt{FLAGS}, \texttt{FLUX\_RADIUS}, \texttt{FLUX\_AUTO}, \texttt{FLUXERR\_AUTO}. Then, some measures obtained with \textsc{a-phot} again on the detection stack are listed: the major semi-axis of the elliptical isophote, the ellipticity and the position angle, and the Kron radius (in pixels). The total fluxes in the 16 photometric bands, computed as described in Section 3.5, are then given, followed by the corresponding 16 uncertainties (all in $\mu$Jy units). Finally, the last column lists the flags discussed in Section 3.5. In the ``optap''-catalogue, the value of the ``optimal'' aperture is also given (in arcseconds). We also release the original raw catalogue containing all the measured aperture fluxes.

We separately release a set of photometric redshift catalogues estimated from the ``optap''-catalogue with local background subtraction. For each source we list the ID, RA and Dec, the spectroscopic redshift when available, the four estimates of the photometric redshift (see Section 5), and the flag assigned to the source as reported also in the ``optap''-catalogue, with the additional indication of a negative sign for the sources lacking information in the bands necessary to identify the Lyman break.

\section{Astrometry}

As explained in Section 3.1, we re-aligned \textrm{HST} mosaics to the NIRCam grid by means of an automatic procedure. Table \ref{astrom} lists the offsets in RA and Dec after the procedure, for all the bands and fields. We point out that the CANDELS EGS and CEERS HDR1 \textrm{HST} data are obtained from the same observations, but the latter has been re-aligned to Gaia-DR3; since we took care to correct all of the images for astrometric accuracy, the two datasets should be perfectly consistent.

\begin{table}
\centering
\caption{The median and MAD of the astrometric offsets of the \textrm{HST} bands.} \label{astrom}
\begin{tabular}{llll}    
\hline
Band & $N_{obj.}$ & $\Delta{RA}$ ('') & $\Delta{DEC}$ ('')  \\ \hline\hline
\multicolumn{4}{c}{ABELL2744} \\
F435W & 1307 & 0.009 $\pm$ 0.02 & 0.000 $\pm$ 0.01 \\ 
F606W & 3309 & 0.001 $\pm$ 0.02 & 0.001 $\pm$ 0.01 \\ 
%F775W & 581 & 0.003 $\pm$ 0.02 & 0.000 $\pm$ 0.02 \\ 
F814W & 3188 & -0.001 $\pm$ 0.02 & -0.001 $\pm$ 0.01 \\ 
F105W & 4595 & 0.003 $\pm$ 0.01 & 0.001 $\pm$ 0.01 \\ 
F125W & 2170 & 0.003 $\pm$ 0.02 & -0.001 $\pm$ 0.01 \\
F140W & 849 & 0.005 $\pm$ 0.02 & 0.003 $\pm$ 0.01 \\ 
F160W & 2054 & 0.003 $\pm$ 0.02 & 0.000 $\pm$ 0.01 \\ %\hline
\multicolumn{4}{c}{CEERS}\\
F606W & 7063 & 0.044 $\pm$ 0.03 & 0.032 $\pm$ 0.02 \\ 
F814W & 8615 & 0.066 $\pm$ 0.02 & 0.041 $\pm$ 0.01 \\
F105W & 1437 & 0.007 $\pm$ 0.05 & 0.010 $\pm$ 0.03 \\ 
F125W & 8470 & 0.027 $\pm$ 0.01 & 0.036 $\pm$ 0.01 \\ 
F160W & 4355 & 0.064 $\pm$ 0.05 & 0.040 $\pm$ 0.02 \\  %\hline
\multicolumn{4}{c}{JADES-GN}\\
F435W & 1144 & 0.036 $\pm$ 0.05 & 0.015 $\pm$ 0.02 \\ 
F606W & 1822 & 0.045 $\pm$ 0.04 & 0.015 $\pm$ 0.02 \\ 
%& F775W & 2595 & 0.045 $\pm$ 0.04 & 0.014 $\pm$ 0.02 \\ 
F814W & 3228 & 0.040 $\pm$ 0.03 & 0.015 $\pm$ 0.01 \\  
F105W & 2800 & 0.041 $\pm$ 0.02 & 0.012 $\pm$ 0.01 \\ 
F125W & 3430 & 0.039 $\pm$ 0.01 & 0.013 $\pm$ 0.00 \\ 
F140W & 1532 & 0.044 $\pm$ 0.01 & 0.015 $\pm$ 0.01 \\ 
F160W & 1436 & 0.041 $\pm$ 0.04 & 0.013 $\pm$ 0.02 \\ %\hline
\multicolumn{4}{c}{JADES-GS}\\
F435W & 13997 & 0.006 $\pm$ 0.03 & -0.015 $\pm$ 0.03 \\ 
F606W & 21408 & 0.004 $\pm$ 0.02 & -0.012 $\pm$ 0.02 \\ 
%& F775W & 17368 & 0.005 $\pm$ 0.02 & -0.013 $\pm$ 0.02 \\ 
F814W & 19677 & 0.000 $\pm$ 0.02 & -0.010 $\pm$ 0.01 \\  
%& F098M & 3953 & 0.006 $\pm$ 0.01 & -0.013 $\pm$ 0.01 \\ 
F105W & 12169 & 0.005 $\pm$ 0.01 & -0.013 $\pm$ 0.01 \\ 
F125W & 18764 & 0.007 $\pm$ 0.01 & -0.013 $\pm$ 0.01 \\ 
F140W & 8542 & 0.008 $\pm$ 0.01 & -0.014 $\pm$ 0.01 \\ 
F160W & 6235 & 0.006 $\pm$ 0.03 & -0.013 $\pm$ 0.02 \\ %\hline
%%& F225W & 627 & 0.007 $\pm$ 0.05 & -0.004 $\pm$ 0.04 \\ 
%%& F275W & 2051 & 0.012 $\pm$ 0.04 & -0.018 $\pm$ 0.04 \\ 
%%& F336W & 4167 & 0.007 $\pm$ 0.04 & -0.019 $\pm$ 0.03 \\ 
%%& F850LP & 19549 & 0.003 $\pm$ 0.02 & -0.013 $\pm$ 0.01 \\ \hline 
\multicolumn{4}{c}{NGDEEP}\\
F435W & 758 & 0.082 $\pm$ 0.04 & -0.058 $\pm$ 0.03 \\ 
F606W & 1194 & 0.077 $\pm$ 0.03 & -0.052 $\pm$ 0.03 \\ 
F775W & 1121 & 0.077 $\pm$ 0.03 & -0.052 $\pm$ 0.03 \\ 
F814W & 1096 & 0.079 $\pm$ 0.03 & -0.053 $\pm$ 0.02 \\  
F105W & 815 & 0.091 $\pm$ 0.01 & -0.049 $\pm$ 0.01 \\ 
F125W & 965 & 0.091 $\pm$ 0.01 & -0.050 $\pm$ 0.01 \\ 
F160W & 506 & 0.089 $\pm$ 0.02 & -0.049 $\pm$ 0.02 \\ %\hline
%%& F850LP & 974 & 0.094 $\pm$ 0.01 & -0.046 $\pm$ 0.01 \\  \hline 
\multicolumn{4}{c}{PRIMER-COSMOS}\\
F435W & 733 & 0.010 $\pm$ 0.04 & 0.012 $\pm$ 0.05 \\ 
F606W & 1279 & -0.011 $\pm$ 0.03 & -0.022 $\pm$ 0.03 \\ 
F814W & 2925 & -0.004 $\pm$ 0.01 & -0.008 $\pm$ 0.01 \\ 
F125W & 4704 & -0.004 $\pm$ 0.01 & -0.011 $\pm$ 0.01 \\
F140W & 2797 & -0.006 $\pm$ 0.01 & -0.006 $\pm$ 0.02 \\
F160W & 3709 & -0.005 $\pm$ 0.02 & -0.011 $\pm$ 0.02 \\ %\hline
\multicolumn{4}{c}{PRIMER-UDS}\\
F435W & 1214 & 0.008 $\pm$ 0.04 & 0.000 $\pm$ 0.04 \\ 
F606W & 3613 & -0.003 $\pm$ 0.03 & 0.010 $\pm$ 0.03 \\ 
F814W & 4779 & -0.003 $\pm$ 0.03 & 0.009 $\pm$ 0.03 \\ 
F125W & 3911 & 0.010 $\pm$ 0.01 & -0.004 $\pm$ 0.01 \\ 
F140W & 1653 & 0.010 $\pm$ 0.02 & -0.004 $\pm$ 0.02 \\ 
F160W & 4915 & 0.010 $\pm$ 0.03 & -0.005 $\pm$ 0.04 \\ \hline
\end{tabular}
\tablefoot{The values are computed by cross-matching $N_{obj}$ sources and corrected before re-projecting the HST images to the JWST grid. For each field, the F160W values are  the coordinate offsets between the original \textrm{HST} F160W and the NIRCam F150W band, while the others refer to the offsets between the given band and F160W band after it was re-aligned to the F150W.}
\end{table}

\onecolumn
\section{Selection of point-like and potentially spurious sources} \label{SGapp}

The PCA technique described in Section 3.5 works as follows. We fed the \texttt{scikit-learn} module \texttt{decomposition.PCA} with three parameters from the \textsc{SExtractor} detection catalogue, i.e. the signal-to-noise ratio S/N obtained as the ratio between \texttt{FLUX\_AUTO} and \texttt{FLUXERR\_AUTO}, the peak surface brightness above background \texttt{MUMAX}, and the half light radius \texttt{FLUX\_RADIUS}. We used a sub-sample of the PRIMER-COSMOS \textsc{SExtractor} detection catalogue to determine the principal components matrix [[0.38283975  0.65126842  0.65519704], [0.92363354 -0.28388812 -0.25750461]] (with explained variance ratio [0.65961785 0.27738227]), and then applied this matrix to the full catalogues of all fields. The loci of PLS and PSS are easily identified in the resulting principal components diagram, and after a rigid anti-clockwise rotation of 66.5° PSS can be singled out by a simple law, 
$0.15<\texttt{rPCA2}<0$ \& \texttt{rPCA2}$<c_{st}$ (where \texttt{rPCAi} indicates the value of the rotated component). PSS can be identified as those in the uppermost region of the diagram, using the law 
$\texttt{rPC2}>c_{sp1} \times \mbox{log}_{10}(1+\texttt{rPC1})+c_{sp2}$ \& $\texttt{rPC1}>c_{sp3}$.
The exact definition of the locus of the two populations slightly varies from field to field, as shown by the values of the constants reported in Table \ref{pca}. 

\begin{figure*}
\includegraphics[width=0.99\textwidth]{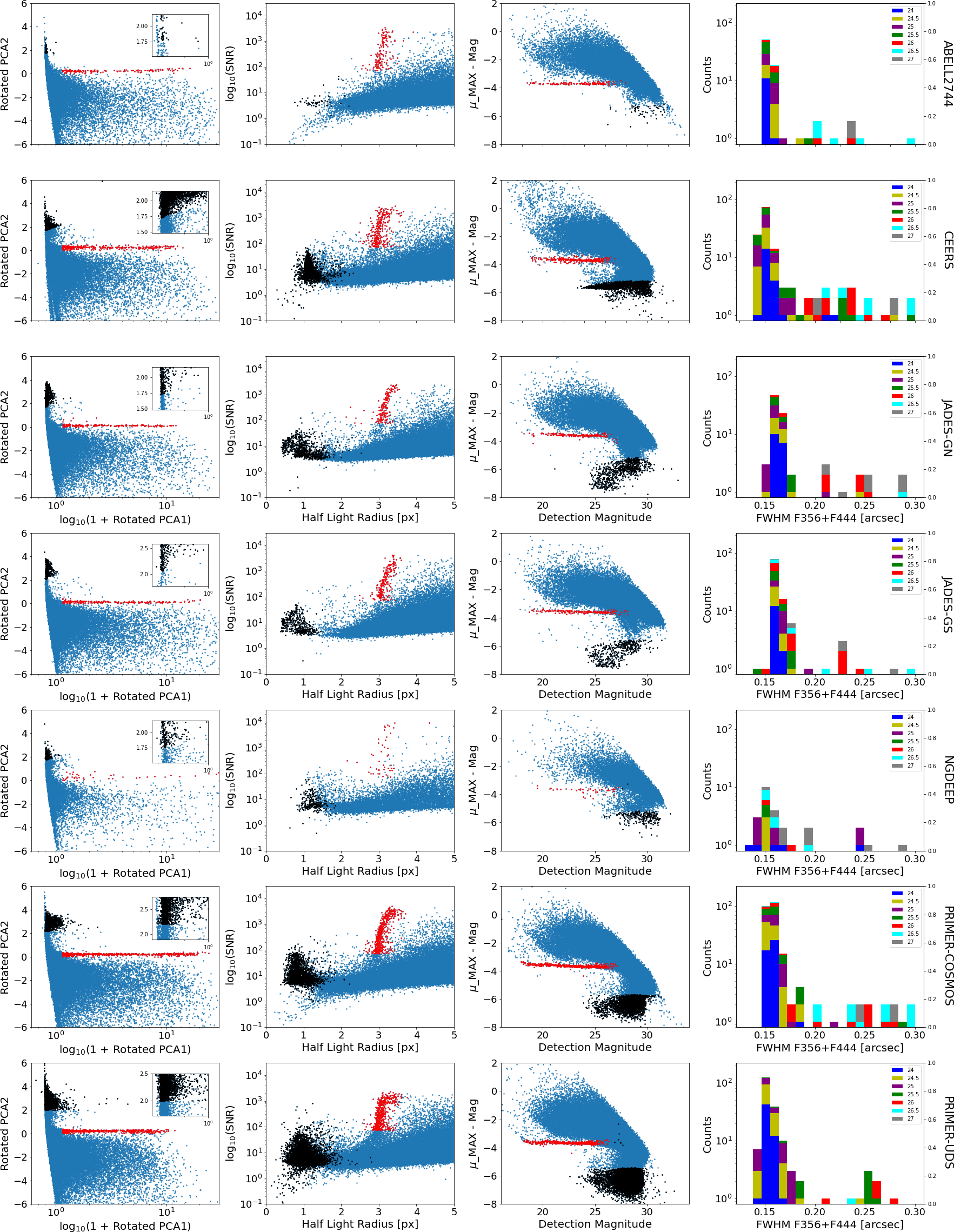}
\captionof{figure}{Star/galaxy separation and flagging of spurious sources. From left to right: as obtained in the PCA plane described in Section 3.5, projected on the hlr vs. S/N plane, projected on the $\mu$MAX-MAG vs. MAG plane. The last column shows the values of the FWHM in the detection F356W+F444W stack, for the sources identified as point-source.} \label{SGappa}
\end{figure*}

We point out that this approach joins two similar and complementary techniques using diagnostic diagrams to identify the loci occupied by point-like and spurious sources, i.e. the S/N vs radius plane and the $\mu$-mag vs mag plane. Fig. \ref{SGappa} shows the PCA plane for all fields in the left column, and the two diagrams corresponding to the other techniques in the central and right columns.

\begin{table}[H]
\caption{Constants used to single out stars and potentially spurious objects in the rotated PCA plane.}
\label{pca}
\centering
\begin{tabular}{lllll}
\hline\hline
Field & $c_{st}$ & $c_{sp1}$ & $c_{sp2}$ & $c_{sp3}$ \\ \hline
ABELL2744 & 0.5 & 0 & 1.75 & -0.2  \\ 
CEERS & 0.5 & 3 & 2 & -10  \\ 
JADES-GN & 0.8 & 3 & 2 & -10  \\ 
JADES-GS & 0.5 & 5 & 2.5 & -10  \\ 
NGDEEP & 0.8 & 3 & 2 & -10  \\ 
PRIMER-COSMOS & 0.5 & 0 & 2.2 & -0.2  \\ 
PRIMER-UDS & 0.5 & 0 & 2 & -10  \\ \hline 
\end{tabular}
\end{table}

% \section{Depth of mosaics} \label{depths}

% The histograms in Fig. \ref{depth} show the 5$\sigma$ limiting magnitudes of the mosaics in apertures of 0.2", computed from the re-scaled RMS maps, as described in Section 3.2.

% \begin{figure*}[hbt!]
% \centering
% \includegraphics[width=0.9\textwidth]{figs/depth_hists_withGN.png} 
% \caption{Histograms of the pixel distributions of limiting magnitudes (total at 5$\sigma$ in 0.2" diameter apertures), computed as described in Section 3.2, for all bands and fields.} \label{depth}
% \end{figure*}

\onecolumn
\section{Comparisons with other photometric catalogues} \label{photcomp}

\begin{figure*}
\centering
\includegraphics[width=0.9\textwidth]{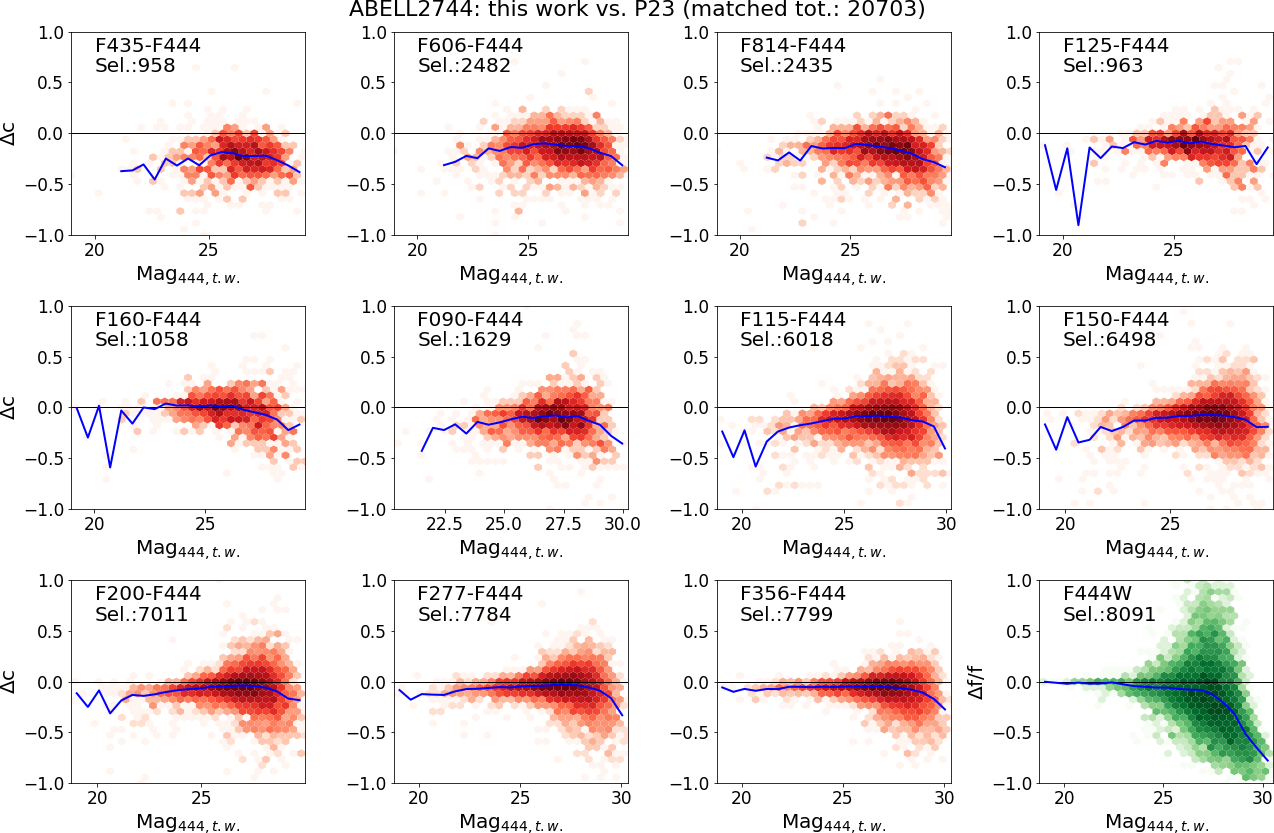} %\label{compflux1} 

\vspace{0.5cm}

\includegraphics[width=0.9\textwidth]{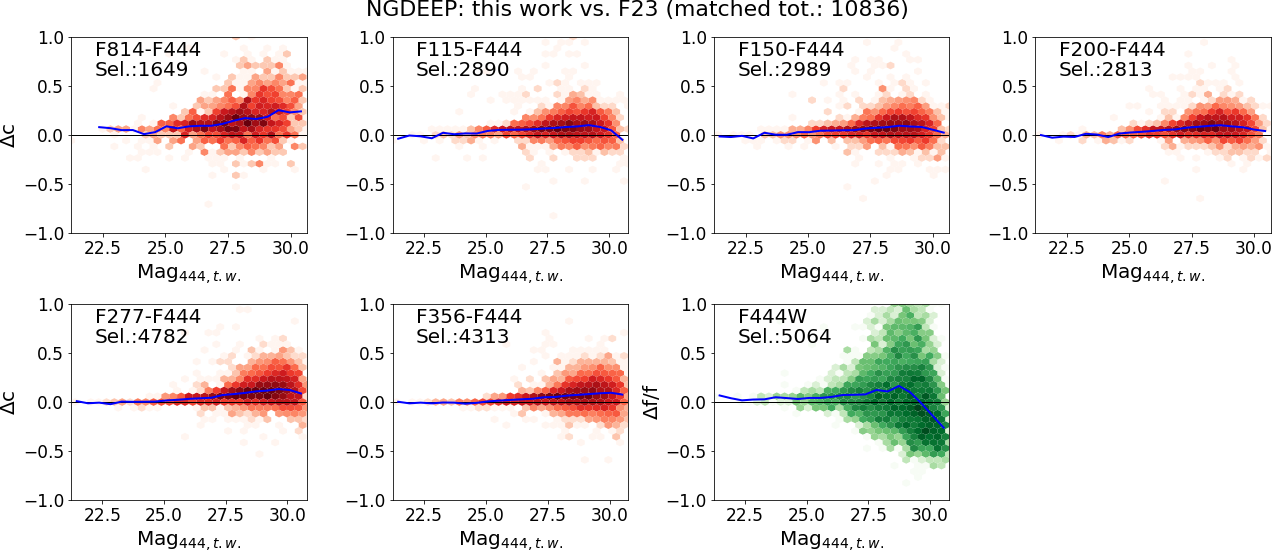} %\label{compflux5}
\caption{Comparison of colours measured in this work and archival catalogues. Shown are relative errors $\Delta c$, i.e. the colours measured in this work minus those in the reference catalogues), vs. the second band magnitude in this work catalogue (e.g. F444W for the F356-F444W colour). Top to bottom: ABELL2744 vs. P23 and NGDEEP vs. Finkelstein's catalogue (priv. comm.). The number of cross-matched sources after excluding those with  S/N $<5$ in any of the two catalogues or flag $\geq200$ in this work is also given in each panel; the blue line is the median of the distribution.} \label{compflux1}

\end{figure*}

\begin{figure*}
\centering
\includegraphics[width=0.9\textwidth]{figs/JWSTcomp_UNCOVER_colors2_V12r.png} %\label{compflux2}

\vspace{0.5cm}

\includegraphics[width=0.9\textwidth]{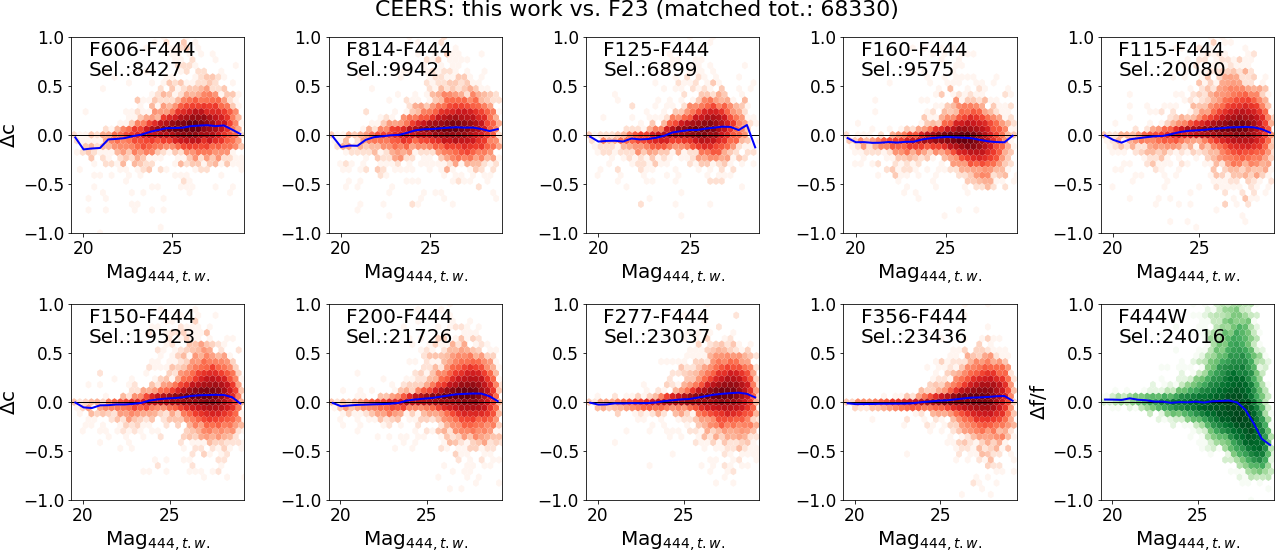} %\label{compflux3}
\caption{Same as Fig. \ref{compflux1}, for (top to bottom) ABELL2744 vs. \citet{Weaver2024} and CEERS vs. \citet{Finkelstein2023a}.} \label{compflux2}
%\end{minipage}%
\end{figure*}

\begin{figure*}
%\begin{minipage}{0.9\linewidth}
\centering
\includegraphics[width=0.9\textwidth]{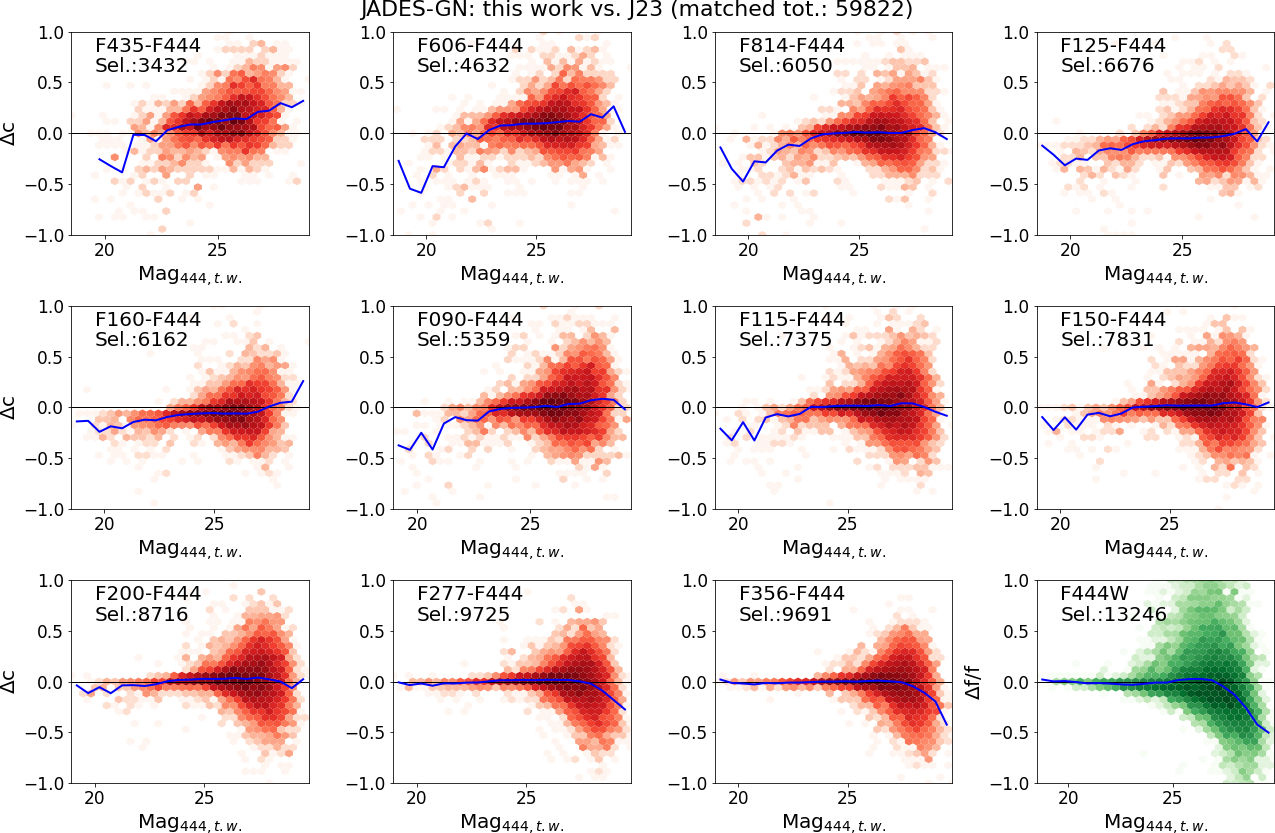} %\label{compflux4a}

\vspace{0.5cm}

\includegraphics[width=0.9\textwidth]{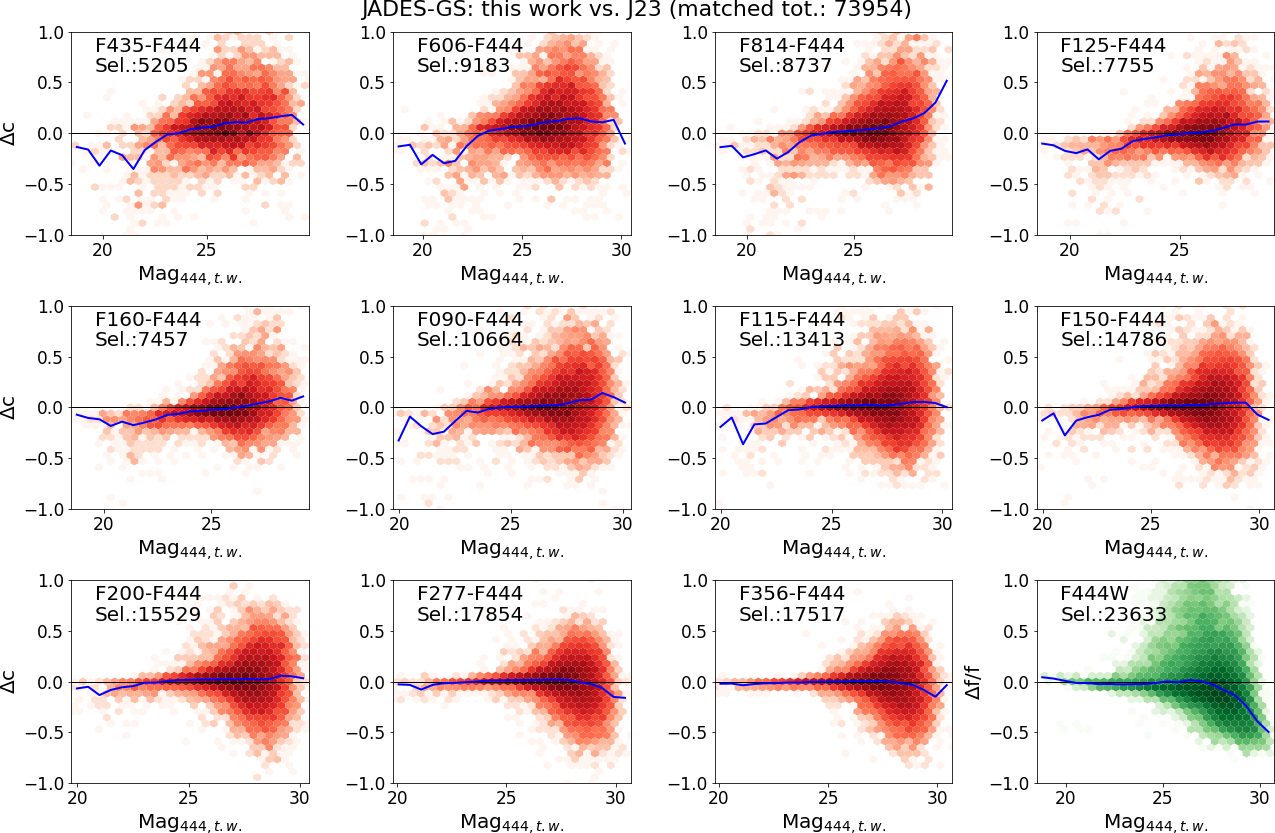} %\label{compflux4}

\caption{Same as Fig. \ref{compflux1}, for JADES-GN and JADES-GS vs. the JADES team catalogues \citep[]{Rieke2023}.}
\label{compflux3}
%\end{minipage}%
\end{figure*}

\begin{figure*}
%\begin{minipage}{0.9\linewidth}
\centering

\includegraphics[width=0.9\textwidth]{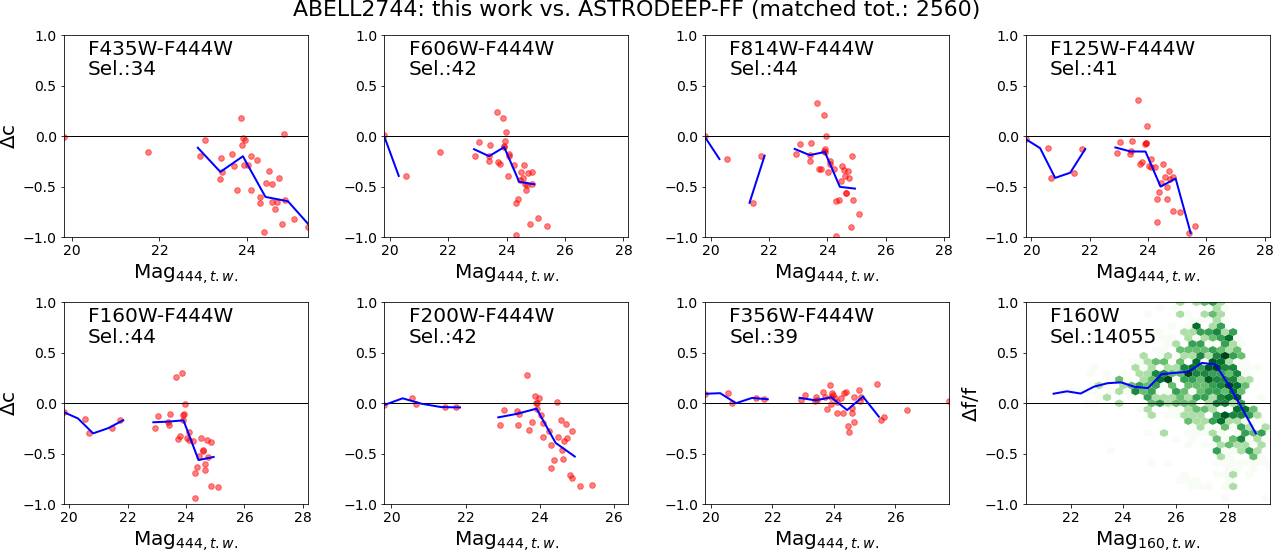} %\label{compflux6}

\vspace{0.5cm}

\includegraphics[width=0.9\textwidth]{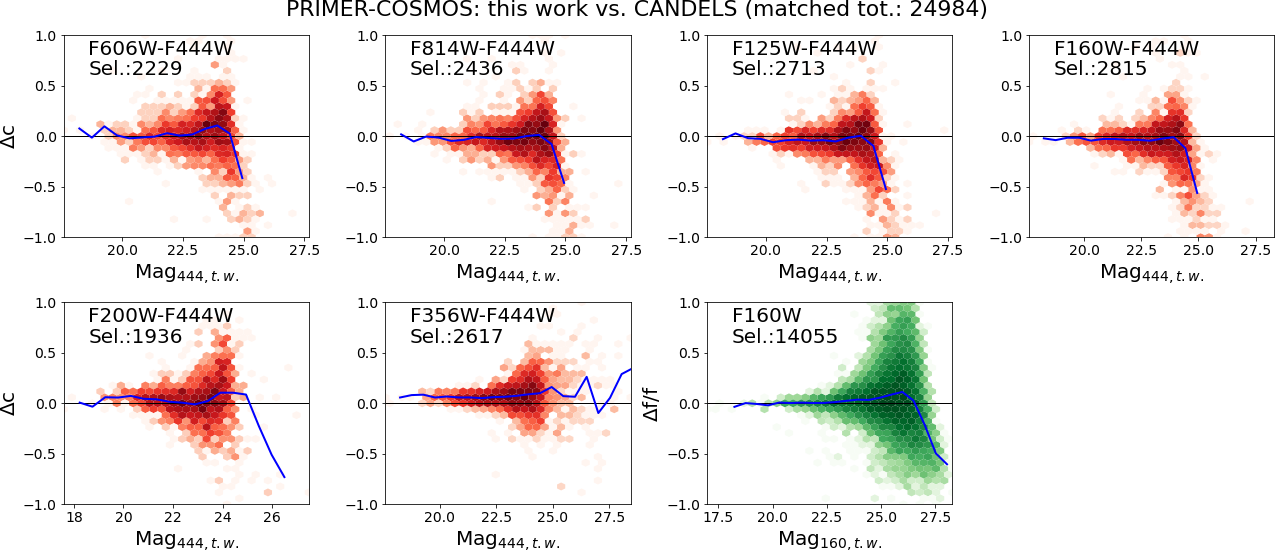} %\label{compflux8}

\caption{Same as Fig. \ref{compflux1}, for ABELL2744 vs. \citet{Merlin2016a} showing individual points rather than a density \texttt{hexbin} plot because of the small number of cross-matched sources, and PRIMER-COSMOS vs. \citet{Nayyeri2017}. The \textrm{HST} colours are transformed into \textrm{JWST} colours by means of the corrections described in Appendix \ref{photcomp}.} \label{compflux4}
%\end{minipage}%
\end{figure*}

\begin{figure*}
%\begin{minipage}{0.9\linewidth}
\centering

\includegraphics[width=0.9\textwidth]{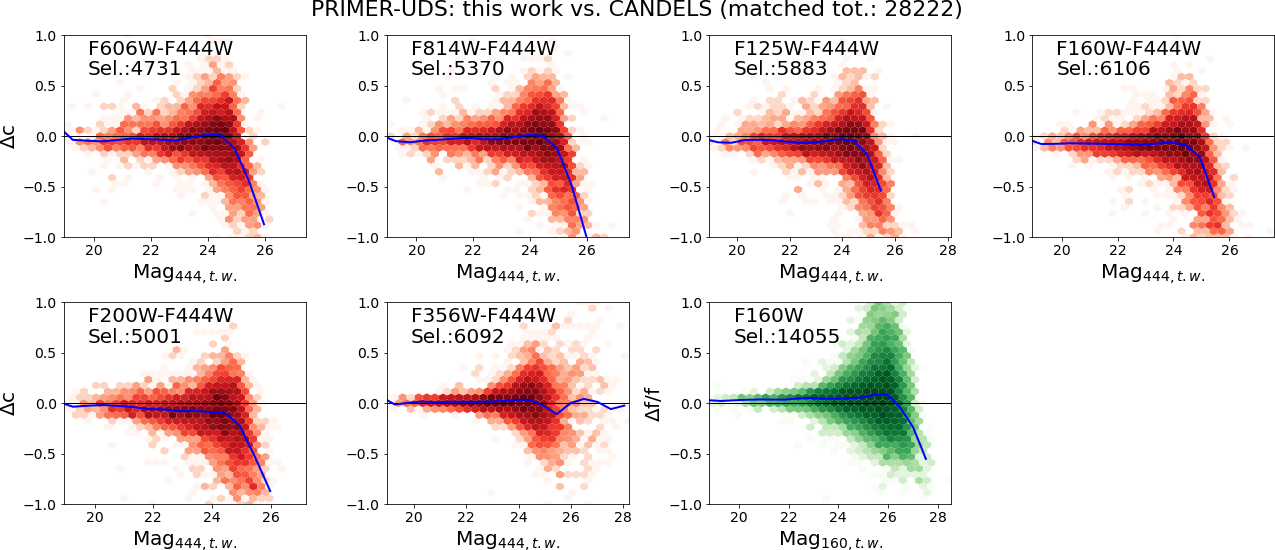} %\label{compflux8}

\vspace{0.5cm}

\includegraphics[width=0.9\textwidth]{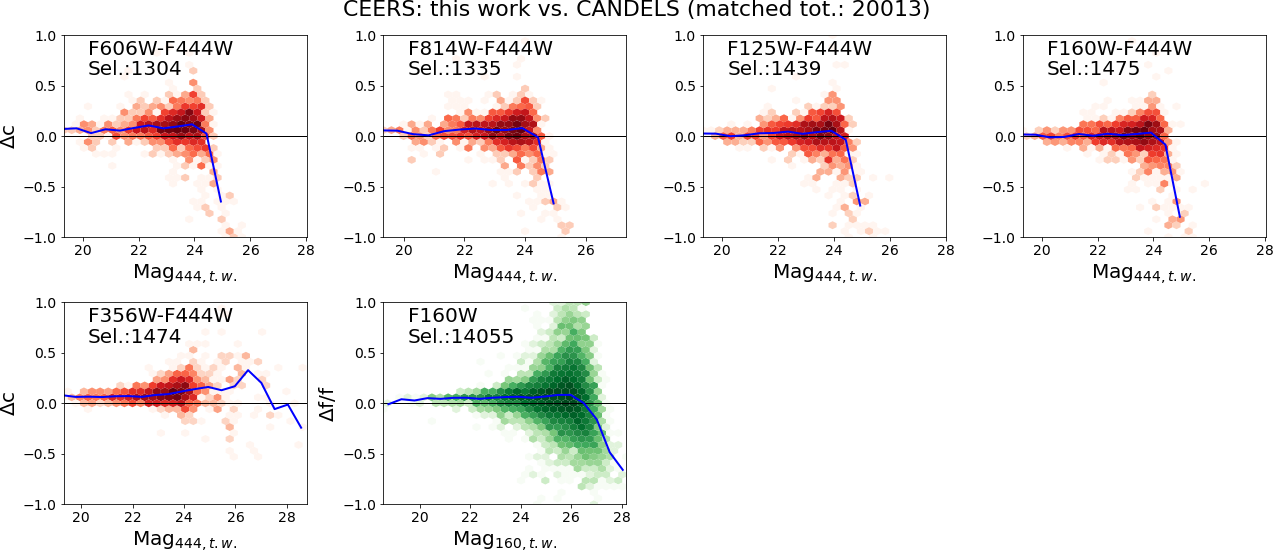} %\label{compflux9}

\caption{Same as Fig. \ref{compflux1}, for PRIMER-UDS vs. \citet{Galametz2013}, and CEERS vs. \citet{Stefanon2017}.} \label{compflux5}
%\end{minipage}%
\end{figure*}

\begin{figure*}
%\begin{minipage}{0.9\linewidth}
  \centering
\includegraphics[width=0.9\textwidth]{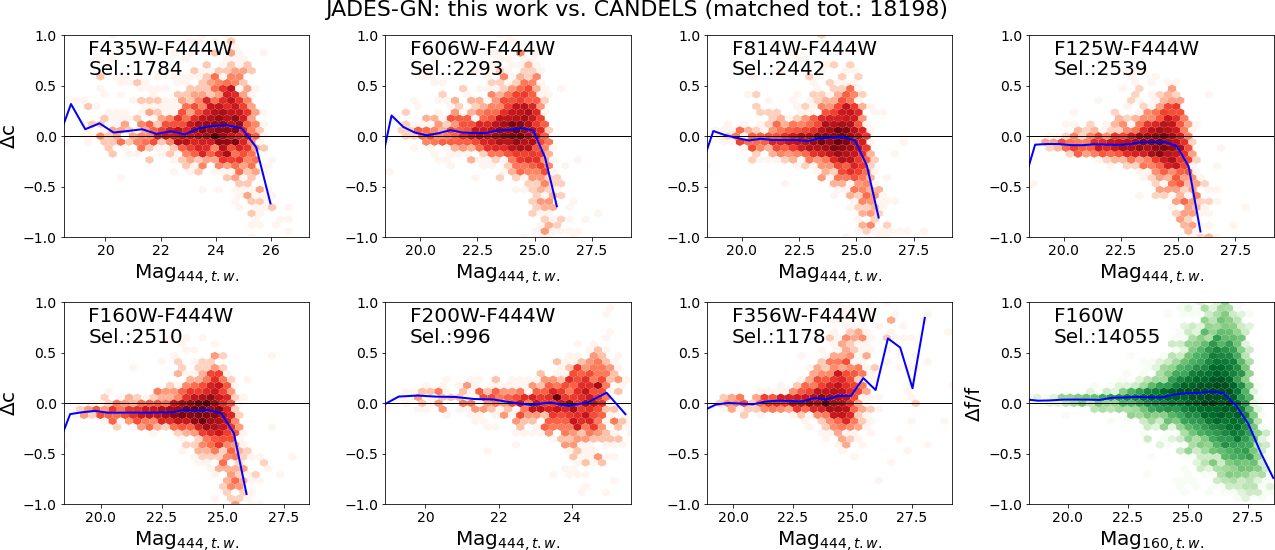} 

\vspace{0.5cm}

\includegraphics[width=0.9\textwidth]{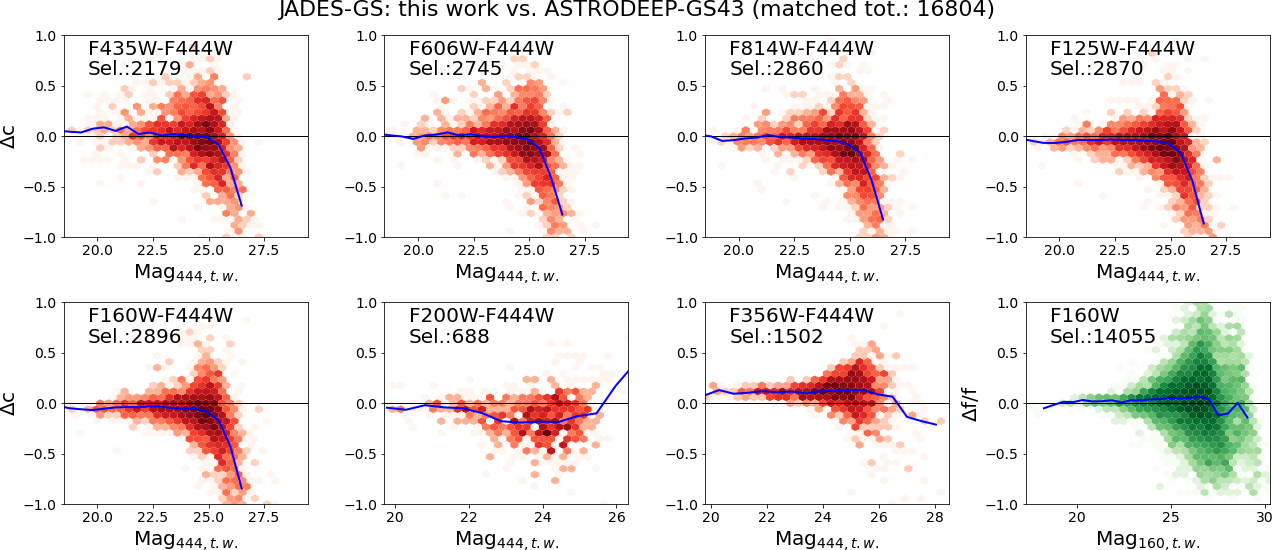} %\label{compflux10}

\vspace{0.5cm}

\includegraphics[width=0.9\textwidth]{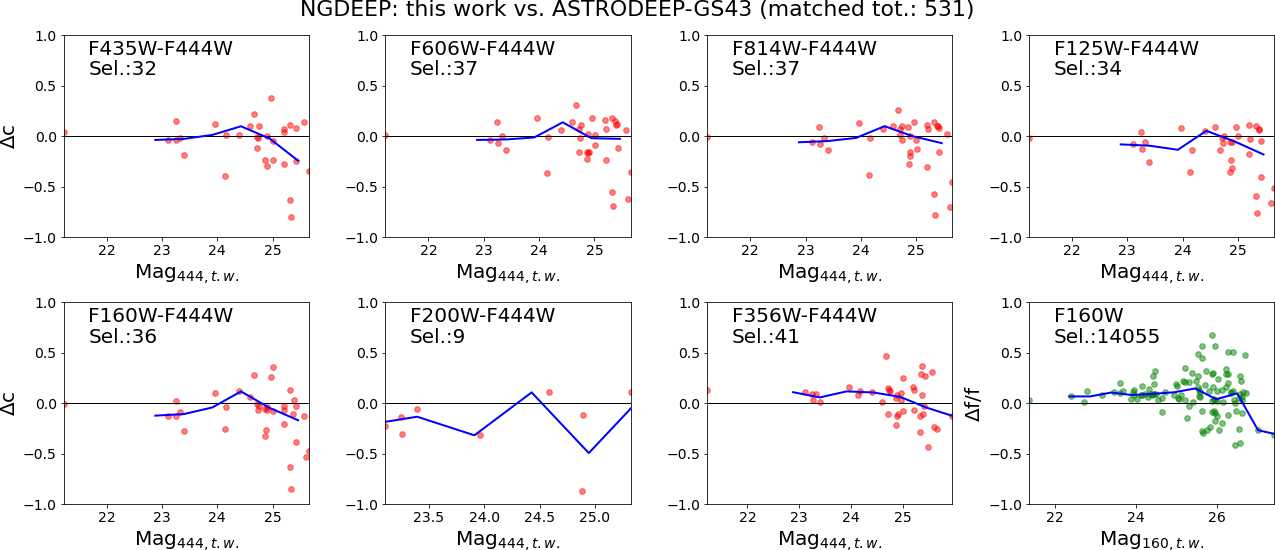} %\label{compflux11}

\caption{Same as Fig. \ref{compflux1}, JADES-GS vs. \citet{Merlin2021} and NGDEEP vs. \citet{Merlin2021}, the latter showing individual points rather than a density \texttt{hexbin} plot because of the small number of cross-matched sources. The \textrm{HST} colours are transformed into \textrm{JWST} colours by means of the corrections described in Appendix \ref{photcomp}.} \label{compflux6}
%\end{minipage}%
\end{figure*}

Fig. \ref{compflux1} to \ref{compflux6} show how the photometry obtained in this work compares to that from other available catalogues. For each field we consider recent NIRCam-based catalogues and archival \textrm{HST} based data, from CANDELS or \textsc{Astrodeep}. For the latter, we used the $Ks$ and IRAC CH1-2 bands as proxies for the F200W, F356W and F444W \textrm{JWST} bands, applying the following colour corrections, based on \citet{Bruzual2003} theoretical templates: $Ks$=F200W-0.06, IRAC1=F356W+0.02, IRAC2=F444W-0.01.

Each panel of the plots shows the comparison for one colour, with the relative difference $\Delta c/c$ (i.e. colour measured in this work minus colour in the reference work, divided by colour in the archival work) as a function of the magnitude of the second band in this work (e.g., F444W magnitude if the colour is F356W-F444W). The blue line is the median of the distribution. We consider sources with S/N$>$5 and flag$<$200. See Section 4 for more details.

%HST

\onecolumn

\section{Photometric redshift validation}

In Fig. \ref{zspec} we show the comparison between the spectroscopic and photometric redshifts, the latter computed as the median of the three runs with (i) \textsc{zphot}, (ii) \textsc{EAzY} with \texttt{eazy\_v1.3}, and (iii) \textsc{EAzY} with \citet{Larson2023} FSPS Set 1 + Set 4 templates, as described in Section 5. 
Table \ref{photozcomp} summarises the statistics for all four runs. 

\begin{figure*}
\centering
\includegraphics[width=0.95\textwidth]{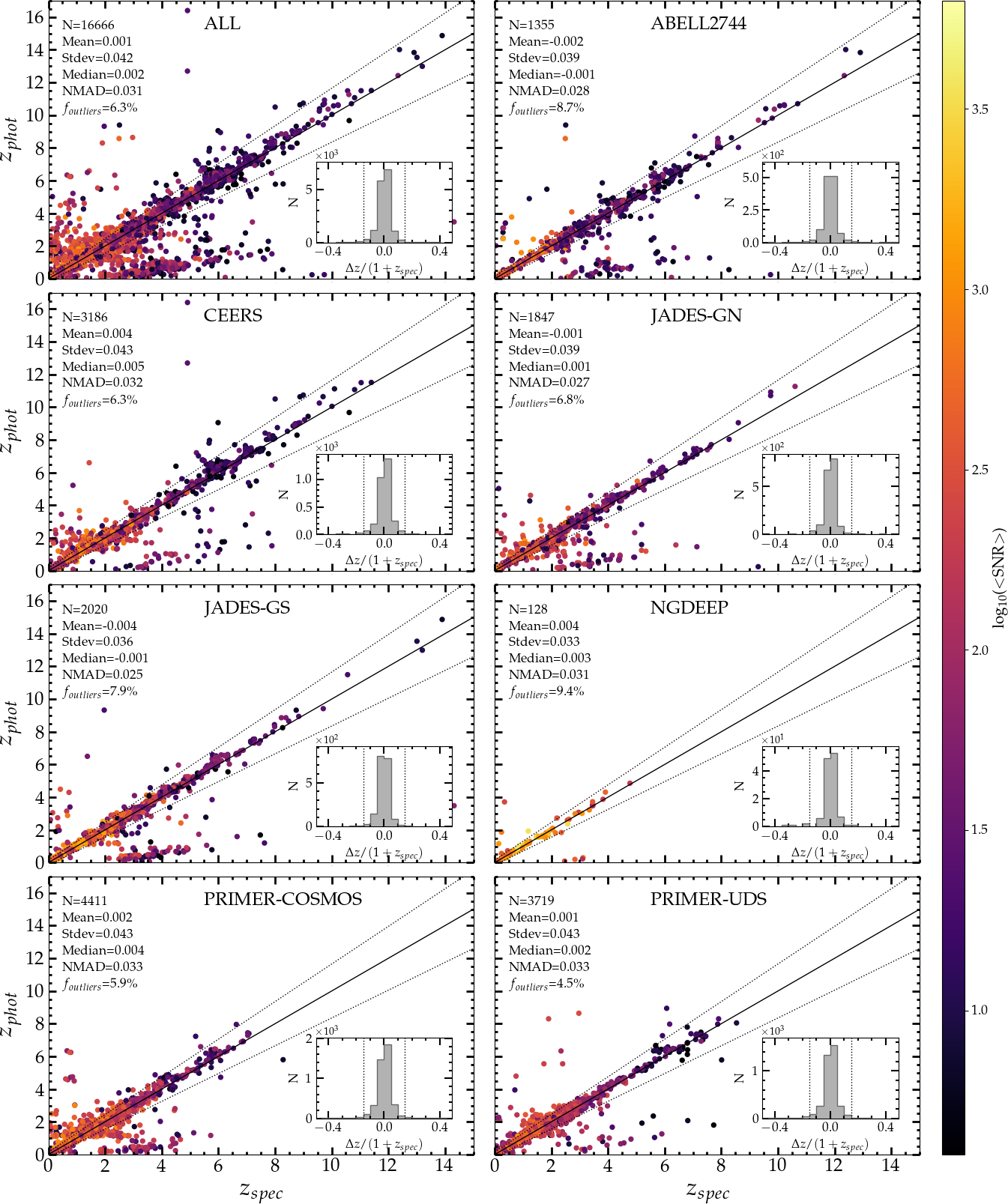}
\caption{Accuracy of photometric redshifts (median of the three runs described in Section 5) on spectroscopic samples, for all fields together (top left panel), and separately for each field. In each panel, N is the number of sources; $f_{outliers}$ is the percentage of sources with $|z_{phot}-z_{spec}|/(1+z_{spec})>0.15$; the other statistics are computed on non-outliers, with NMAD=$1.48\times \mbox{median}[|z_{phot}-z_{spec}|/(1+z_{spec})]$. colour coding is $<\mbox{S/N}>=(\mbox{S/N}_{\mbox{F444W}}+\mbox{S/N}_{\mbox{F356W}}+\mbox{S/N}_{\mbox{F277W}}+\mbox{S/N}_{\mbox{F200W}})/4$.}\label{zspec}
\end{figure*}

\begin{table*}[hbt!]
\caption{Comparative statistics for photometric vs. spectroscopic redshifts.} %The rows named ``median'' show the values obtained taking for each source the median of the four runs as the estimate of the photometric redshift.}
\label{photozcomp}
\centering
\begin{tabular}{lcccccc}
\hline\hline
 & N & 100$\times$mean & 100$\times$stdev & 100$\times$median & 100$\times$NMAD & $\eta$ \\ \hline 
\multicolumn{7}{c}{ABELL2744} \\ 
\textsc{zphot} & 1355 & 0.57 & 4.28 & 0.47 & 3.33 & 10.04 \\
\textsc{EAzY} v1.3 & 1354 & -0.81 & 4.31 & -0.46 & 3.36 & 13.81 \\ 
\textsc{EAzY} Larson Lya & 1338 & -0.15 & 4.64 & -0.28 & 3.29 & 13.75 \\  
\textsc{EAzY} Larson LyaRed & 1338 & -0.32 & 4.70 & -0.33 & 3.46 & 14.28 \\ 
\multicolumn{7}{c}{CEERS} \\  
\textsc{zphot} & 3186 & 1.17 & 4.63 & 0.89 & 3.64 & 7.97 \\  
\textsc{EAzY} v1.3 & 3182 & 0.38 & 4.55 & 0.41 & 3.59 & 7.54 \\ 
\textsc{EAzY} Larson Lya & 3180 & -0.40 & 4.96 & -0.26 & 4.04 & 9.03 \\ 
\textsc{EAzY} Larson LyaRed & 3180 & -0.41 & 4.97 & -0.30 & 4.08 & 9.18 \\  
\multicolumn{7}{c}{JADES-GN} \\ 
\textsc{zphot} & 1848 & 0.81 & 4.25 & 0.75 & 3.47 & 6.71 \\ 
\textsc{EAzY} v1.3 & 1846 & -0.24 & 4.30 & 0.01 & 2.98 & 8.88 \\ 
\textsc{EAzY} Larson Lya & 1841 & -0.74 & 4.08 & -0.53 & 2.89 & 8.53 \\ 
\textsc{EAzY} Larson LyaRed & 1841 & -0.77 & 4.10 & -0.55 & 2.93 & 8.75 \\  
\multicolumn{7}{c}{JADES-GS} \\ 
\textsc{zphot} & 2021 & 0.41 & 4.01 & 0.32 & 3.02 & 5.79 \\ 
\textsc{EAzY} v1.3 & 2019 & -0.57 & 4.05 & -0.35 & 2.91 & 10.10 \\ 
\textsc{EAzY} Larson Lya & 2017 & -0.75 & 3.80 & -0.47 & 2.81 & 9.92 \\ 
\textsc{EAzY} Larson LyaRed & 2017 & -0.85 & 3.83 & -0.54 & 2.84 & 10.21 \\ 
\multicolumn{7}{c}{NGDEEP} \\
\textsc{zphot} & 128 & 1.46 & 4.08 & 0.92 & 3.10 & 10.16 \\ 
\textsc{EAzY} v1.3 & 128 & -0.10 & 4.26 & 0.21 & 3.62 & 9.38 \\ 
\textsc{EAzY} Larson Lya & 128 & -0.29 & 4.16 & -0.14 & 3.76 & 15.62 \\ 
\textsc{EAzY} Larson LyaRed & 128 & -0.29 & 4.18 & -0.10 & 3.80 & 16.41 \\ 
\multicolumn{7}{c}{PRIMER-COSMOS} \\ 
\textsc{zphot} & 4413 & 1.37 & 4.61 & 1.32 & 4.15 & 6.80 \\ 
\textsc{EAzY} v1.3 & 4408 & -0.03 & 4.59 & 0.10 & 3.64 & 7.17 \\  
\textsc{EAzY} Larson Lya & 4399 & -0.63 & 4.69 & -0.36 & 3.85 & 8.98 \\ 
\textsc{EAzY} Larson LyaRed & 4399 & -0.65 & 4.71 & -0.37 & 3.89 & 9.09 \\ 
\multicolumn{7}{c}{PRIMER-UDS} \\ 
\textsc{zphot} & 3721 & 1.03 & 4.50 & 0.95 & 3.80 & 5.80 \\ 
\textsc{EAzY} v1.3 & 3719 & -0.05 & 4.55 & 0.09 & 3.65 & 5.22 \\ 
\textsc{EAzY} Larson Lya & 3713 & -0.64 & 4.86 & -0.43 & 3.85 & 7.41 \\ 
\textsc{EAzY} Larson LyaRed & 3713 & -0.65 & 4.86 & -0.46 & 3.87 & 7.57 \\ \hline
\end{tabular}
\tablefoot{ The listed values are: number N of matched sources (the number can vary due to failings in the fitting procedures of the different codes); mean, standard deviation, median and NMAD of the quantity $|z_{phot}-z_{spec}|/(1+z_{spec})$, all multiplied by 100 to make the differences clearer; and percentage of outliers $f_{outliers}$ as described in Section 5, for each run using different software and/or templates, and for each field.}
\end{table*}

\end{appendix}

\end{document}